\documentclass{jfm}
\usepackage{hyperref}
\usepackage{graphicx}
\usepackage{epstopdf, epsfig}
\usepackage{amsmath,bm}
\usepackage{subcaption}

\usepackage{siunitx}
\usepackage{booktabs}
\usepackage[ruled,vlined]{algorithm2e}
\usepackage{esint}
\usepackage{cleveref}
\usepackage{soul}
\usepackage{xcolor}
\usepackage{ltablex}

\definecolor{revision}{rgb}{1, 0, 0}
\definecolor{ForestGreen}{RGB}{34,139,34}

\usepackage[commentmarkup=footnote]{changes}
\definechangesauthor[name=Lusseyran, color=orange]{FL}
\definechangesauthor[name=Noack, color=red]{BN}

\crefname{section}{§}{§§}
\Crefname{section}{§}{§§}

\shorttitle{Stabilization with gradient-enriched machine learning control}
\shortauthor{G. Y. Cornejo Maceda, Y. Li, F. Lusseyran, M. Morzy\'{n}ski and B. R. Noack}

\title{Stabilization of the fluidic pinball with gradient-enriched machine learning control}

\author{Guy Y. Cornejo Maceda\aff{1},
  Yiqing Li\aff{2},
  Fran\c{c}ois Lusseyran\aff{1},
  Marek Morzy\'{n}ski\aff{3}
  \and Bernd R. Noack\aff{2,}\aff{4}\corresp{\email{bernd.noack@hit.edu.cn}}}

\affiliation{
\aff{1} Universit\'{e} Paris-Saclay, CNRS, Laboratoire Interdisciplinaire des Sciences du Num\'{e}rique, 91400, Orsay, France.
\aff{2} Center for Turbulence Control,
Harbin Institute of Technology (Shenzhen), Room 312, Building C,
University Town, Xili,
Shenzhen 518058, People’s Republic of China.
\aff{3} Department of Virtual Engineering, Pozna\'{n} University of Technology,
Jana Pawla II 24, PL 60-965 Pozna\'{n}, Poland.
\aff{4} Institut f\"{u}r Str\"{o}mungsmechanik und Technische Akustik (ISTA),
Technische Universit\"{a}t Berlin, M\"{u}ller-Breslau-Straße 8, D-10623 Berlin, Germany.}

\begin{document}
%Term definitions
%\newacronym{mlc}{MLC}{$\>$ Machine Learning Control}
%\newacronym{gmlc}{gMLC}{\quad Gradient-enriched Machine Learning Control}

\maketitle
%\glsunsetall

\begin{abstract}
We stabilize the flow past a cluster 
of three rotating cylinders---the fluidic pinball---with automated gradient-enriched machine learning algorithms.
The control laws command the rotation speed of each cylinder
in an open- and closed-loop manner.
These laws are optimized with respect to the average distance 
from the target steady solution
in three successively richer search spaces. 
First, stabilization is pursued with steady symmetric forcing.
Second, we allow for asymmetric steady forcing. 
And third, we determine an optimal feedback controller
employing nine velocity probes downstream.
As expected, the control performance increases 
with every generalization of the search space. 
Surprisingly, both open- and closed-loop optimal controllers 
include an asymmetric forcing, 
which surpasses symmetric forcing.
Intriguingly, the best performance is achieved 
by a combination of phasor control and asymmetric steady forcing.
We hypothesize that asymmetric forcing is typical for pitchfork bifurcated dynamics 
of nominally symmetric configurations.
Key enablers are automated machine learning algorithms augmented with gradient search: 
explorative gradient method for the open-loop parameter optimization and 
a gradient-enriched machine learning control (gMLC) for the feedback optimization.
gMLC learns the control law significantly faster 
than previously employed genetic programming control.
The gMLC source code is freely available online.
\end{abstract}

%------------------------------------------------------------------------
%\listofchanges
%------------------------------------------------------------------------
%\tableofcontents

\section{Introduction}\label{sec:introduction}

We stabilize the wake behind a fluidic pinball
using a hierarchy of model-free self-learning control methods
from a one-parametric study of open-loop control
to a gradient-enriched machine learning feedback control.
Flow control is at the heart of many engineering applications.
Traffic alone profits from flow control
via drag reduction of transport vehicles \citep{Choi2008arfm}, 
lift increase of wings \citep{Semaan2016jfm},
mixing control for more efficient combustion \citep{Dowling2005arfm}, 
and noise reduction \citep{Jordan2013arfm}.

The control logic is a critical component for performance increases
after the actuators and sensors have been deployed.
The hardware is typically determined from engineering wisdom \citep{Cattafesta2011arfm}.
The control law may be designed with a rich arsenal of mathematical methods. 
Control theory offers powerful methods for control design 
with large success for model-based stabilization of low-Reynolds number flows or simple first and second order dynamics \citep{Rowley2006arfm}.
Transport-related engineering applications are at high Reynolds numbers 
and thus associated with turbulent flows.
So far, turbulence has eluded most attempts for model-based control albeit for few simple exceptions \citep{Brunton2015amr}.
Examples relate to first and second order dynamics,
e.g., the quasi-steady response to quasi-steady actuation \citep{Pfeiffer2012aiaa}, 
opposition control near walls \citep{Choi1994jfm,Fukagata2003}, 
stabilizing phasor control of oscillations \citep{Pastoor2008jfm}, 
and  two-frequency crosstalk \citep{Glezer2005aiaaj,Luchtenburg2009jfm}.
In general, control design is challenged by the high-dimensionality of the dynamics,
the nonlinearity with many frequency crosstalk mechanisms,
and the large time-delay between actuation and sensing.

Hence, most closed-loop control studies of turbulence 
resort to a model-free approach.
A simple example is extremum seeking \citep{Gelbert2012} for online tuning of one or few actuation parameters,
like amplitude and frequency of periodic actuation.
More complex examples involve high-dimensional parameter optimization
with methods of machine learning,
like evolutionary strategies \citep{Koumoutsakos2001aiaaj}
and genetic algorithms \citep{Benard2016ef}.
Even regression problems for nonlinear feedback laws
have been learned by genetic programming \citep{Ren2020jh} 
and reinforcement learning \citep{Rabault2019jfm}.

Genetic programming control (GPC) has been pioneered by \citet{Dracopoulos1997book} over 20 years ago
and has been proven to be particularly successful for nonlinear feedback turbulence control in  experiments.
Examples include the drag reduction of the Ahmed body \citep{Li2018am} and  the same obstacle under yaw angle \citep{Li2019prf},
mixing layer control \citep{Parezanovic2016jfm},
separation control of a turbulent boundary layer \citep{Debien2016ef},
recirculation zone reduction behind a backward facing step \citep{Gautier2015jfm},
and  jet mixing enhancement \citep{Zhou2020jfm}, just to name a few.
GPC has consistently outperformed existing optimized control approaches,
often with unexpected frequency crosstalk mechanisms \citep{Noack2019springer}.
GPC has a  powerful capability to find new mechanisms (exploration)
and populate the best minima (exploitation).
Yet, the exploitation is inefficient leading 
to increasing redundant testing of similar control laws 
 with poor convergence to the minimum.
This challenge is well known and will be addressed in this study.

As benchmark control problem, 
we chose the fluidic pinball,
the flow around three parallel cylinders one radius apart from each other
\citep{Noack2016jfm,Deng2020jfm,ChenAlam2020jfm}.
The triangle of centers points in the direction of the flow.
The actuation is performed by rotating each cylinder independently.
The flow is monitored by 9 velocity probes downstream.
The control goal is the complete stabilization of the unstable symmetric steady Navier-Stokes solution.
This choice is motivated by several reasons.
First, already the unforced fluidic pinball shows a surprisingly rich dynamics.
With increasing Reynolds number 
the steady wake becomes successively
unstable in a Hopf bifurcation, 
a pitchfork bifurcation,
another Hopf bifurcation
before, eventually, a chaotic state is reached.
Second, the cylinder rotations may encapsulate the most common wake stabilization approaches,
like Coanda forcing \citep{Geropp2000ef},
base bleed \citep{Wood1964jras,Bearman1967aq},
low-frequency forcing \citep{Pastoor2008jfm},
high-frequency forcing \citep{Thiria2006jfm},
phasor control \citep{Roussopoulos1993jfm}, 
and circulation control \citep{Cortelezzi1994jfm}.
Third, 
the rich unforced and controlled dynamics mimics nonlinear behaviour of turbulence
while the computation of the two-dimensional flow is manageable on workstations.
To summarize, the fluidic pinball is an attractive all-weather plant for non-trivial multiple-input multiple-output control dynamics.

This study focuses on the stabilization of the unstable symmetric steady solution of the fluidic pinball 
in the pitchfork regime, i.e., for asymmetric vortex shedding.
This goal is pursued under  symmetric steady actuation,
general non-symmetric steady actuation and general nonlinear feedback control.
We aim to physically explore the actuation mechanisms in a rich search space
and to efficiently exploit the performance gains from gradient-based approaches.
This multi-objective optimization leads to innovations of hitherto employed 
parameter optimizations and regression solvers as a beneficial side effect.

The manuscript is organized as follows.
\S~\ref{Sec:Plant} introduces the fluidic pinball problem 
and the corresponding direct numerical simulation.
 \S~\ref{Sec:Methods} reviews and augments machine learning control strategies.
In \S~\ref{Sec:Results}, a hierarchy of increasingly more complex control laws is optimized for wake stabilization.
\S~\ref{Sec:Discussion} discusses design aspects of the proposed methodology.
\S~\ref{Sec:Conclusions} summarizes the results and indicates directions for future research.
Table~\ref{tab:acronyms} lists all the acronyms used in the manuscript.

\begin{table}
\begin{center}
\def~{\hphantom{0}}
\begin{tabular}{p{1.75cm}p{8cm}}
%  \hline
\textbf{EGM} & Explorative Gradient Method\\
\textbf{gMLC} & Gradient-enriched Machine Learning Control\\
\textbf{GPC} & Genetic Programming Control\\
\textbf{LGP} & Linear Genetic Programming\\
\textbf{LHS} & Latin Hypercube Sampling\\
\textbf{MC} & Monte Carlo\\
\textbf{MIMO} & Multiple-Input Multiple-Output\\
\textbf{MLC} & Machine Learning Control\\
\textbf{PSD} & Power Spectral Density\\
\end{tabular}
\caption{\label{tab:acronyms}Table of acronyms.}
\end{center}
\end{table}
 %-------------------------------------------------------------------------------------------------------------------------------------------------------------

%----------------------------------------------------------------------

\section{The fluidic pinball---A benchmark flow control problem}
\label{Sec:Plant}
In this section, 
we describe the fluid system studied for the control optimization---the fluidic pinball.
First we present the fluidic pinball configuration and the unsteady 2D Navier-Stokes solver in \S~\ref{sec:config_numerical_solver}, 
then the unforced flow spatio-temporal dynamics in \S~\ref{sec:flow_characteristics} 
and finally the control problem for the fluidic pinball in \S~\ref{sec:control_objective}.

\subsection{Configuration and numerical solver} 
\label{sec:config_numerical_solver}
The test case is a two-dimensional uniform flow past a cluster of three cylinders of same diameter $D$.
The center of the cylinders form an equilateral triangle pointing upstream.
The flow is controlled by the independent rotation of the cylinders along their axis.
The rotation of the cylinders enables the steering of incoming fluid particles, 
like a pinball machine.
Thus, we refer this configuration as the fluidic pinball.
%Over the past decades, the flow over a cluster of three cylinders has been experimentally studied in regards of heat transfer, fluid-structure interactions and multiple frequencies interactions (\ref{Price1984,Sayers1987,Lam1988,Tatsuno1998,Bansal2017}).
In our study, we choose the side length of the equilateral triangle equal to be $1.5D$.
The distance of one radius gives rise to an interesting 
flip-flopping dynamics \citep{ChenAlam2020jfm}.

The flow is described in a Cartesian coordinate system, 
where the origin is located midway between the two rearward cylinders. 
The $x$-axis is parallel to the streamwise direction and 
the $y$-axis is orthogonal to the cylinder axis.
The velocity field is denoted by $\bm{u}=(u,v)$ and the pressure field by $p$. 
Here, $u$ and $v$ are, respectively, the streamwise and transverse components of the velocity.
We consider a Newtonian fluid of constant density $\rho$ and kinematic viscosity $\nu$.
For the direct numerical simulation, 
the unsteady incompressible viscous Navier-Stokes equations 
are non-dimensionalized with cylinder diameter $D$, 
the incoming velocity $U_{\infty}$ and the fluid density $\rho$.
The corresponding Reynolds number is $\Rey_{D} = U_{\infty}D\slash \nu$.
Throughout this study, only $\Rey_D =100$ is considered.

The computational domain $\Omega$ is a rectangle bounded 
by $[-6,20]\times[-6,6]$ and excludes the interior of the cylinders:
\begin{equation*}
\Omega = \{[x,y]^\intercal \in \mathcal{R}^2 \colon [x,y]^\intercal \in [-6,20]\times[-6,6] \land (x-x_i)^2+(y-y_i)^2 \geq 1/4, i=1, 2, 3 \}.
\end{equation*}
Here, 
$[x_i,y_i]^\intercal$ with $i=1,2,3$, are the coordinates of the cylinder centers, 
starting from the front cylinder and numbered in mathematically positive direction,
\begin{equation*}	
\begin{array}{clccl}
x_1 = & -3/2\cos(30^{\circ}) & &y_1= & ~0,\\
x_2 = & 0 & &y_2= & -3/4,\\
x_3 = & 0 & &y_3= & \quad 3/4.
\end{array}
\end{equation*}
The computational domain $\Omega$ is discretized on an unstructured grid comprising 4225 triangles and 8633 nodes.
The grid is optimized to provide a balance between computation speed and accuracy.
Grid independence of the direct Navier-Stokes solutions
has been established by \citet{Deng2020jfm}.

The boundary conditions for the inflow, upper and lower boundaries are $U_{\infty}=\bm{e}_x$ while a stress-free condition is assumed for the outflow boundary.
The control of the fluidic pinball is carried out by the rotation of the cylinders.
A non-slip condition is adopted on the cylinders: 
the flow adopts the circumferential velocities of the front, bottom and top cylinder
specified by  $b_1 = U_F$, $b_2 = U_B$ and $b_3 = U_T$.
The actuation command comprises these velocities, $\bm{b} = [b_1,b_2,b_3]^\intercal$.
A positive (negative) value of the actuation command 
corresponds to counter-clockwise (clockwise) rotation of the cylinders along their axis.
%A parametric study showed that our solver fails to converge for rotation speeds greater than 6 in absolute value.
%Therefore an upper and lower limit of 5 has been set to the rotation speed.
The numerical integration of the Navier-Stokes equations is carried by an in-house solver using a fully implicit Finite-Element Method.
The time integration is performed with an iterative Newton-Raphson-like approach.
The chosen time step of 0.1 corresponds to  about 1\% of the characteristic shedding period.
The method is  third order accurate in time and space and employs a pseudo-pressure formulation. The solver has been employed in recent fluidic pinball investigations
for reduced-order modeling \citep{Deng2020jfm,Noack2016jfm} and for control \citep{Ishar2019jfm}.
We refer to \citet{Noack2003jfm,Noack2016jfm} for further information on the numerical method.
The code is accessible on GitLab on email request.

%% Figure : Snapshots---------------------------------------------------
\begin{figure}
\centering
\begin{subfigure}{.45\textwidth}
  \centering
\includegraphics[width=1.\textwidth]{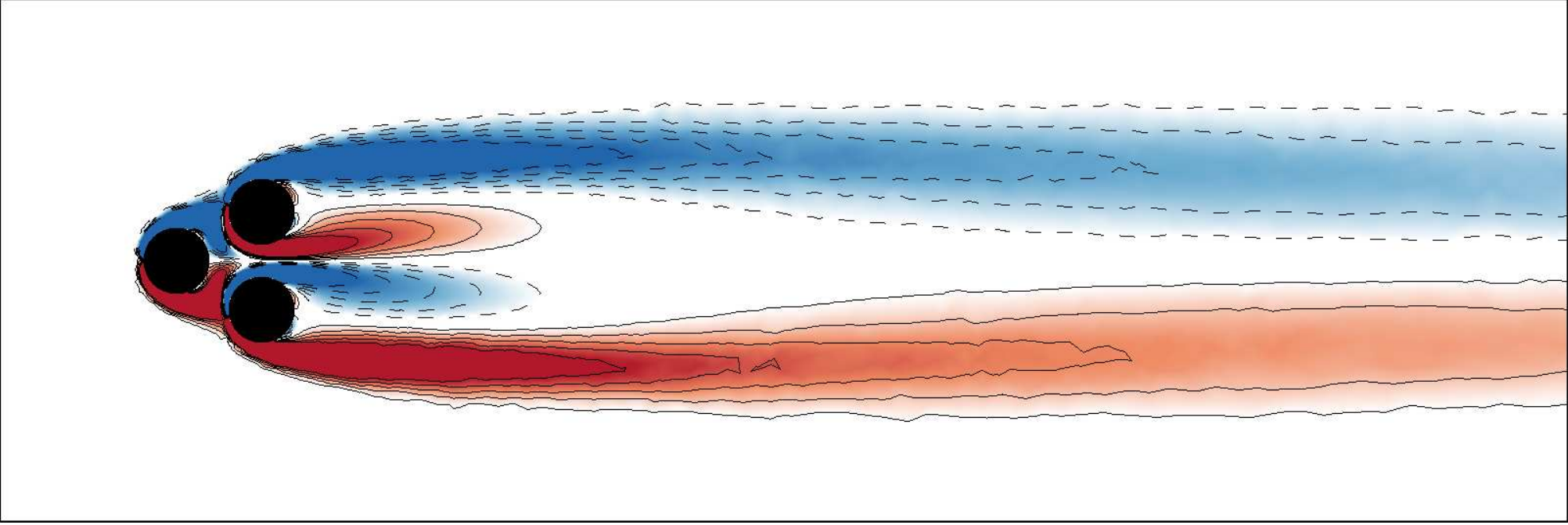}
\caption{\label{fig:steady_solution} Symmetric steady solution.}
\end{subfigure}%
\hfil
\begin{subfigure}{.45\textwidth}
  \centering
\includegraphics[width=1.\textwidth]{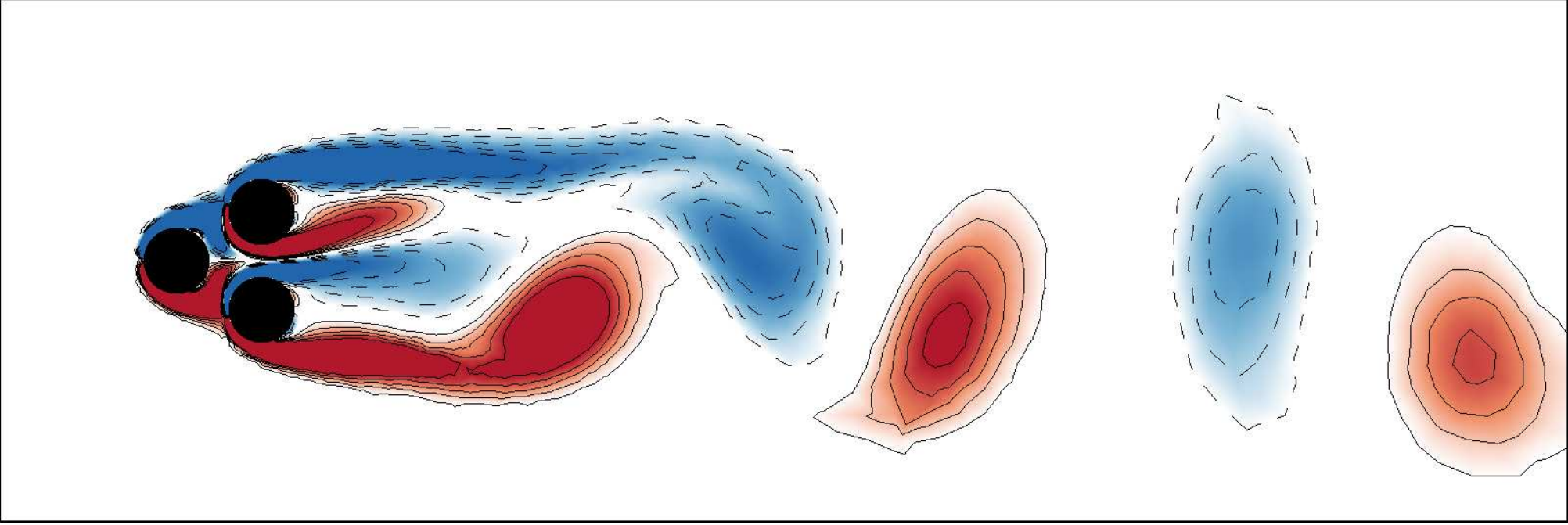}
\caption{\label{fig:natural_flow}Unforced flow at $t=400$.}
\end{subfigure}
\caption{\label{fig:unforced_flow}Vorticity fields for the unforced fluidic pinball at $\Rey_D=100$. 
Blue (red) regions bounded by dashed lines represent negative (positive) vorticity. 
Darker regions indicate higher values of vorticity magnitude.}
\end{figure}
%-----------------------------------------------------------------------
The initial condition for the numerical simulations is the symmetric steady solution.
The symmetrical steady solution is computed with a Newton-Raphson method on the steady Navier-Stokes.
An initial short and small rotation of the front cylinder is used to kick-start the transient to natural vortex shedding in the first period \citep{Deng2020jfm}. 
This rotation has a circumferential velocity 
of $+0.5$ at $t<6.25$ and of $-0.5$ at $6.25<t<12.5$.
The transient regime lasts around 400 convective time units.
Figure \ref{fig:unforced_flow} shows the vorticity field for the symmetric steady solution and the natural unforced flow after 400 convective units.
The snapshot at $t=400$ in figure \ref{fig:natural_flow} will be the initial condition for all the following simulations.
%The natural flow has been then simulated over 1000 convective time units.

%***********************************************************************
\subsection{Flow characteristics} 
\label{sec:flow_characteristics}
%% Figures : Natural flow characteristics ------------------------------
\begin{figure}
\centering
\begin{subfigure}{.45\textwidth}
  \centering
\includegraphics[width=1.\textwidth]{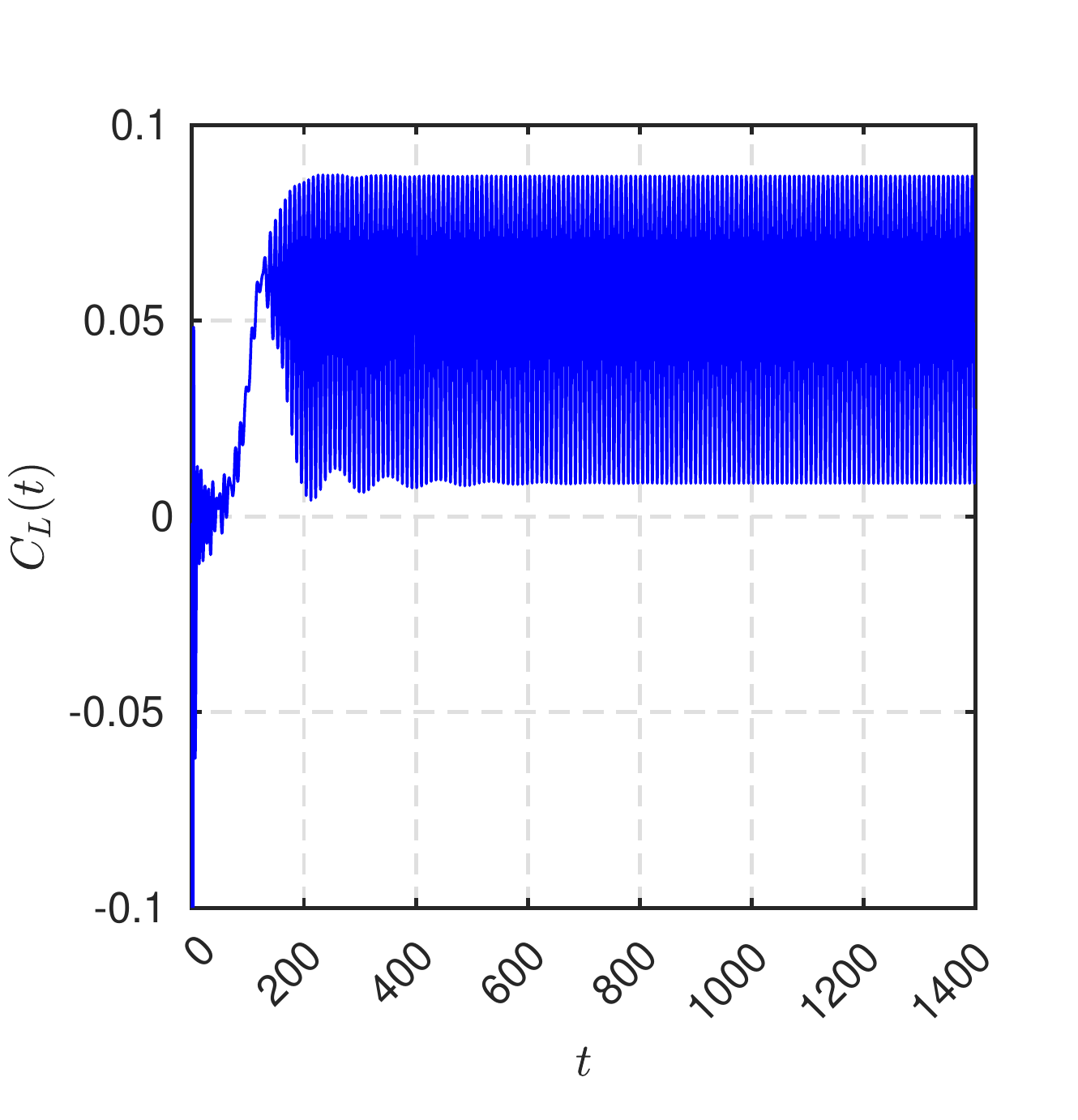}
\caption{\label{fig:CL_natural}}
\end{subfigure}%
\hfil
\begin{subfigure}{.45\textwidth}
  \centering
\includegraphics[width=1.\textwidth]{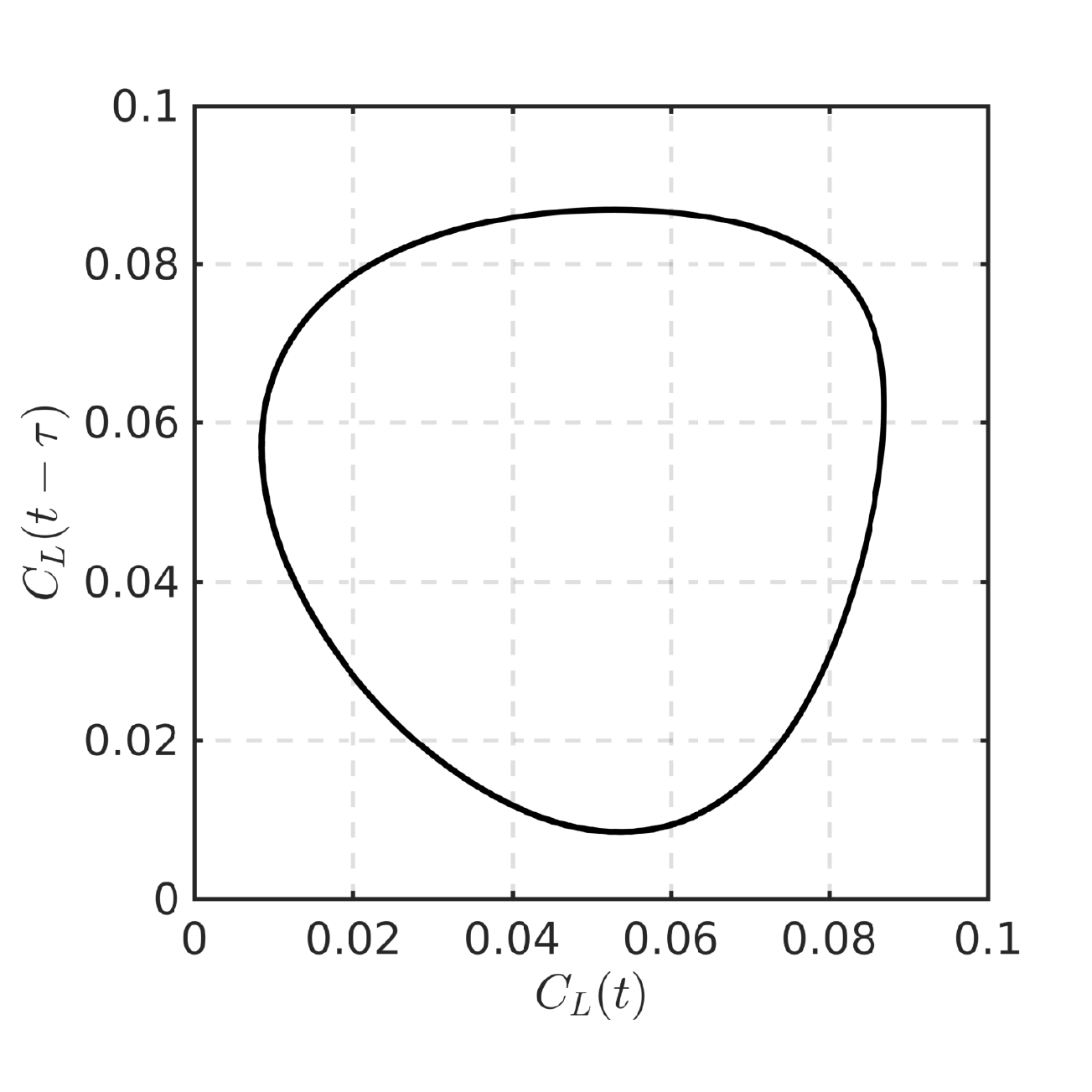}
\caption{\label{fig:PP_natural}}
\end{subfigure}

\begin{subfigure}{.45\textwidth}
  \centering
\includegraphics[width=1.\textwidth]{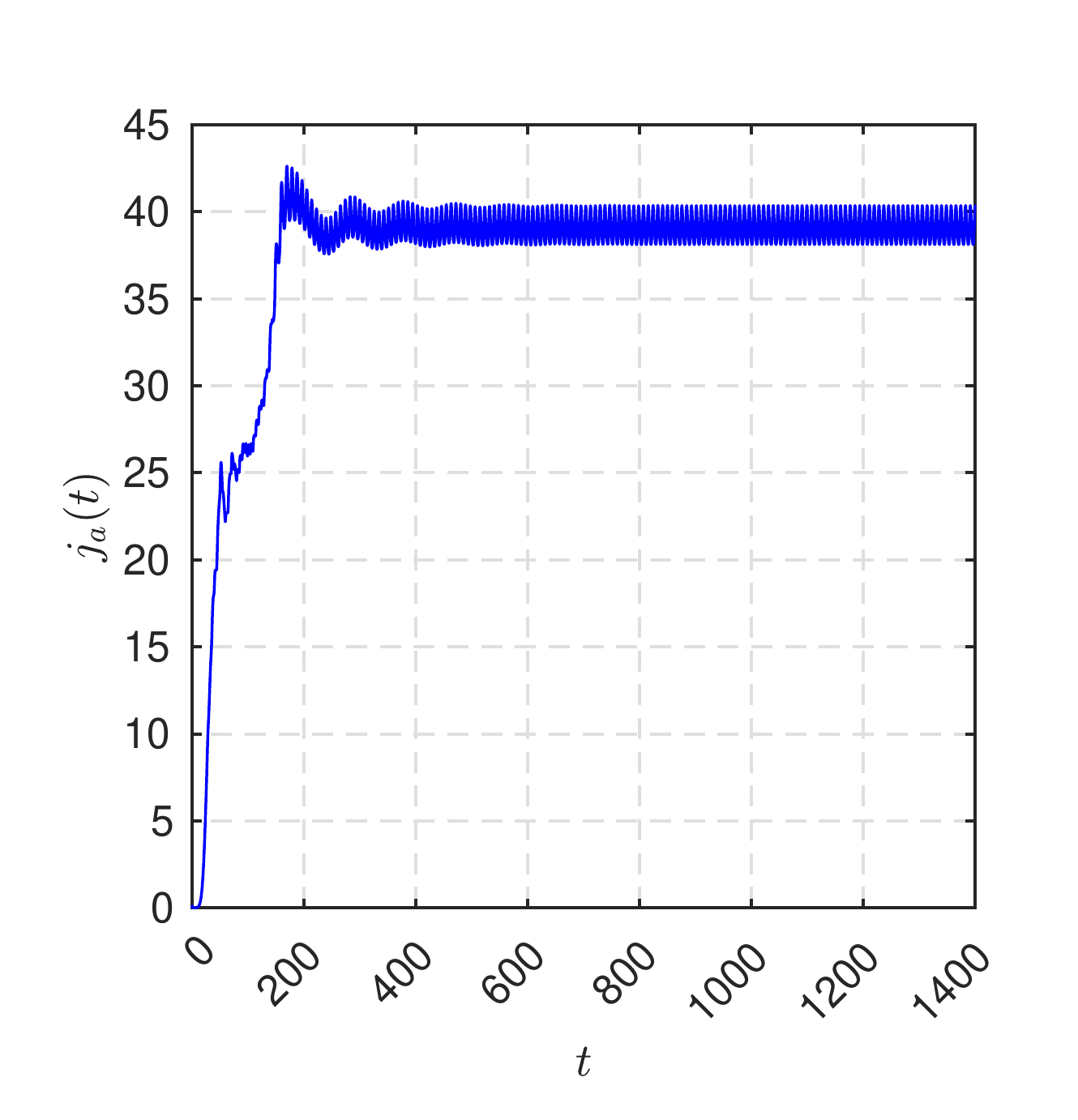}
\caption{\label{fig:DSS_natural}}
\end{subfigure}%
\hfil
\begin{subfigure}{.45\textwidth}
  \centering
\includegraphics[width=1.\textwidth]{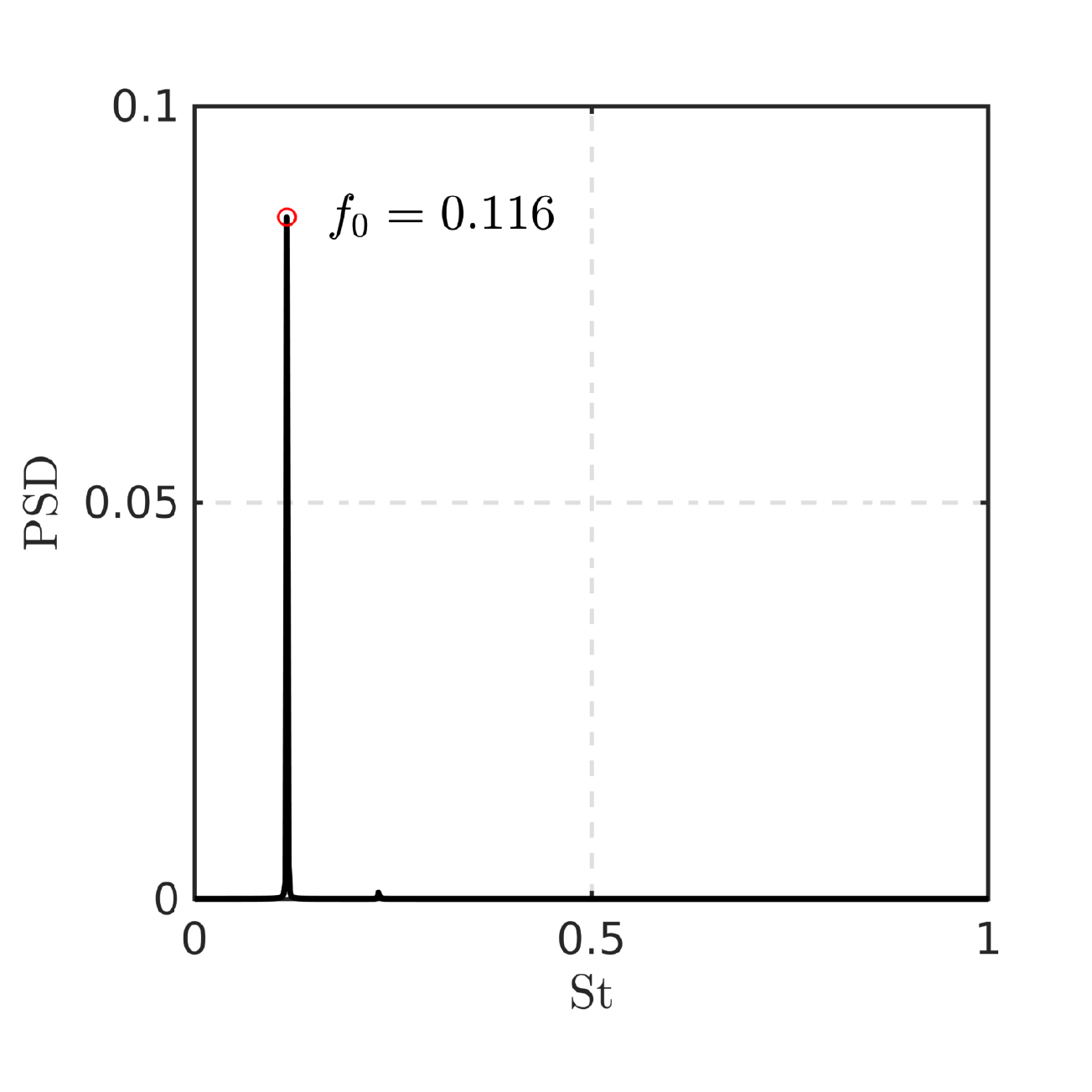}
\caption{\label{fig:PSD_natural}}
\end{subfigure}
\caption{\label{fig:natural_characteristics}Characteristics of the unforced natural flow starting from the steady solution ($t=0$).
The transient spans until $t \approx 400$. (a) Time evolution of the lift coefficient $C_L$, (b) phase portrait, (c)  time evolution of the instantaneous cost function $j_a$ and (d) Power Spectral Density (PSD) showing the natural frequency $f_0=0.116$ and its first harmonic. The phase portrait is computed during the post-transient regime $t \in [900,1400]$ and the PSD is computed over the last 1000 convective time units, $t \in [400,1400]$. 
%The transient regime in red is common with all the following simulations.
}
\end{figure}
%-----------------------------------------------------------------------
The fluidic pinball is a geometrically simple configuration 
that comprises key features of real-life flows 
such as successive bifurcations and frequency crosstalk between modes.
\citet{Deng2020jfm} shows that the unforced fluidic pinball undergoes successive bifurcations with increasing Reynolds number before reaching a chaotic regime.
The first Hopf bifurcation at Reynolds number $\Rey\approx 18$ 
breaks the symmetry in the flow and initiates the von K\'arm\'an vortex shedding.
The second bifurcation at Reynolds number $\Rey \approx 68$ 
is of pitchfork type and gives rise to a transverse deflection of jet-like flow
 between the two rearward cylinders.
The bi-stability of the jet deflection has been reported by \citet{Deng2020jfm}.
At a Reynolds number $\Rey = 100$
the jet deflection is rapid and occurs before the vortex shedding is fully established.
Figure~\ref{fig:CL_natural} shows an increase of the lift coefficient $C_L$
before  oscillations set in and the lift coefficient converges 
against a periodic oscillation around a slightly reduced mean value.
Those bifurcations are a consequence of  multiple instabilities 
present in the flow: there are two shear instabilities, 
on the top and bottom cylinder, 
and a jet bi-stability originating from the gap between the two back cylinders.
The shear-layer instabilities synchronize to a von K\'arm\'an vortex shedding.

% Natural flow snapshots ---------------------------------------------
\begin{figure}
\centering
\begin{subfigure}{.45\textwidth}
  \centering
\includegraphics[width=1.\textwidth]{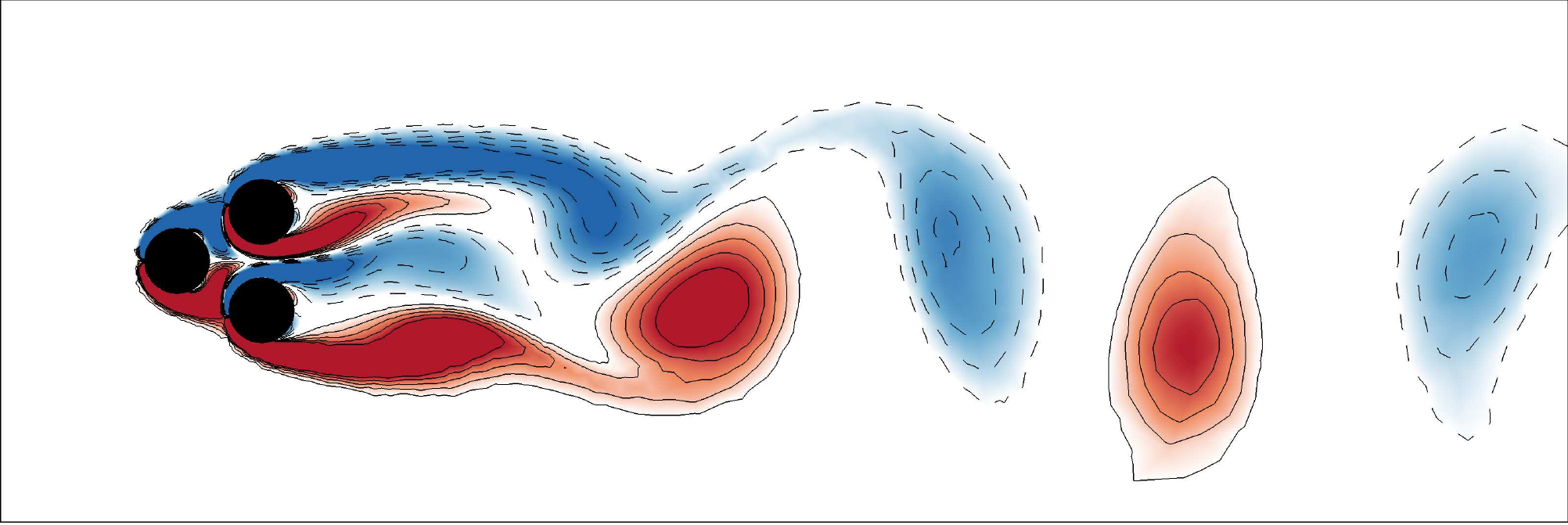}
\caption{\label{fig:nat_T1}$t+T_0/8$}
\end{subfigure}%
\hspace{0.5cm}
\begin{subfigure}{.45\textwidth}
  \centering
\includegraphics[width=1.\textwidth]{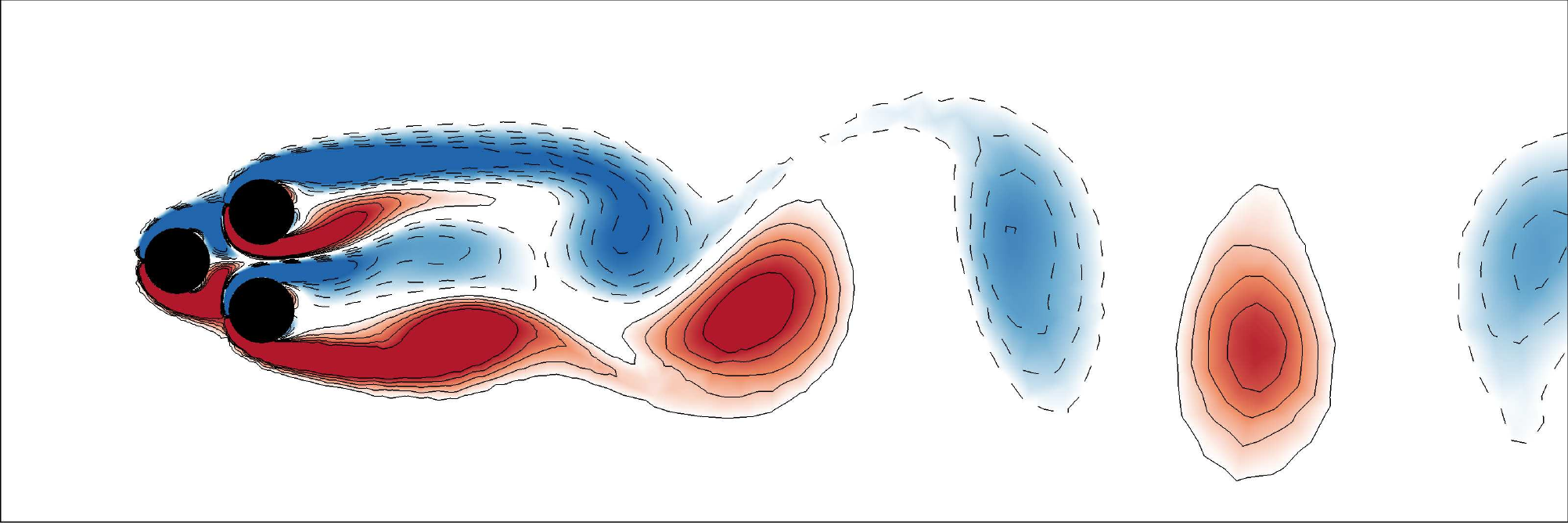}
\caption{\label{fig:nat_T2}$t+2T_0/8$}
\end{subfigure}

\begin{subfigure}{.45\textwidth}
  \centering
\includegraphics[width=1.\textwidth]{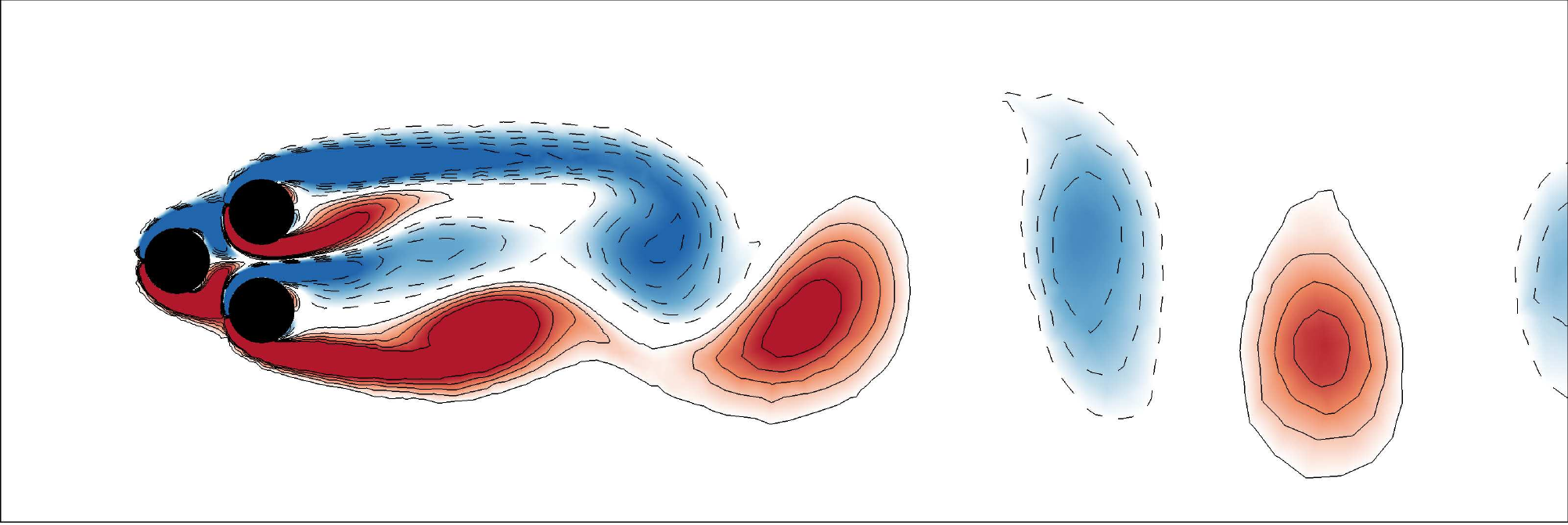}
\caption{\label{fig:nat_T3}$t+3T_0/8$}
\end{subfigure}%
\hspace{0.5cm}
\begin{subfigure}{.45\textwidth}
  \centering
\includegraphics[width=1.\textwidth]{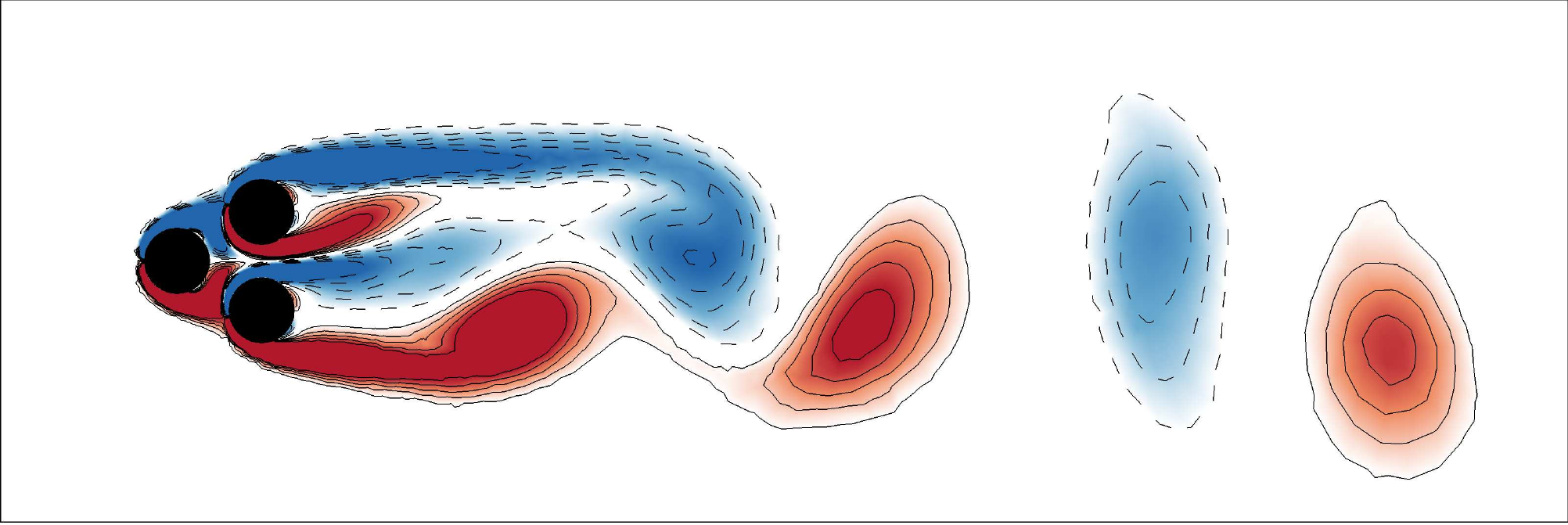}
\caption{\label{fig:nat_T4}$t+4T_0/8$}
\end{subfigure}

\begin{subfigure}{.45\textwidth}
  \centering
\includegraphics[width=1.\textwidth]{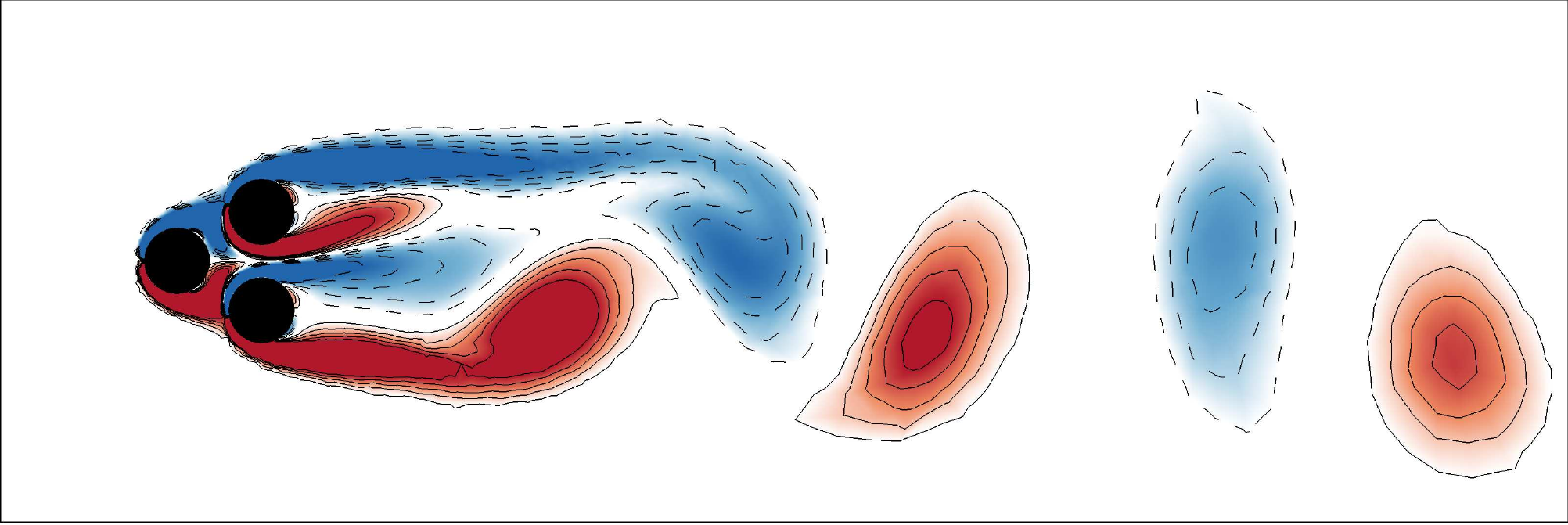}
\caption{\label{fig:nat_T5}$t+5T_0/8$}
\end{subfigure}%
\hspace{0.5cm}
\begin{subfigure}{.45\textwidth}
  \centering
\includegraphics[width=1.\textwidth]{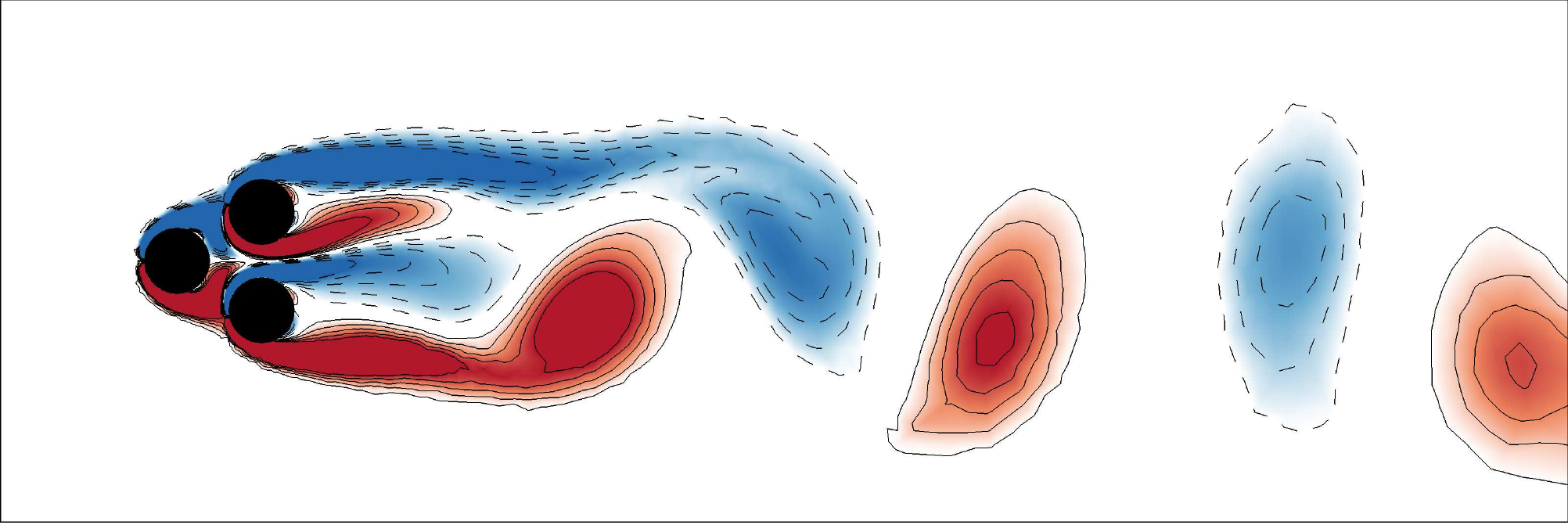}
\caption{\label{fig:nat_T6}$t+6T_0/8$}
\end{subfigure}

\begin{subfigure}{.45\textwidth}
  \centering
\includegraphics[width=1.\textwidth]{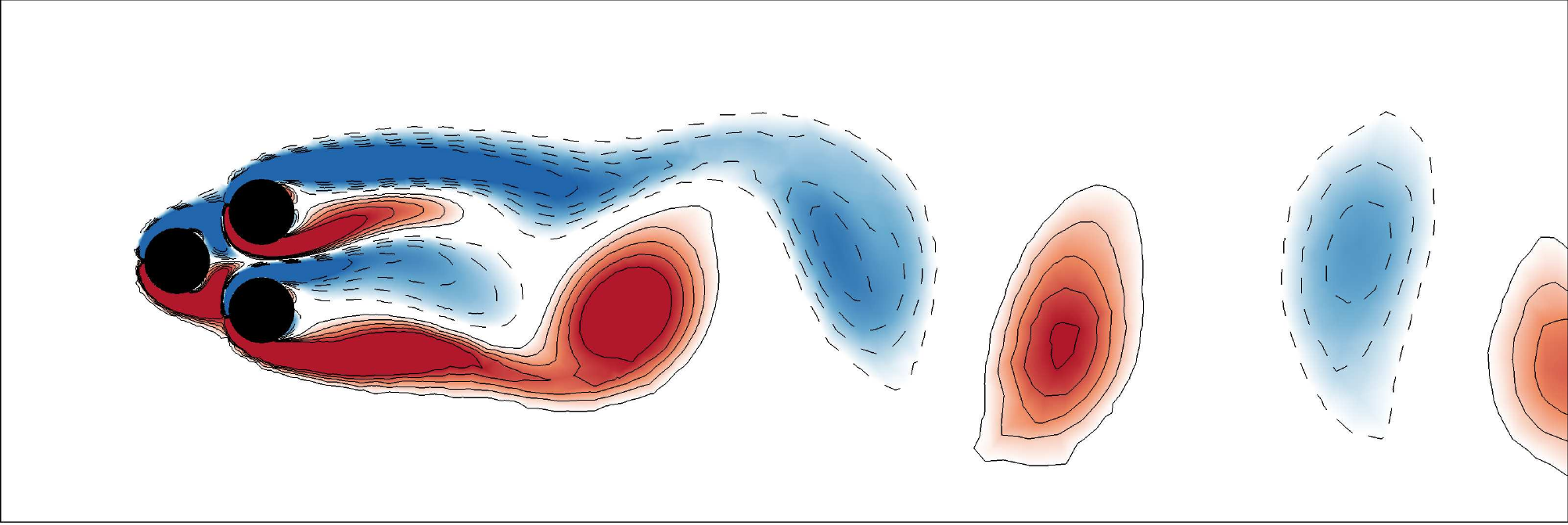}
\caption{\label{fig:nat_T7}$t+7T_0/8$}
\end{subfigure}%
\hspace{0.5cm}
\begin{subfigure}{.45\textwidth}
  \centering
\includegraphics[width=1.\textwidth]{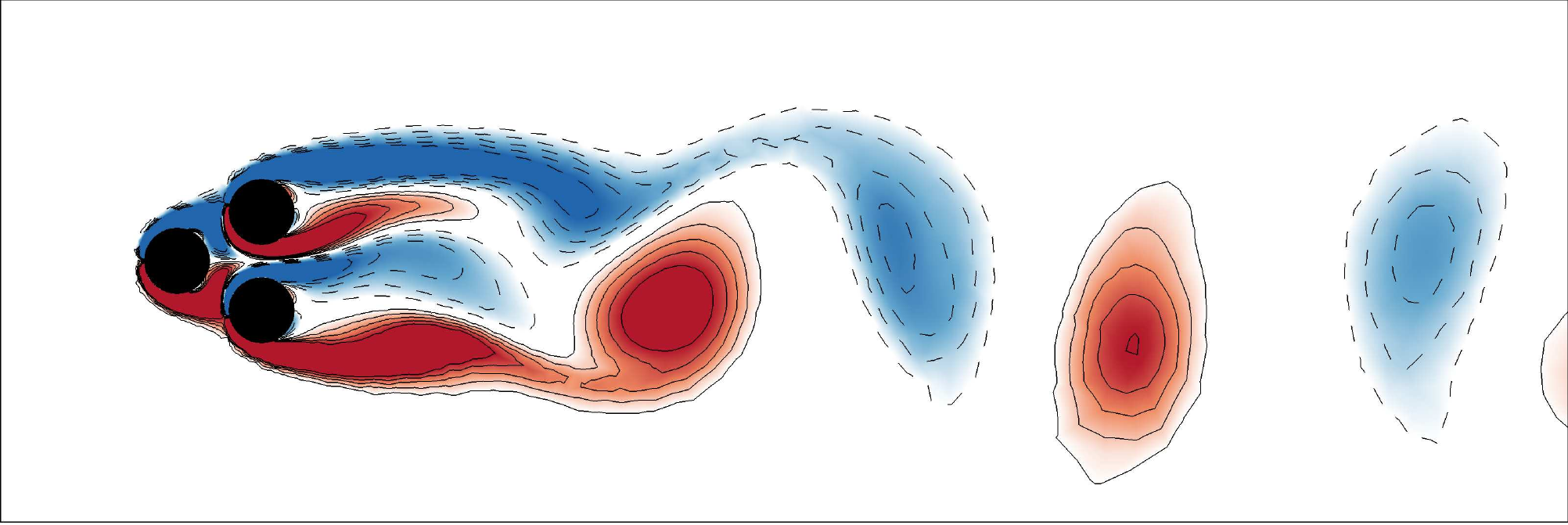}
\caption{\label{fig:nat_T8}$t+T_0$}
\end{subfigure}

\begin{subfigure}{.45\textwidth}
  \centering
\includegraphics[width=1.\textwidth]{Figures/Snapshots/SteadySolution}
\caption{\label{fig:sss}Symmetric steady solution}
\end{subfigure}%
\hspace{0.5cm}
\begin{subfigure}{.45\textwidth}
  \centering
\includegraphics[width=1.\textwidth]{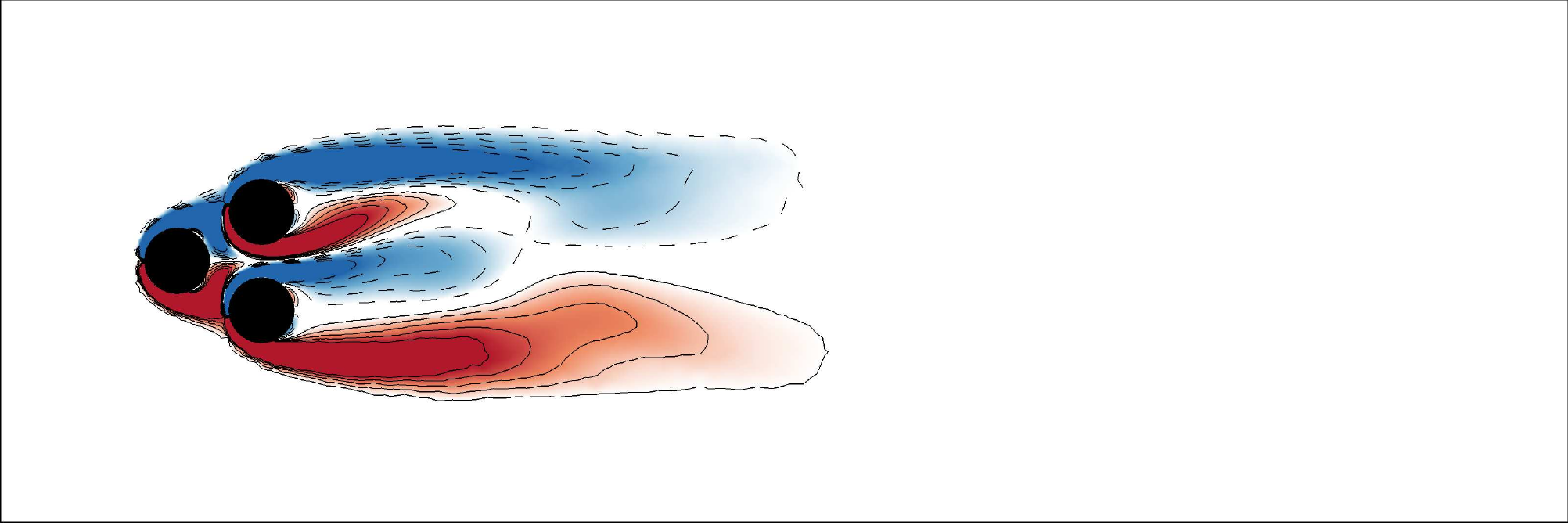}
\caption{\label{fig:mean_nat}Mean field}
\end{subfigure}
\caption{\label{fig:natural_snap}Vorticity fields of the unforced flow. (a)-(f) Time evolution of the vorticity field in the last period of the  simulation,
(i) the objective symmetric steady solution and (j) the mean field of the unforced flow.
The color code is the same as figure~\ref{fig:unforced_flow}.
$T_0$ is the natural period associated to the natural frequency $f_0$.
The mean field has been computed by averaging the flow over 100 periods.}
\end{figure}
%-----------------------------------------------------------------------
Figure~\ref{fig:natural_characteristics} illustrates the dynamics of the unforced flow 
from the unstable steady symmetric solution to the post-transient  periodic flow.
The phase portrait in figure~\ref{fig:PP_natural} and the power spectral density (PSD) in figure~\ref{fig:PSD_natural} show a periodic regime with frequency $f_0=0.116$ and its harmonic.
Figure~\ref{fig:CL_natural} shows that the mean value of the lift coefficient $C_L$ is not null.
This is due to the deflection of the jet behind the two rearward cylinders during the post-transient regime.
During this regime, the deflection of the jet stays on one side as it is illustrated in figure~\ref{fig:nat_T1}-\ref{fig:nat_T8} over one period and in figure~\ref{fig:natural_snap}j in the mean field.
This deflection explains the asymmetry of the lift coefficient $C_L$.
Indeed, the upward oriented jet increases the pressure on the lower part of the top cylinder leading to an increase of the lift coefficient.
In figure~\ref{fig:CL_natural}, the initial downward spike on the lift coefficient is due to the initial kick.
The unforced natural flow is our reference simulation for future comparisons.

Thanks to the rotation of the cylinders, 
the fluidic pinball is capable of reproducing six actuation mechanisms 
inspired from wake stabilization literature and exploiting distinct physics. 
Examples of those mechanisms can be found in \citet{Ishar2019jfm}.
First, the wake can be stabilized by shaping the wake region 
more aerodynamically---also called fluidic boat tailing. 
Here, shear layer is vectored towards the center region with passive devices, 
like vanes \citep{Fluegel1930stg} or active control through Coanda blowing \citep{Geropp1995patent,Geropp2000ef,Barros2016jfm}.
In the case of the fluidic pinball, 
we can mimic this effect by a counter-rotating rearward cylinders 
which accelerates the boundary layer and delays separation.
This fluidic boat tailing is typically associated with significant drag reduction.
Second, the two rearward cylinders can also rotate oppositely ejecting a fluid jet on the centerline.
Thus, interaction between the upper and lower shear layer is suppressed, 
preventing the development of a von K\'arm\'an vortex in the vicinity of the cylinders.
Such base-bleeding mechanism has a similar physical effect as a splitter plate behind a bluff body and has been proved to be an effective means for wake stabilization \citep{Wood1964jras,Bearman1967aq}.
Third, phasor control can be performed by estimating the oscillation phase 
and feeding it back with a phase shift and gain \citep{Protas2004pf}.
Fourth, unified rotation of the three cylinders in the same direction 
gives rise to higher velocities, and thus larger vorticity, on one side at the expense of the other side, destroying the vortex shedding. 
This effect relates to the Magnus effect and stagnation point control \citep{Seifert2012review}.
Fifth, high-frequency forcing can be effected by symmetric periodic oscillation of the rearward cylinders.
With a vigorous cylinder rotation \citep{Thiria2006jfm}, the upper and lower shear layer are re-energized, reducing the transverse wake profile gradients and thus the instability of the flow.
Thus, the effective eddy viscosity in the von K\'arm\'an vortices increases, adding a damping effect.
Sixth and finally,
a symmetrical forcing at a lower frequency than the natural vortex shedding may stabilize the wake \citep{Pastoor2008jfm}.
This is due to the mismatch between the anti-symmetric vortex shedding and the forced symmetric dynamics whose clock-work is distinctly out of sync with the shedding period.
High- and low-frequency forcing lead to frequency crosstalk between actuation and vortex shedding
over the mean flows,
as described by  low-dimensional generalized mean-field model \citep{Luchtenburg2009jfm}.

The fluidic pinball is an interesting Multiple-Input Multiple-Output (MIMO) control benchmark.
The configuration 
exhibits well-known wake stabilization mechanisms in physics.
From a dynamical perspective, nonlinear frequency crosstalk can easily be enforced.
In addition, even long-term simulations can easily be performed on a laptop within an hour.

%***********************************************************************
\subsection{Control objective and optimization problem} 
\label{sec:control_objective}
Several control objectives related to the suppression or reduction of undesired forces 
can be considered for the fluidic pinball.
We can reduce the net drag power, 
increase the recirculation bubble length, 
reduce lift fluctuations or even mitigate the total fluctuation energy.
%A parametric study confronting symmetric constant actuations has been carried out and is presented in figure \ref{fig:ParametricStudy}.
%\begin{figure}
%\caption{\label{fig:ParametricStudy} Parametric study on the fluidic pinball}
%\end{figure}

In this study, we aim to stabilize the unstable steady symmetric Navier-Stokes solution at $\Rey_D=100$.
The associated objectives are $J_a$, quantifying the closeness to the symmetric steady solution 
and $J_b$, the actuation power.
%Thus, the performance of the actuation command $\bm{b}=[b_1,b_2,b_3]^\intercal$ is characterized 
%with the cost function $J$:
%\begin{equation}
%\label{eq:cost_function}
%J(\bm{b}) = J_a(\bm{b})
%\end{equation}
%and $J^{\circ}$  that takes into account the actuation penalization term :
%\begin{equation}
%J^{\circ}(\bm{b}) = J_a(\bm{b}) + \gamma J_b(\bm{b})
%\end{equation}
%The penalization coefficient $\gamma$ tunes the balance between the components of the objective function.
The cost $J_a$ is defined as the temporal average of the residual fluctuation energy 
of the actuated flow field $\bm{u}_{\bm{b}}$ 
with respect to the symmetric steady flow $\bm{u}_s$:
\begin{equation}
J_a=\frac{1}{T_{ev}}\int_{t_0}^{t_0+T_{ev}}  j_a(t)  \:\mathrm{d} t
\end{equation}
with the  instantaneous cost function 
\begin{equation}
 j_a(t) =\Vert \bm{u}_{\bm{b}}(t)-\bm{u}_s \Vert_\Omega^2
\end{equation}
 based on the $L_2$-norm
\begin{equation}
\Vert \bm{u} \Vert_\Omega = \sqrt{ \iint\limits_{\Omega}  {u}^2 + {v}^2\:\mathrm{d} \bm{x}}.
\end{equation}
The control is activated at $t_0=400$ convective time units after the starting kick on the steady solution.
 Thus,  we have a fully established post-transient regime.
The cost function is evaluated until $T_{\rm ev}=1400$ convective time units.
Thus, the time average is effected over 1000 convective time units 
to make sure that the transient regime has far less weight as compared to the actuated regime.
Yet,  a faster stabilizing response to actuation is clearly desirable and factors positively into the cost.

$J_b$ is naturally chosen as a measurement of the actuation energy investment.
Evidently, a low actuation energy is desirable.
The actuation power 
is computed as the power of the torque applied by the fluid on the cylinders.
% is computed as the cumulative power of the torque applied by the fluid on the cylinders.
$J_b$ is the time-averaged actuation power over $T_{\rm ev} = 1000$ time units:
\begin{equation}
%J_b(\bm{b})=\frac{1}{T_{ev}}\int_{t_0}^{t_0+T_{\rm ev}}  -\boldsymbol{\tau} \cdot  (\bm{b}\slash R)  \:\mathrm{d} t
J_b(\bm{b})=\frac{1}{T_{ev}}\int_{t_0}^{t_0+T_{\rm ev}}  \sum_{i=1}^{3}\mathcal{P}_{\rm{act},i} \:\mathrm{d} t
\end{equation}
where $\mathcal{P}_{\rm{act},i}$ is the actuation power supplied integrated over cylinder $i$:
$$\mathcal{P}_{\rm{act},i} = -  \oiint  b_i  F^{\theta}_{si} \:\mathrm{d} s $$
where $\left( F^{\theta}_{si} \mathrm{d}s \right)$ is the azimuthal component of the local fluid forces applied to cylinder $i$.
The negative sign denotes that the power is supplied and not received by the cylinders.
The numerical value of $J_b$ may be compared 
with the unforced drag coefficient $c_D=3.57$ 
which is also the non-dimensionalized parasitic drag power.
%with $\boldsymbol{\tau} = (\tau_1,\tau_2,\tau_3)$, $\tau_i = \oint R |\bm F_i \wedge \bm{x}_s| \: \mathrm{d}\bm{s}$ being the torque applied by the fluid forces $\bm F_i$ on the cylinder $i$. The surface vector is denoted $\mathrm{d}\bm{s}$ and points outwards. $(\bm{b} \slash R)$ stands for the angular velocity vector of the cylinders.

In this study, optimization is based on the cost function $J=J_a$ 
and the actuation investment $J_b$ is evaluated separately.
We refrain from a cost function $J$ 
which employs the objective function $J_a$ 
and penalizes the actuation investment $J_b$ with suitable weight $\gamma$, i.e.,
$J = J_a + \gamma J_b$.
The procedure has three reasons.
First, the distance between two flows and actuation energy 
belong to two different worlds, kinematics and dynamics. 
Any choice of the penalization parameter $\gamma$ 
will be subjective and implicate a sensitivity discussion.
Moreover, a strong penalization would constraint the search space 
and may rule out relevant actuation mechanisms.
In this study, we look for stabilization mechanisms rather than the most power-efficient solutions.
Second, the complete stabilization of the steady solution 
would lead to a vanishing actuation $\bm{b} \equiv 0$ and thus vanishing energy $J_b$.
Thus, the optimization problem without actuation energy can be expected to be well-posed.
Third, a Pareto front of $J_a$, $J_b$ reveals 
how much actuation power is required 
for which closeness to the steady solution.
Using Pareto optimality, 
there is no need to decide in advance on the subjective weight $\gamma$. 
Foreshadowing the results, 
the best performance $J_a$ turns out to be achieved with the least actuation energy $J_b$.
This result corroborates a posteriori the decision not to include actuation energy in the cost.

The instantaneous cost function $j_a$ of the unforced flow is shown in figure~\ref{fig:DSS_natural}.
We notice a slight overshoot around $t=200$ before converging to a post-transient fluctuating regime.
The post-transient regime shows the expected periodic behaviour from von K\'arm\'an vortex shedding.
The cost averaged over 1000 convective time units is $J_0=39.08$ and serves as reference to actuation success.

To reach the steady  symmetric solution, 
the flow is controlled by the rotation of the three cylinders.
The actuation command $\bm{b}=[b_1,b_2,b_3]^\intercal$ 
is determined by control law $\bm{K}$.
This control law may operate open-loop or closed-loop with flow input.
Considered open-loop actuations are steady or harmonic oscillation around a vanishing mean.
Considered feedback  includes velocity sensor signals in the wake.
Thus, in the most general formulation, the control law reads
\begin{equation}
\bm{b}(t)=\bm{K}(\bm{h}(t), \bm{s}(t))
\end{equation}
with $\bm{h}(t)$ and $\bm{s}(t)$ being vectors comprising respectively time dependent harmonic functions and sensor signals.
The sensor signals include the instantaneous velocity signals 
as well as three recorded values over one period 
as elaborated in the result section \S~\ref{sec:results_gMLC}.
In the following, $N_b$ represents the number of actuators, 
$N_h$ for the number of time-dependent functions and $N_s$ for the number of sensor signals.
% Evidently if $\bm{K}$ is independent of time, the actuation command is constant\added[id=FL]{?? ? ??}.
Then  optimal control problem determines the control law which minimizes the cost:
\begin{equation}
\label{eq:control_problem}
\bm{K}^* = \underset{\bm{K}\in \mathcal{K}}{\operatorname{arg\,min}} \; J(\bm{K})
\end{equation}
with $\mathcal{K}:X \mapsto Y$ being the space of control laws.
Here, $X$ is the input space, e.g., sensor signals and $Y$ is the output for actuation commands.
In general, \eqref{eq:control_problem}  is  a challenging  non-convex optimization problem.

%-----------------------------------------------------------------------
\section{Control optimization framework}
\label{Sec:Methods}
In this section, 
we present the control optimization for stabilizing the fluidic pinball.
This constitutes a challenging nonlinear non-convex optimization problem
in which the possibility of several local minima must be expected.
Hence, we specifically address how to explore new minima 
while keeping the convergence rate
and efficiency of gradient-based approaches.
In \S~\ref{sec:disc_exploration_exploitation}, the principles of exploration and exploitation 
are discussed for parameter and control law optimization.
Then, the employed algorithms are described:
the Explorative Gradient Method (EGM) for parametric optimization (\S~\ref{sec:EGM}) 
and the gradient-enriched Machine Learning Control (gMLC) for control law optimization (\S~\ref{sec:gMLC}).

%***********************************************************************
\subsection{Optimization principles---Exploration versus exploitation} \label{sec:disc_exploration_exploitation}
The two algorithms, EGM and gMLC, 
enable model-free control optimization. 
These algorithms combine the advantages of exploitation and exploration.
Exploitation is based on a downhill simplex method 
with the best performing of all tested control laws,
also called `individuals'.
The goal is to `slide down' the best identified minimum.

Exploration is performed with another algorithm
using all previously tested individuals.
The goal is to find potentially new and better minima, 
ideally the global minimum.
The method for exploration depends on the search space.
For a low-dimensional parameter space, 
a space-filling version of the Latin Hypercube Sampling (LHS) 
guarantees optimal geometric coverage of the search space.
For a high-dimensional function space, 
genetic programming is found to be efficient.

EGM and gMLC start with an initial set of individuals to be evaluated.
Then, exploitive and explorative phases iterate
until a convergence criterion is reached.
The iteration hedges against several worst-case scenarios.
The control landscape may have only a single minimum accessible 
from any other point by steepest descend. 
In this case, exploration is often inefficient,
although it might help in avoiding slow marches 
through long shallow valleys \citep{LiA2020jfm}.
The control landscape may also have many minima 
accessible by gradient-based searches.
In this case, exploitation is likely to 
incrementally improve performance in suboptimal minima
and the search strategy should have a significant investment in exploration.
The minima of the control landscape may also have narrow basins of attractions 
for gradient-based iterations and extended plateaus.
This is another scenario where iteration 
between exploitation and exploration is advised.

Many optimizers balance exploration and exploitation
and gradually shift from the former to the latter.
This strategy sounds reasoning 
but is not a good hedge against 
the described worst case scenarios 
where almost all exploitative 
or almost all explorative algorithms are doomed to fail.

Note that the chances of exploration landing close to a new better minimum are small.
Yet, the explorative phases may find new basins of attractions for successful gradient-based descends.
This is another argument for the alternating execution of exploration and exploitation.

Finally, we note that the proposed explorative-exploitive schemes 
allows that  both kinds of iterations may be adjusted to the control landscape.
For instance, LHS in a high-dimensional search space 
will initially explore only the boundary 
and  may better be replaced by Monte-Carlo or a genetic algorithm.
We refer to \citet{LiA2020jfm} for a thorough comparison of EGM
and five common optimizers
and to  \citet{Duriez2016book} for genetic programming control.
The next two sections detail both optimizers, EGM and gMLC.

%***********************************************************************
\subsection{Parameter optimization with the explorative gradient method} \label{sec:EGM}
The Explorative Gradient Method (EGM) 
optimizes $N_p$ parameters 
$\bm{b} = \left[ b_1, \ldots, b_{N_p} \right]^\intercal$
with respect to cost $J(\bm{b})$ 
and  comprises exploration and exploitation phases.
In the context of parameter optimization, we do not differentiate between the control law $\bm{K}=\hbox{const}$ and the associated actuation command $\bm{b}=\bm{K}$.
The search space, or actuation domain, is  a compact subset $\mathcal{B}$ of $\mathbb{R}^{N_p}$,
typically defined by upper and lower bounds for each parameter.
The exploration phase is based on a space-filling variant of Latin hypercube sampling (LHS) \citep{McKay1979} whereas the exploitation phase is carried out by Nelder-Mead's \emph{downhill simplex} \citep{Nelder1965}.

The first $N_p+1$ initial individuals $\bm{b}_m$, $m=1,\ldots, N_p+1$ 
define the first `amoeba' of the downhill simplex method.
The first individual $\bm{b}_1$ is typically placed at the center of $\mathcal{B}$.
The $N_p$ remaining vertices are slightly displaced along the $b_m$ axes.
In other words, 
$\bm{b}_m = \bm{b}_1 + h_m \bm{e}_{m-1}$ for $m=2, \ldots, N_p+1$. 
Here, $\bm{e}_m := \left[ \delta_{m,1},...,\delta_{m,N_p} \right]^\intercal$ 
is the unit vector in the $m$th direction 
and $h_m$ is the corresponding step size. 
The increment $h_m$ is chosen to be small compared to the range of the corresponding dimension.

The exploitation phase employs the downhill simplex method.
This method is robust and widely used for data-driven optimization
in low and moderate-dimensional search spaces that requires neither analytical expression of the cost function nor local gradient information.
The new individual is a linear combination of the simplex individuals
and follows a geometric reasoning.
The vertex with the worst performance is replaced
by a point reflected   
at the centroid of the opposite side of the simplex.
This step leads to a mirror-symmetric version of the simplex
where the new vertex has the best performance
if the cost function depends linearly on the input.
Subsequent operations, like expansion, single contraction, 
and global shrinking ensure that iterations exploit a favourable downhill behaviour 
and avoid getting stuck  by nonlinearities.
We refer to \citet{LiA2020jfm} for a detailed description.

The explorative phase of  EGM  is inspired by the LHS method.
LHS aims to fill the complete domain $\mathcal{B}$ optimally.
The pre-defined number $m$ of individuals 
maximizes the minimum distance of its neighbours:
$$ 
\left  \{ \bm{b}^{\rm LHS}_1, \ldots, \bm{b}^{\rm LHS}_m 
\right \}:= \underset{\bm{b}_1,\ldots,\bm{b}_m \in \mathcal{B}}{\arg \max}  \quad 
         \min\limits_{i \in \{ 1, \ldots, m-1 \}, \atop  j \in \{ i +1, \ldots, m \}} \left \Vert \bm{b}_i - \bm{b}_j \right \Vert .$$
Here, $\Vert \cdot \Vert$ denotes the Euclidean norm.
The number of individuals has to be determined in advance and cannot be augmented.
This static feature is incompatible with the iterative nature of the EGM algorithm.
Thus, we resort to a recursive `greedy' version.
Let $\bm{b}^{\bullet}_1$ be the first individual.
Then, $\bm{b}^{\bullet}_2$ maximizes the distance from $\bm{b}^{\bullet}_1$,
$$ \bm{b}^{\bullet}_2 := \underset{\bm{b} \in \mathcal{B}}{\arg \max} \Vert \bm{b} - \bm{b}^{\bullet}_1 \Vert .$$
The $m$th individual maximizes the minimum distance to all previous individuals,
$$ \bm{b}^{\bullet}_m := \underset{\bm{b} \in \mathcal{B}}{\arg \max}  \quad 
               \underset{i \in \{ 1, \ldots, m-1 \}}{\min} \Vert \bm{b} - \bm{b}^{\bullet}_i \Vert .
$$
This recursive definition allows adding explorative phases 
from any given set of individuals.

Exploitation and exploration are iteratively continued
until the stopping criterion is reached.
In our study, 
the stopping criterion is the total number of cost function evaluations,
i.e., a given budget of simulations.
This criterion is validated after the run by checking the convergence of the performance.
The \emph{Explorative Gradient Method} (EGM) phases are summarized in algorithm~\ref{algo:EGM}.

\begin{algorithm}
%\DontPrintSemicolon
\SetKwBlock{ExplorPhase}{Exploration phase---Latin hypercube sampling}{end}
\SetKwBlock{ExploiPhase}{Exploitation phase---Downhill simplex}{end}
\SetAlgoLined
\KwResult{$\bm b^*$, the best individual}
Initialize the $N_p+1$ individuals of the dataset $\mathcal{B}_I$\;
Test all the individuals\;
Build the simplex $\mathcal{S}$ by taking the $N_p+1$ best individuals\;
 \While{Stopping criterion is not reached}{
 \ExplorPhase{
   Select $\bm b^{\rm LHS}$ by solving:
$$ \bm{b}^{\rm LHS} := \underset{\bm b \in \mathcal{B}}{\arg \max}  \quad 
               \underset{\bm b_i \in \mathcal{B}_I}{\min} \Vert \bm{b} - \bm{b}_i \Vert
$$\\
Test $\bm b^{\rm LHS}$\;
Augment dataset: $\mathcal{B}_I :=  \mathcal{B}_I \cup \left\{ \bm b^{\rm LHS} \right\}$ \;
Update simplex: replace the worst individual of $\mathcal{S}$ by $\bm b^{\rm LHS}$\;
}
 \ExploiPhase{
% Take the $N_p+1$ best individuals of $\mathcal{B}_I$ for the simplex: $ \bm b_1,\ldots,\bm b_{N_p+1}$\;   
 Sort and relabel $\mathcal{S}$ such as: $J^S_1 \le J^S_2 \le \ldots \le J^S_{N_p+1}$\;
   Compute the centroid 
   $\bm{c} = \frac{1}{N_p}\sum_{i=1}^{N_p}\bm{b}_i$ of $\mathcal{S}$ excluding $\bm b_{N_p+1}$\;
   \textbf{Reflection}: compute and test $\bm b_r:=\bm{c} +(\bm{c}- \bm b_{N_p+1})$\;
  \uIf{$J^S_1 < J^S_r < J^S_{N_p+1}$}{
  Update simplex: $\bm b_{N_p+1}:= \bm b_r$\;
  }
  \uElseIf{$J^S_r < J^S_1$}{
   \textbf{Expansion}: compute and test $\bm b_e := \bm{c} + 2\ (\bm{c} - \bm b_{N_p+1})$ \;
    Update simplex: $\bm b_{N_p+1}:= {\rm min} \left\{ \bm b_r, \bm b_e\right \}$\;
  }
  \ElseIf{$J^S_{N_p+1} \leq J^S_r$}{     
    \textbf{Contraction}: compute and test $\bm b_c:= 1/2\ (\bm{c} + \bm b_{N_p+1})$\;
    \uIf{$J^S_c < J^S_{N_p+1}$}{
  Update simplex: $\bm b_{N_p+1}:= \bm b_c$\;
  }
      \Else{
        \textbf{Shrink}: compute and test $\bm b_{s,i}:=1/2\ (\bm b_1+\bm b_i),  i=2,\ldots, N_p+1$\;
        Update simplex: $\bm b_i:= \bm b_{s,i},  i=2,\ldots, N_p+1$\;
      }
  }
  Augment dataset: add all the new individuals to $\mathcal{B}_I$\;
  }
 }
 \caption{\label{algo:EGM}Explorative Gradient Method}
\end{algorithm}

%\begin{description}
%\item[\textbf{Step 1 --- Initialization}:]\ 
%Choose $N_{p+1}$ individuals to initialize the \textrm{downhill simplex} algorithm.
%\item[\textbf{Step 2 --- Exploration}:]\ 
%Compute a new individual thanks to LHS and evaluate its cost. 
%If the cost of the new individual is lower than the worst one in the \textrm{simplex}, 
%replace the latter by the LHS individual.
%\item[\textbf{Step 3 --- Exploitation}:]\ 
%Perform one downhil simplex iteration with the updated $N_{p+1}$ individuals of the \textrm{simplex}.
%\item[\textbf{Step 4 --- Stop}:]\ 
%If the stopping criterion is reached terminate the algorithm. 
%Otherwise go to \textbf{Step 2}.
%\end{description}
% BRN20201006: Note that JFM has an "algorithms" environment for pseudo code!

%***********************************************************************
\subsection{Multiple-input multiple-output control optimization 
with gradient-enriched machine learning control} \label{sec:gMLC}

In this section, we cure a challenge of 
linear genetic programming control---the suboptimal exploitation of gradient information.
Starting point is machine learning control (MLC)
based on linear genetic programming (LGP).
MLC optimizes a control law without assuming a polynomial or other structure
of the mapping from input to output.
The only assumption is that the law can be expressed 
by a finite number of mathematical operations
with a finite memory, i.e., is computable.
The optimization process relies on a stochastic recombination of the control laws, also called \emph{evolution}.
MLC has been amazingly efficient in outperforming
existing optimal control laws---often with surprising frequency crosstalk mechanisms---in dozens of experiments \citep{Noack2019springer}.
MLC demonstrates a good exploration of actuation mechanisms
but a slow convergence to an optimum 
despite an increasing testing of redundant similar control laws.

The proposed gradient-enriched MLC
departs in two aspects from MLC.
First, 
the concept of evolution from generation to generation is not adopted.
The genetic operations include \emph{all} tested individuals.
One can argue that the neglection of previous generations might imply loss of important information.
Second, the exploitation is accelerated 
by \textrm{downhill subplex} iteration \citep{Rowan1990phd}.
The best $k+1$ individuals are chosen to define a $k$-dimensional subspace
% GYCM20201010: I changed K-> k, because K was to close to the control law \bm{K}.
and a downhill simplex algorithm optimizes the control law in this subspace.

MLC and gMLC share a representation of the control laws
used for LGP \citep{Brameier2006book}.
The individuals are considered as little computer programs,
using a finite number $N_{\rm inst}$ of instructions, 
a given register of variables 
and a set of constants. 
The instructions employ
 basic operations ($+$, $-$, $\times$, $\div$, $\cos$, $\sin$, $\tanh$, etc.) 
using  inputs ($h_i$ time-dependent functions and $s_i$ sensor signals)
and yielding the control commands as outputs.
A matrix representation conveniently comprises the operations of each individual.
Every row describes one instruction.
The first two columns define the register indices of the arguments,
the third column the index of the operation
and the fourth column the output register.
Before execution, all registers are zeroed.
Then, the first registers are initialized with the input arguments,
while the output is read from the last registers after the execution of all instructions.
This leads to a $N_{\rm inst}\times 4$ matrix 
representing the control law $\bm{K}$. 
We refer to \citet{Li2018am} for details.

The algorithm begins with a Monte Carlo initialization of 
$N_{\rm MC}$ individuals, i.e., the indices of the matrix.
The cost of these  randomly  generated functions are evaluated in the plant.
The number of individuals $N_{\rm MC}$ needs to balance exploration and cost.
Too few individuals may lead to descend in a suboptimal local minimum.
Too many individuals may lead to unnecessary inefficient testing,
as  Monte Carlo sampling is purely explorative.

Once the initial individuals are evaluated, 
an exploration phase is carried out.
New individuals are generated thanks to crossover and mutation operations.
Thus, this phase is also referred as evolution phase.
These operations are performed on the matrix representation of the individuals.
As for MLC, crossover combines two individuals by exchanging lines in their matrix representation, whereas mutation randomly replaces values of some lines by new ones.
In this approach, we no longer consider a population 
but the database of all the individuals evaluated so far.
Thus, we no longer need the replication and elitism operators of MLC.
This choice is justified by the fact that we want to learn as much as possible from what we already know and avoid reevaluating individuals.
To perform the crossover and mutation operation, 
individuals are selected from the database thanks to a tournament selection.
A tournament selection of size 7 for a population of 100 individuals is used in \citet{Duriez2016book}.
That means that for a population of 100 individuals, 7 individuals are selected randomly and the among the 7, the best one is chosen for the crossover or mutation operation.
For gMLC, as the individuals are selected among all the evaluated individuals, the tournament size is properly scaled at each call to preserve the $7 \slash 100$ ratio between the tournament size and the size of the database.
The crossover and mutation operation are repeated randomly following $P_c$, the crossover probability, and $P_m$, the mutation probability, until $N_G$ individuals are generated.
The probabilities $P_c$ and $P_m$ are such as $P_m+P_c=1$.
%The evolution phase comprising crossover and mutation is considered as an exploration for two reasons:
%\begin{itemize}
%\item firstly, the mutation operation creates effectively new matrices and thus control laws with different structures and inputs;
%\item secondly, even though the crossover operation recombines matrices and thus can be considered as an exploitation operation, the new individuals may have radically different structure.
%\end{itemize}

Once the evolution phase is achieved, 
$N_G$ new individuals are generated thanks to \textrm{downhill subplex} iterations.
% The same number of individuals as in the exploration phase is generated to assure a balance between exploration and exploration.
%% Downhill subplex
Being in an infinite dimension function space, 
Nelder-Mead's \textrm{downhill simplex} is impractical as an exploitation tool.
Thus, we propose a variant of \textrm{downhill simplex} inspired by \citet{Rowan1990phd},
commonly called  \textit{downhill subplex}.
Just as \textrm{downhill simplex}, the strength of this approach is to exploit local gradients to explore the search space.
In the original approach of \citet{Rowan1990phd}, 
\textrm{downhill simplex} is applied to several orthogonal subspaces.
However, in order to limit the number of cost function evaluations, 
we apply \textrm{downhill simplex} to only one subspace.
This subspace is initialized by selecting $N_{\rm sub}$ individuals.
Two ways to build the subspace after the Monte Carlo process are listed below:
\begin{itemize}
\item \textbf{Choose the best individual}: 
select the best $N_{\rm sub}$ individuals evaluated so-far in the whole database.
\item \textbf{Individuals near a minimum}: select the best individual evaluated so-far and the $N_{\rm sub}-1$ individuals closest to the best one.
\end{itemize}
The first approach has the benefit to comprise several minima candidates, 
whereas the second one is bound to lead to a minimum in the neighborhood of the best individual and relies on a given metric.
Once the subspace is built, the next steps are similar to the \textrm{downhill simplex} method.
As subplex and simplex are essentially the same algorithm applied to different spaces, we will not designate them differently.

Following the situation, \textrm{downhill subplex} may call $1$ (only reflection), $2$ (expansion or single contraction) or $N_{\rm sub}+1$ (shrink) times the cost function.
Several iterations of \textrm{downhill subplex} are repeated 
until at least $N_G$ individuals are generated.
In this study, the same number of individuals generated with the evolution phase and the \textrm{downhill subplex} phase is chosen to balance exploration and exploitation.

If the stopping criterion is reached, 
the most efficient individual in the database is given back.
Otherwise, we restart a new cycle by generating new individuals with a new evolution phase, combining and modifying individuals derived by evolution and \textrm{downhill subplex}.
%However, the \textrm{downhill subplex} builds new control laws thanks to linear interpolations of the control laws in the  basis $\mathcal{B}$.
However, the individuals built thanks to \textrm{downhill subplex} are linear combination of the original $N_{\rm sub}$ individuals.
These new individuals do not have a matrix representation which is necessary to generate new individuals with genetic operators in the exploitation phase.
To overcome this problem, we introduce a new phase to compute 
a matrix representation for the linearly-combined control laws.
The matrix representation is computed by solving a regression problem of the first kind, 
similar to a function fitting problem, for all the linearly-combined control laws.
First, each control law $\bm{K}_i$ is evaluated on randomly sampled inputs $\bm{s}_{\rm rand}$.
The resulting output $\bm{K}_i (\bm{s}_{\mathrm{rand}})$
 is used to solve a secondary optimization problem:
% The objective is to find a matrix representation of the control law $\bm{K}_i$ on the randomly sampled inputs $\bm{s}_{\rm rand}$.
\begin{equation} \label{eq:second_opti_problem}
    \bm{K_M}^* = \underset{\bm{K_M}}{\operatorname{arg\,min}} \; \Vert (\bm{K_M}(\bm{s}_{\mathrm{rand}})-\bm{K}_i (\bm{s}_{\mathrm{rand}}))\Vert^2
\end{equation}
where $\Vert \cdot \Vert$ denotes the Euclidean norm.
This optimization problem is a function fitting problem that we solve with linear genetic programming.
The LGP parameters are the same used for the gMLC so the computed individuals are compatible with the ones in the database.
The best fitting control law $ \bm{K_M}^\ast$ has then a matrix representation 
and is used as a substitute for the original linear combination of  control laws.
The substitutes are then employed for the evolution phase 
even though they may not be perfect substitutes of the original control laws.
Indeed, following the stopping criterion and population size of the secondary LGP optimization, the control law substitutes may not be able to reproduce all the characteristics of the linearly-combined control laws.
An accurate but costly representation may not be needed as the control laws will be recombined afterwards.
Moreover, the introduction of some error may be beneficial to improve the exploration phase and enrich our database.%, in the same way simulated annealing relies in stochastic exploration \citep{}.

Once the matrix representations are computed, a new cycle may begin with a new evolution phase.
In this phase, if any individual has a better performance than the $N_{\rm sub}$ individuals in the simplex, then the least performing individuals among the $N_{\rm sub}$ individuals are replaced.
Thus, each evolution phase replaces elements in the simplex, allowing exploration beyond the initial subspace.
Then, the optimization continues with the exploitation phase on the updated $N_{\rm sub}$ individuals.

%This process is repeated until the stopping criterion is reached.
%The \emph{Gradient-enriched Machine Learning Control (gMLC)} is summarized by pseudo code in algorithm~\ref{algo:gMLC}.

\begin{algorithm}
%\DontPrintSemicolon
\SetKwBlock{ExplorPhase}{Exploration phase---Evolution}{end}
\SetKwBlock{ExploiPhase}{Exploitation phase---Downhill subplex}{end}
\SetAlgoLined
\KwResult{$\bm K^*$, the best individual }
Monte Carlo initialization: generate $N_{\rm MC}$ individuals\;
Test all the individuals\;
Build the subplex $\mathcal{S}$ by taking the $N_{\rm sub}$ best individuals\;
%Sort the individuals by performance $J_1 \le J_2 \ldots \le J_{\rm sub}$\;
 \While{Stopping criterion is not reached}{
\ExplorPhase{
Generate and test $N_G$ individuals from all the individuals evaluated so far thanks to crossover and mutation\;
   Update subplex $\mathcal{S}$: choose the $N_{\rm sub}$ best individuals among the new $N_G$ individuals and the $N_{\rm sub}$ subplex individuals\;
   }
 \ExploiPhase{
 \While{The number of subplex individuals generated $<$ $N_G$}{
    	Perform a downhill subplex iteration 
    	in the subspace spanned by linear combinations 
    	of $N_{\rm sub}$ subplex control laws
    	\newline (Downhill simplex method like in algorithm \ref{algo:EGM})\;
   }
\textbf{Reconstruction phase---Linear genetic programming}
 
Compute a matrix representation for each new downhill subplex individual (replace linearly-combined individuals by matrices using LGP);
 }
 }
 \caption{\label{algo:gMLC}Gradient-enriched Machine Learning Control}
\end{algorithm}

\begin{figure}
  \centerline{\includegraphics[width=0.75\linewidth]{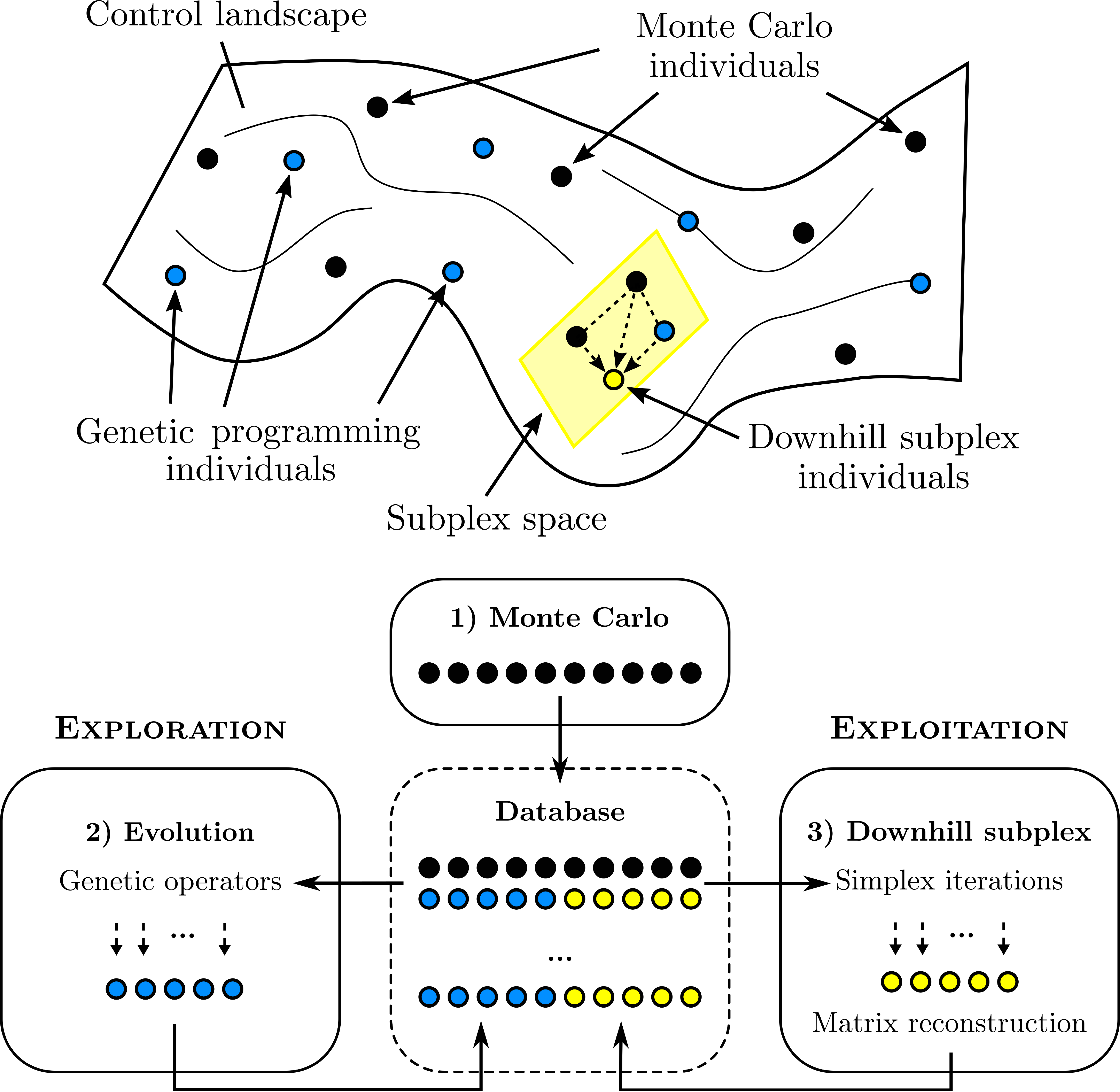}}
  \caption{Schematic of the gradient-enriched MLC algorithm (bottom) and distribution of individuals in the search space (top) .
  First (1),  Monte Carlo initialization performs a first coarse exploration of the search space.
  Second (2), 
  further exploration is performed thanks to genetic programming.
  Individuals are selected in the whole dataset and combined thanks to genetic operators to generate new individuals (blue dots).
%  The evolution step requires that the individuals have a matrix representation to be operated.
Then the database is augmented with the new individuals.
  Third (3), exploitation focuses on a subspace (represented in yellow) of finite dimension where downhill simplex iterations builds new individuals by linear combination (yellow dots).
%A matrix representation is rebuild for the interpolated individuals thanks to linear genetic programming (3b.). 
A matrix representation is computed for the downhill subplex individuals thanks to linear genetic programming , allowing the downhill subplex individuals to be included in the database.
  }
\label{fig:gMLC_principle}
\end{figure}\begin{figure}
  \centerline{\includegraphics[width=\linewidth]{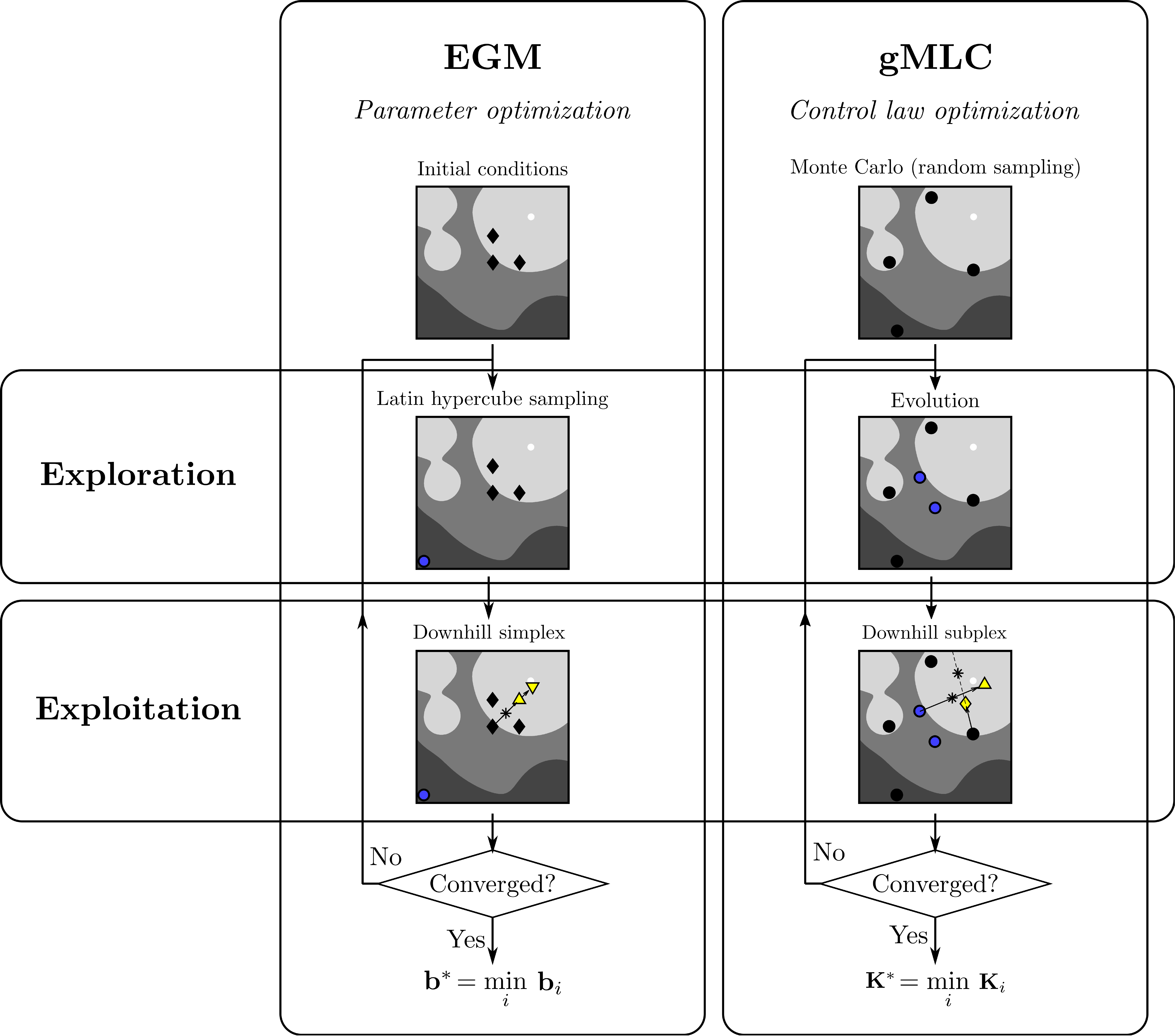}}
  \caption{Summary of the explorative gradient method (EGM) (left column) and gradient-enriched machine learning control (gMLC) (right column).
The level plots are a schematic representation of the control landscape.
Darker regions depict poor performances and light regions depict good performances.
Three minima are shown, two on the top left and the global one in the top right.
The map represents an affine space (of finite dimension) for EGM and a Hilbert function space for gMLC.
The initialization step is depicted with black diamonds for EGM and black dots for gMLC.
The individuals generated thanks to an exploration phase are represented by blue dots.
Exploration is carried out with LHS for EGM and evolution with genetic operators (crossover and mutation) for gMLC.
The individuals generated thanks to an exploitation phase are represented in yellow.
For EGM, downhill simplex steps are carried out.
The associated level plot depicts one iteration of downhill simplex: the reflected individual (yellow triangle) and the expanded individual (reversed yellow triangle), the star is the centroid of the two best black diamonds.
For gMLC, the simplex steps are carried out in a subspace (downhill subplex) of finite dimension.
The associated level plot depicts two distinct simplex steps: first, a reflection step (yellow triangle) with the two best black dots and the best blue dot; then a contraction step (yellow diamond) with the same black dots and the newly evaluated yellow triangle.
The stars are the centroids for each step.
This process is repeated until the stopping criterion is reached.
In this figure, only one iteration of the loop is depicted.
The reconstruction phase is not depicted for the sake of clarity.}
\label{fig:algorithms}
\end{figure}
Figure~\ref{fig:gMLC_principle} illustrates 
the initialization, exploration and exploitation of gMLC.
The exploration is based on LGP. 
Also the exploitation requires LGP.
In the downhill simplex method, 
the individuals are linear combinations of the subplex basis
and are finally approximated as matrices.
This process is repeated until the stopping criterion is reached.
The \emph{Gradient-enriched Machine Learning Control (gMLC)} is summarized by pseudo code in algorithm~\ref{algo:gMLC}.
The source code is freely available at \url{https://github.com/gycm134/gMLC}.
Finally, figure~\ref{fig:algorithms} summarizes the exploration and exploitation phases for EGM and gMLC.

%It is worthy to note that even though the evolution phase comprises an exploitation process (crossover), its efficiency is lesser than \textrm{simplex}'s linear combination to explore the neighborhood of an individual.

%-----------------------------------------------------------------------
\section{Flow stabilization}\label{Sec:Results}
In this section, 
we stabilize the fluidic pinball with optimized control laws 
in increasingly more general search spaces.
First (\S~\ref{sec:results_paramstudy}), 
we consider  symmetric steady actuation
with a parametric study reduced to one parameter $b_2=-b_3=\hbox{const}$. 
Then (\S~\ref{sec:results_EGM}), 
we optimize steady actuation allowing also for non-symmetric forcing, 
i.e., 3 independent inputs $b_1$, $b_2$, $b_3$.
Finally (\S~\ref{sec:results_gMLC}), 
we optimize sensor-based feedback from 9 downstream sensor signals 
driving the 3 cylinder rotations.
Evidently, the three search spaces are successive generalizations.

%***********************************************************************
\subsection{Symmetric steady actuation---Parametric study} \label{sec:results_paramstudy}

%-----------------------------------------------------------------------
\begin{figure}
  \centerline{\includegraphics[width=\linewidth]{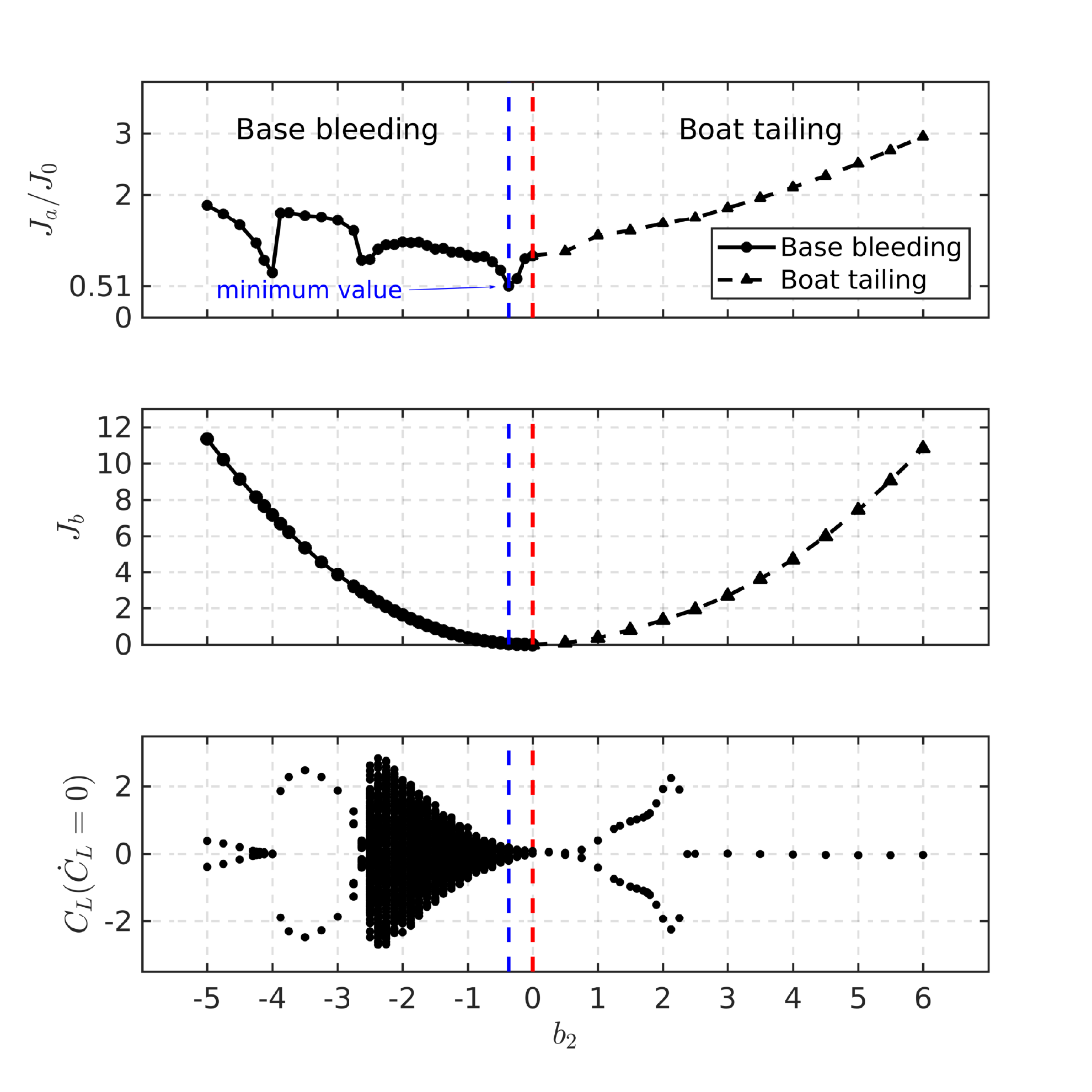}}
  \caption{Parametric study for symmetric steady forcing.
$b_2=-b_3$ is the velocity of the bottom cylinder.
The normalized distance to the steady solution $J_a/J_0$ (top) and the actuation power $J_b$ (middle) are plotted as a function of $b_2$. 
The bifurcation diagram (bottom) 
comprises all  local maximum and minimum lift values.
The vertical red dashed line corresponds to $b_2=0$  and separates the base bleeding and the boat tailing configurations.
The global minimum of $J_a/J_0$ is reached at $b_2=-0.375$,  as indicated by a vertical blue dashed line.}
\label{fig:Constant}
\end{figure}
This section describes the behaviour 
of the fluidic pinball under a symmetric steady actuation.
In this configuration, 
only the two rearward cylinders rotate at equal but opposite rotation speeds, $b_2=-b_3$.
When $b_2$ is positive, 
the rearward cylinders accelerate the outer boundary layers and suck near-wake fluid upstream.
This forcing delays separation, mimics Coanda forcing 
and leads to a fluidic boat tailing.
When $b_2$ is negative, 
the cylinders eject fluid in the near wake like in base bleed
and oppose the outer boundary-layer velocities.
Figure \ref{fig:Constant} shows the evolution of $J_a/J_0$ (top), $J_b$ (middle) and the bifurcation diagram (bottom) as a function of $b_2$.

We limited our study to $b_2 \in [-5, \> 6]$.
%At the extremal velocity,
%the actuation requires much more power than for the parasitic drag power.
% GYCM: We do not consider drag power in this study.
The trends are resolved with a discretization step of $0.25$ 
and a finer resolution in the ranges $[-2.5,0]$ and $[1,2]$.
For each parameter, the cost $J_a$ and actuation power $J_b$ have been computed over 1000 convective time units.
The bifurcation diagram has been built by detecting the extrema of the lift coefficient over the last 600 convective time units.
The bifurcation diagram reveals five regimes:
%-----------------------------------------------------------------------
\begin{description}
    \item[Regime $b_2<-4$:]\ the lift amplitude decreases to zero and the cost decreases to the first minimum.
    \item[Regime $-4<b_2<-2.5$:]\ the extremal lift values increase and decrease to zero again.
    The cost approaches another local minimum near $b \approx -2.5$.
    \item[Regime $-2.54<b_2<0$:]\  a period doubling cascade is observed for decreasing $b_2$ leading to a chaotic regime.
    At $b_2 \approx 0.375$, the cost assumes it global minimum with residual fluctuation of the lift coefficient.
    \item[Regime $0<b_2<2.375$:]\ the cost and the extremal lift values monotonically increase.
    \item[Regime $2.375<b_2$:]\ the Coanda forcing completely stabilizes a symmetric steady solution. The cost increases with the rotation speed.
\end{description}
%-----------------------------------------------------------------------
%It should be noted that the forced symmetric steady solution for $b_2>2.375$ is different from the unforced symmetric steady solution.% as we can see in figure~.
Interestingly, the boat tailing discontinuity at $b_2=2.375$ does not appear in the graph of the cost function $J_a/J_0$. This continuity, even in the derivative, corresponds to a continuous passage  from a periodic symmetrical solution to a stationary solution which is itself symmetrical.
 As the value of the cost function indicates, 
 this stationary solution is quite far from the unforced symmetric steady solution.
The global minimum of $J_a/J_0=0.51$ is reached near $b_2=-0.375$, 
i.e., for a base bleeding configuration,  corresponding to a small actuation power $J_b= \num{0.0490}$,
roughly 0.1\% of the $J_0$. % We do not talk about parasite drag

\begin{figure}
\centering
\begin{subfigure}{.45\textwidth}
  \centering
\includegraphics[width=1.\textwidth]{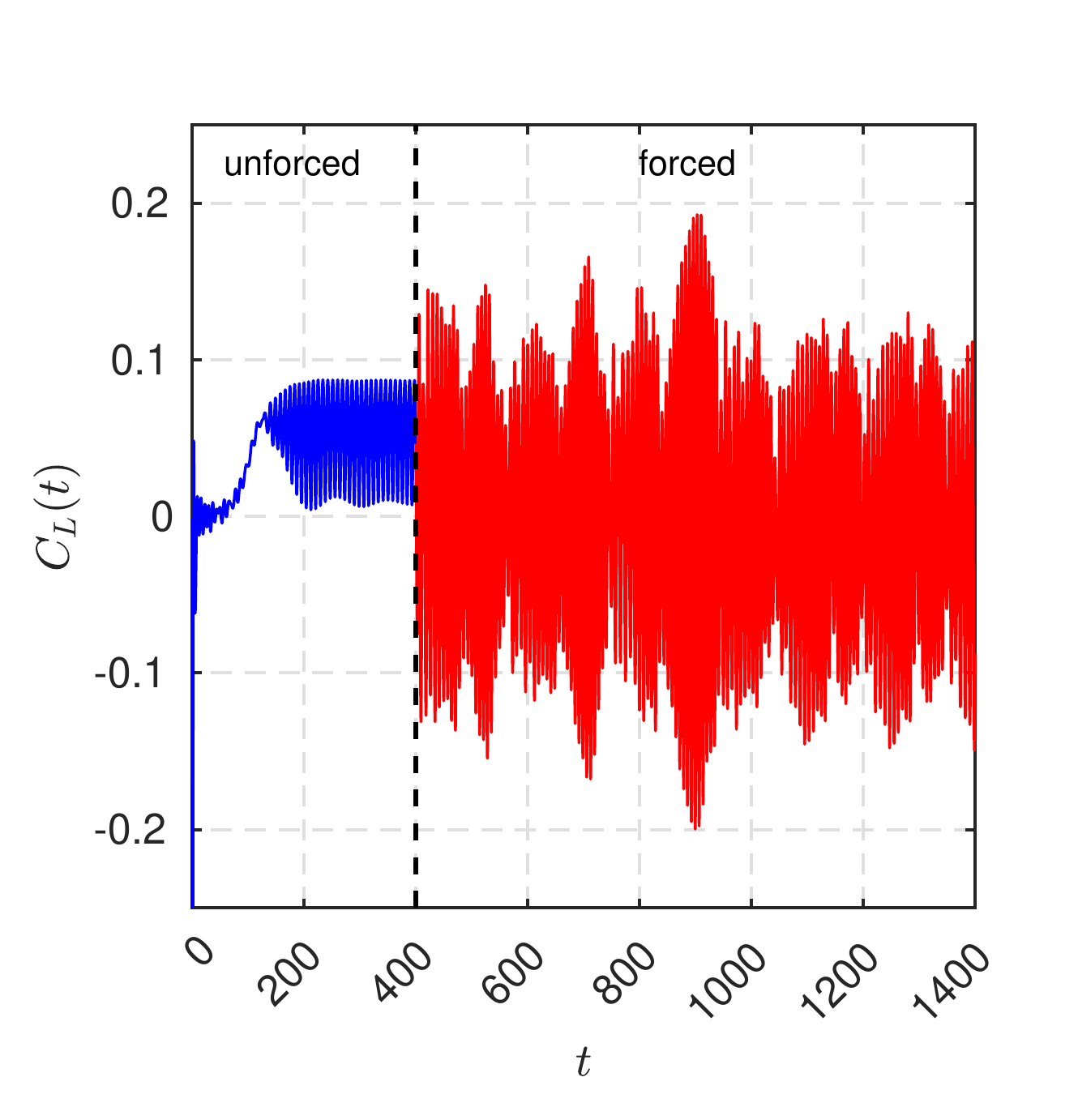}
\caption{\label{fig:CL_BB}}
\end{subfigure}%
\hfil
\begin{subfigure}{.45\textwidth}
  \centering
\includegraphics[width=1.\textwidth]{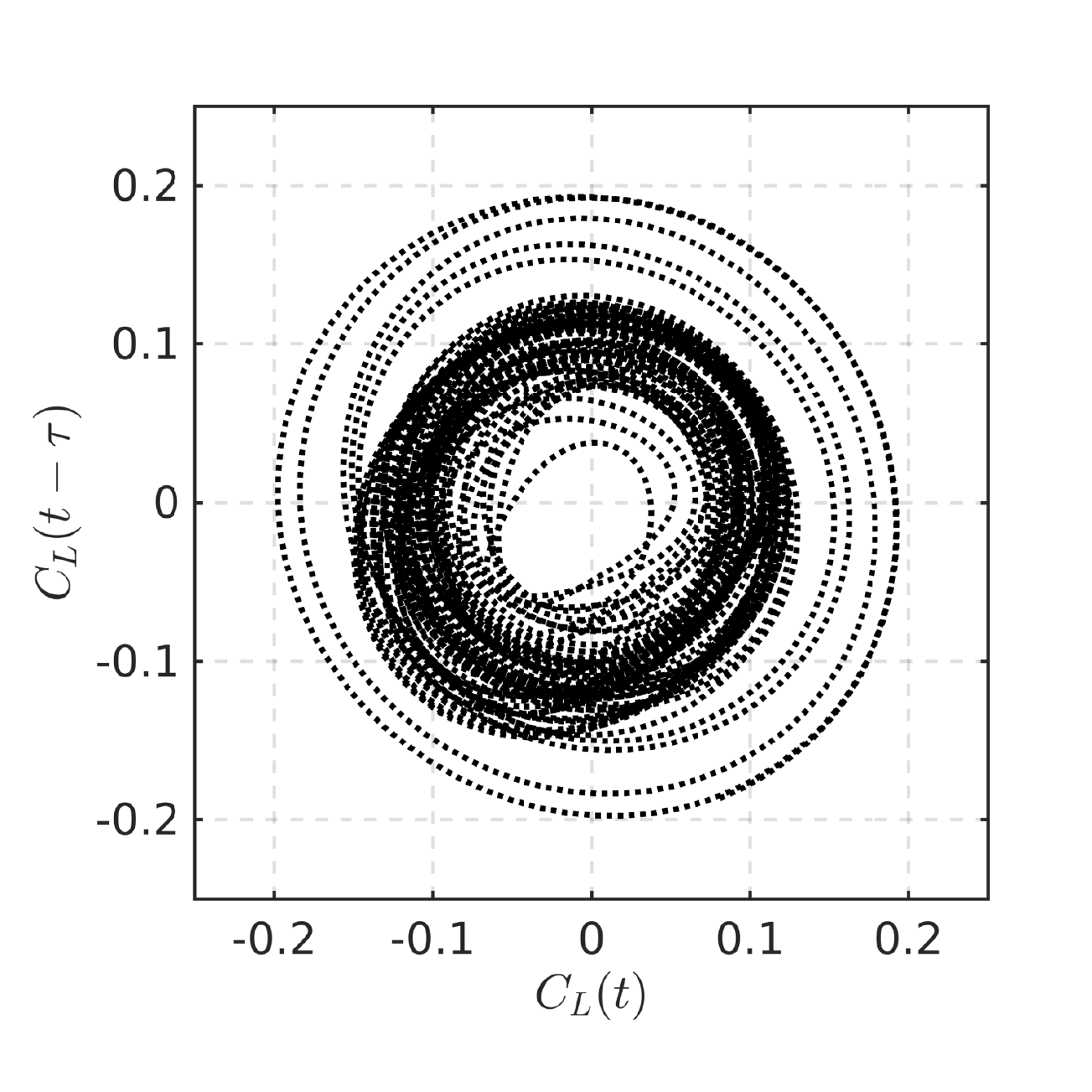}
\caption{\label{fig:PP_BB}}
\end{subfigure}

\begin{subfigure}{.45\textwidth}
  \centering
\includegraphics[width=1.\textwidth]{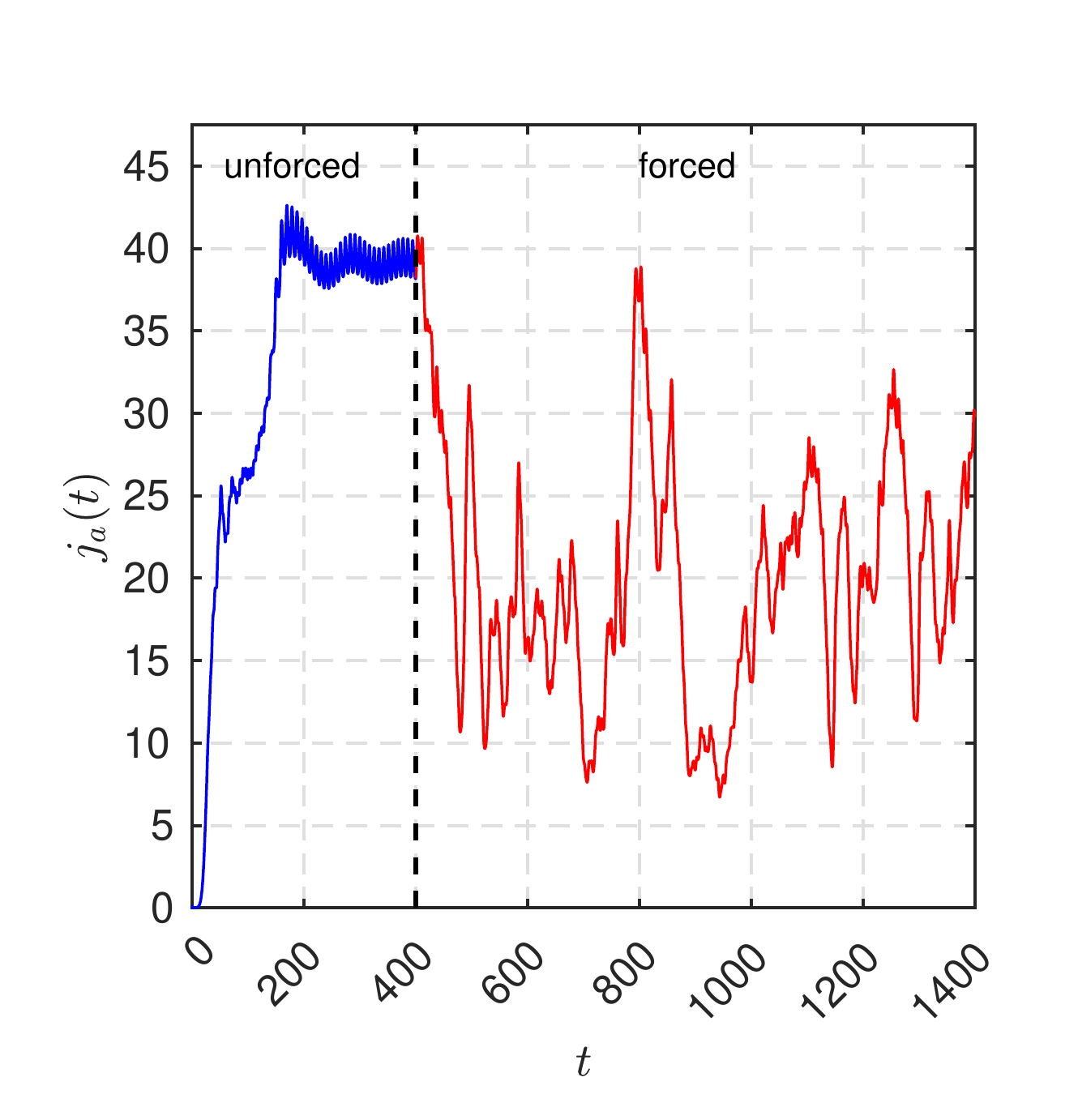}
\caption{\label{fig:DSS_BB}}
\end{subfigure}%
\hfil
\begin{subfigure}{.45\textwidth}
  \centering
\includegraphics[width=1.\textwidth]{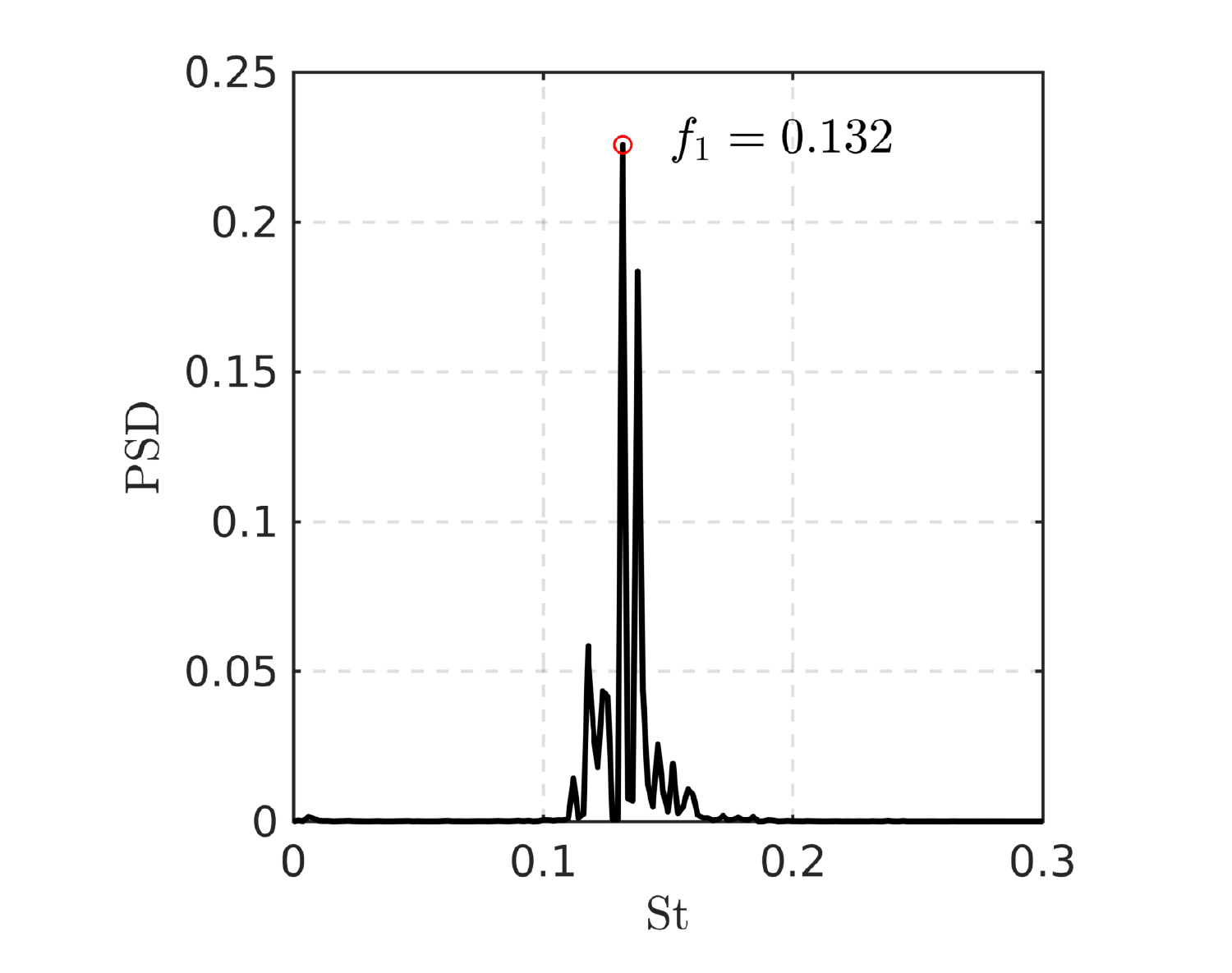}
\caption{\label{fig:PSD_BB}}
\end{subfigure}
\caption{\label{fig:BB_description}Characteristics of the best base bleeding solution. (a) Time evolution of the lift coefficient $C_L$, (b) phase portrait (c) time evolution of instantaneous cost function $j_a$ and (d) Power Spectral Density (PSD) showing a broad spectral peak at $f_1=0.132$. 
The control starts at $t=400$.
The unforced phase is depicted in blue and the forced one in red.
The phase portrait is computed over $t \in [900,1400]$ and the PSD is computed on the forced regime $t \in [400,1400]$.}
\end{figure}

\begin{figure}
\centering
\begin{subfigure}{.45\textwidth}
  \centering
\includegraphics[width=1.\textwidth]{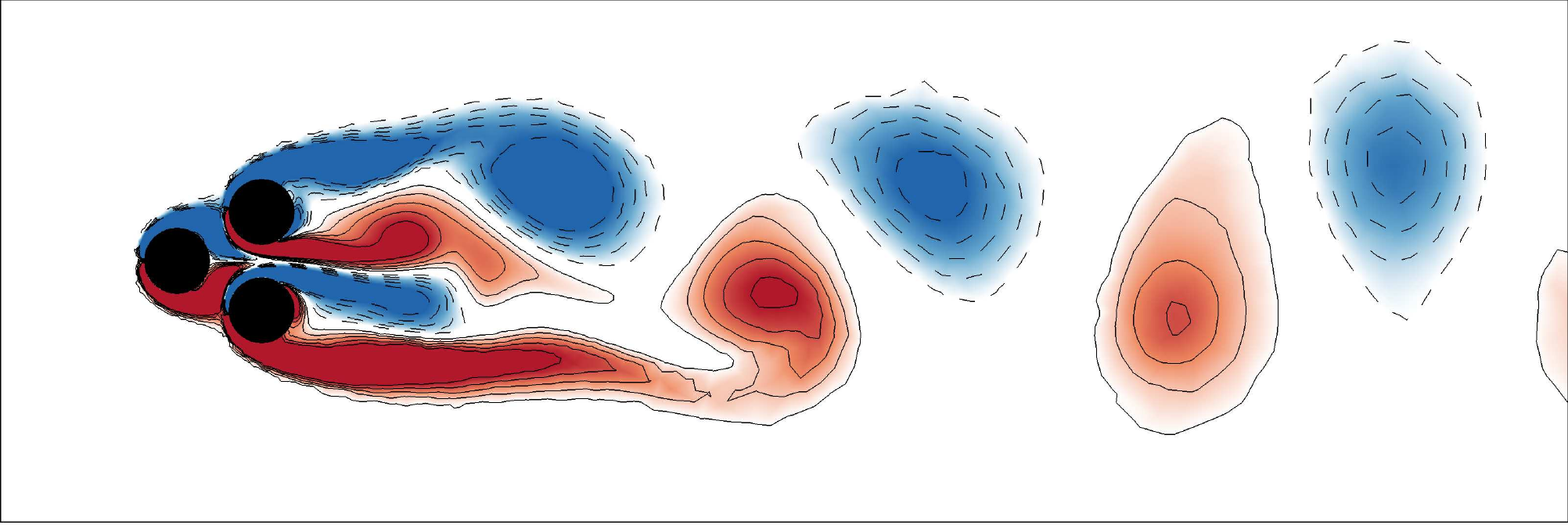}
\caption{\label{fig:Param_T1}$t+T_1/8$}
\end{subfigure}%
\hspace{0.5cm}
\begin{subfigure}{.45\textwidth}
  \centering
\includegraphics[width=1.\textwidth]{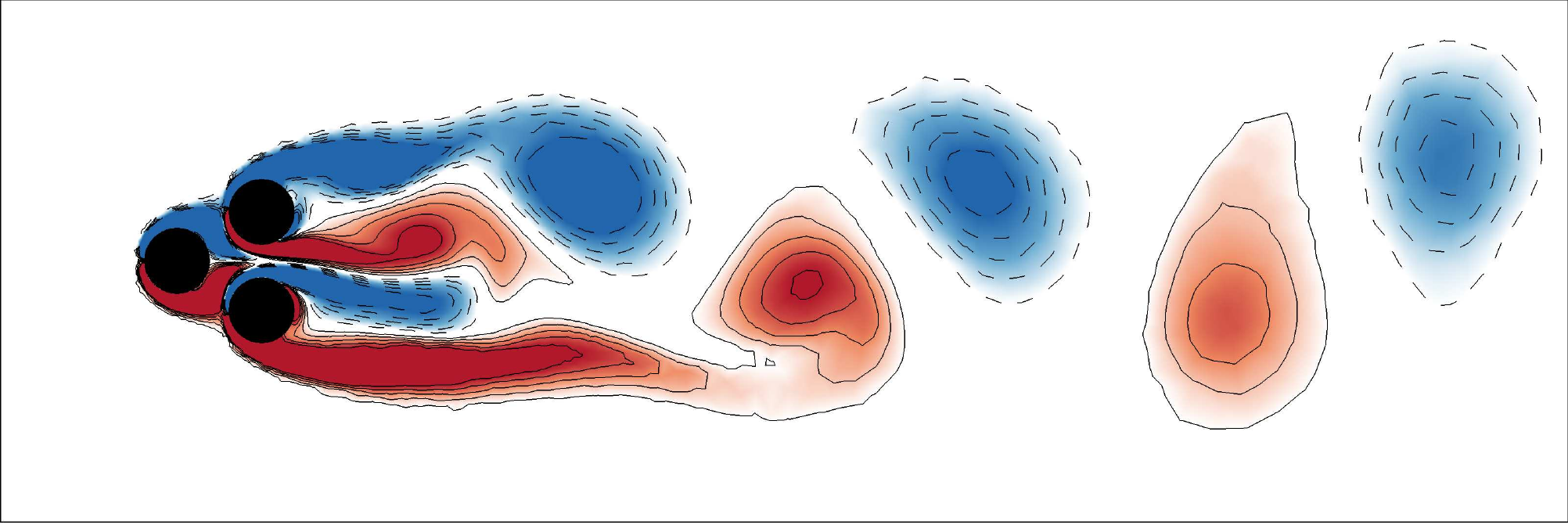}
\caption{\label{fig:Param_T2}$t+2T_1/8$}
\end{subfigure}

\begin{subfigure}{.45\textwidth}
  \centering
\includegraphics[width=1.\textwidth]{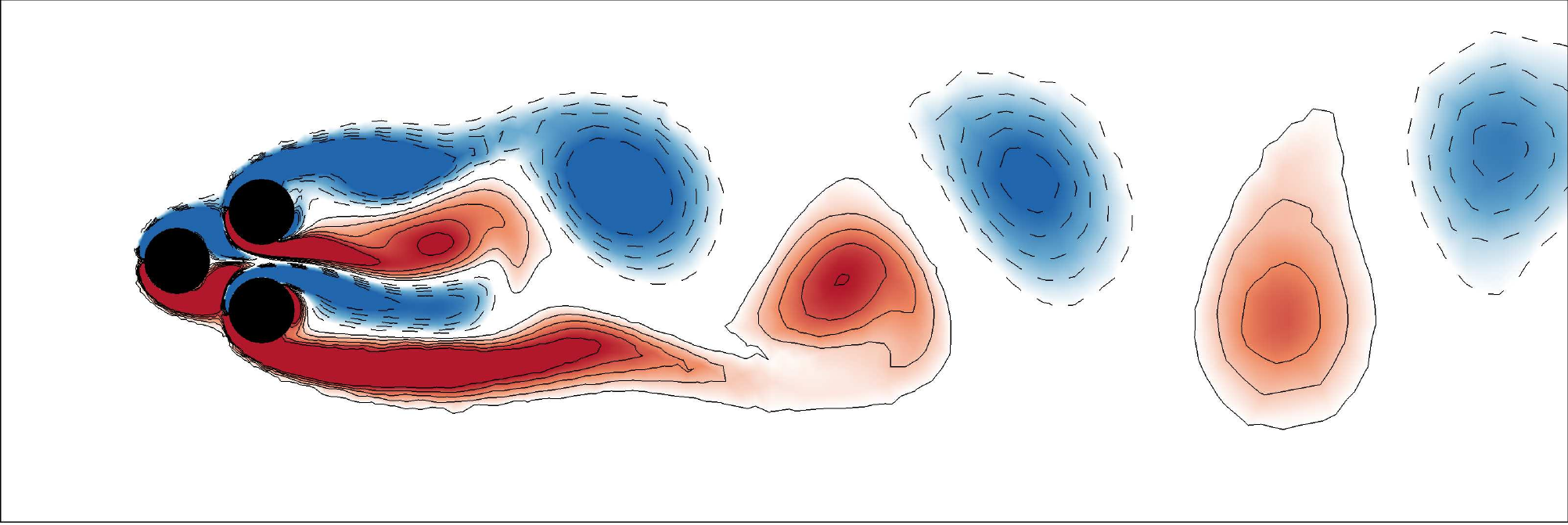}
\caption{\label{fig:Param_T3}$t+3T_1/8$}
\end{subfigure}%
\hspace{0.5cm}
\begin{subfigure}{.45\textwidth}
  \centering
\includegraphics[width=1.\textwidth]{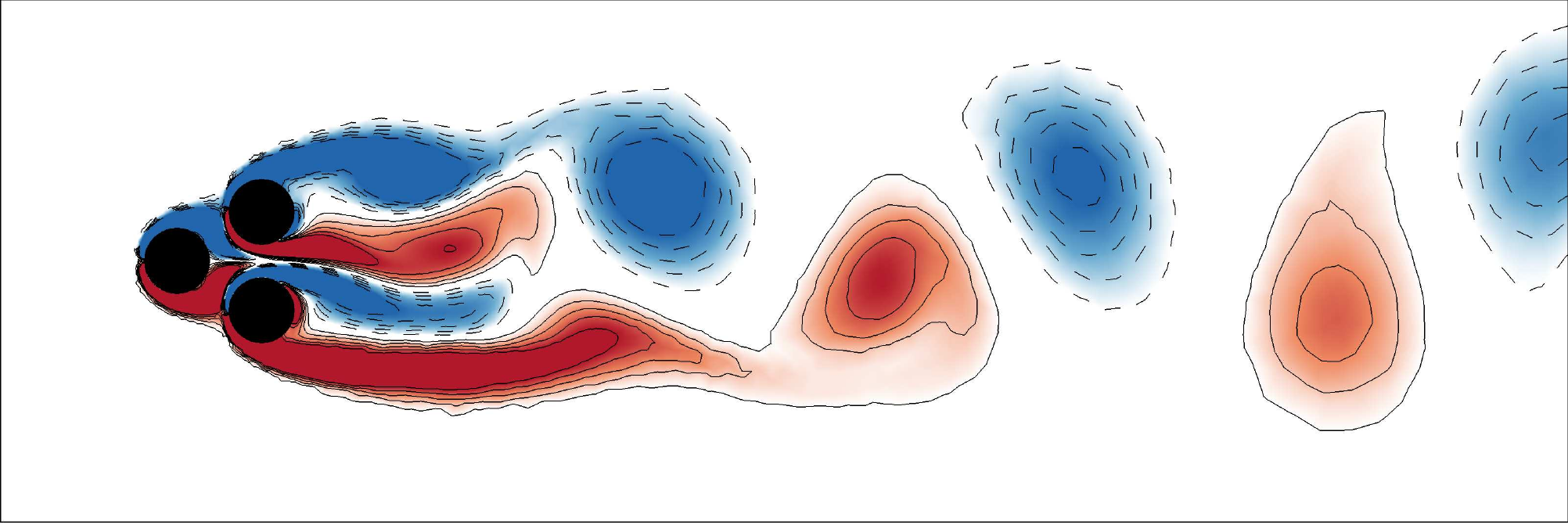}
\caption{\label{fig:Param_T4}$t+4T_1/8$}
\end{subfigure}

\begin{subfigure}{.45\textwidth}
  \centering
\includegraphics[width=1.\textwidth]{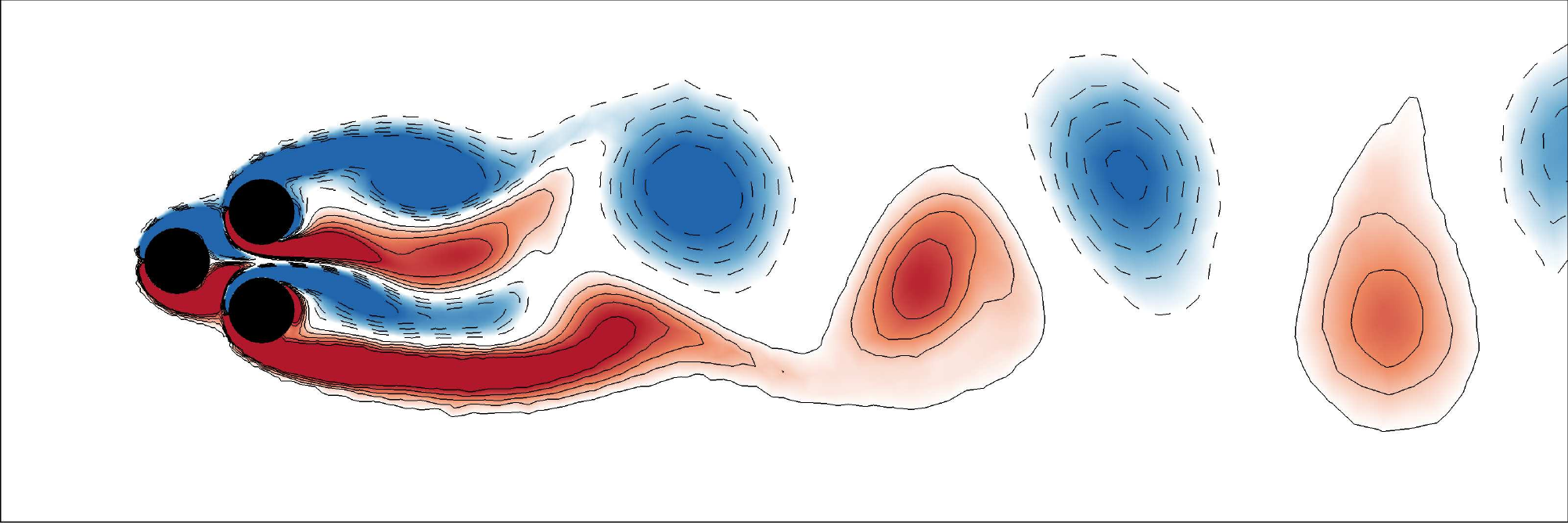}
\caption{\label{fig:Param_T5}$t+5T_1/8$}
\end{subfigure}%
\hspace{0.5cm}
\begin{subfigure}{.45\textwidth}
  \centering
\includegraphics[width=1.\textwidth]{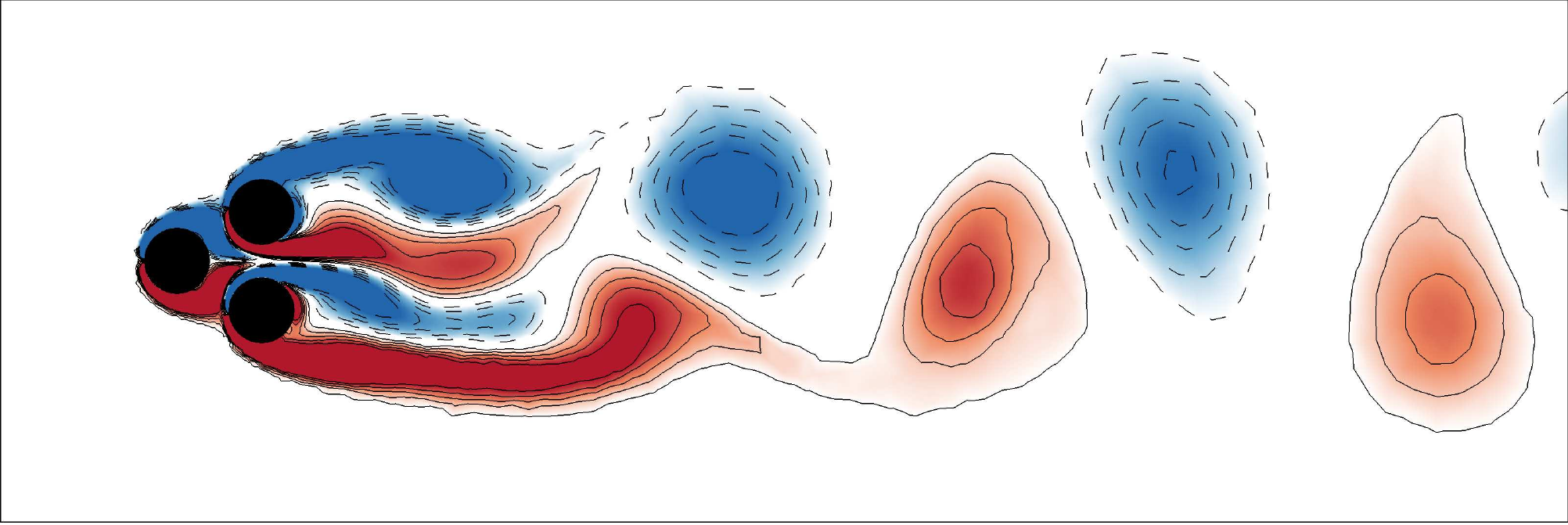}
\caption{\label{fig:Param_T6}$t+6T_1/8$}
\end{subfigure}

\begin{subfigure}{.45\textwidth}
  \centering
\includegraphics[width=1.\textwidth]{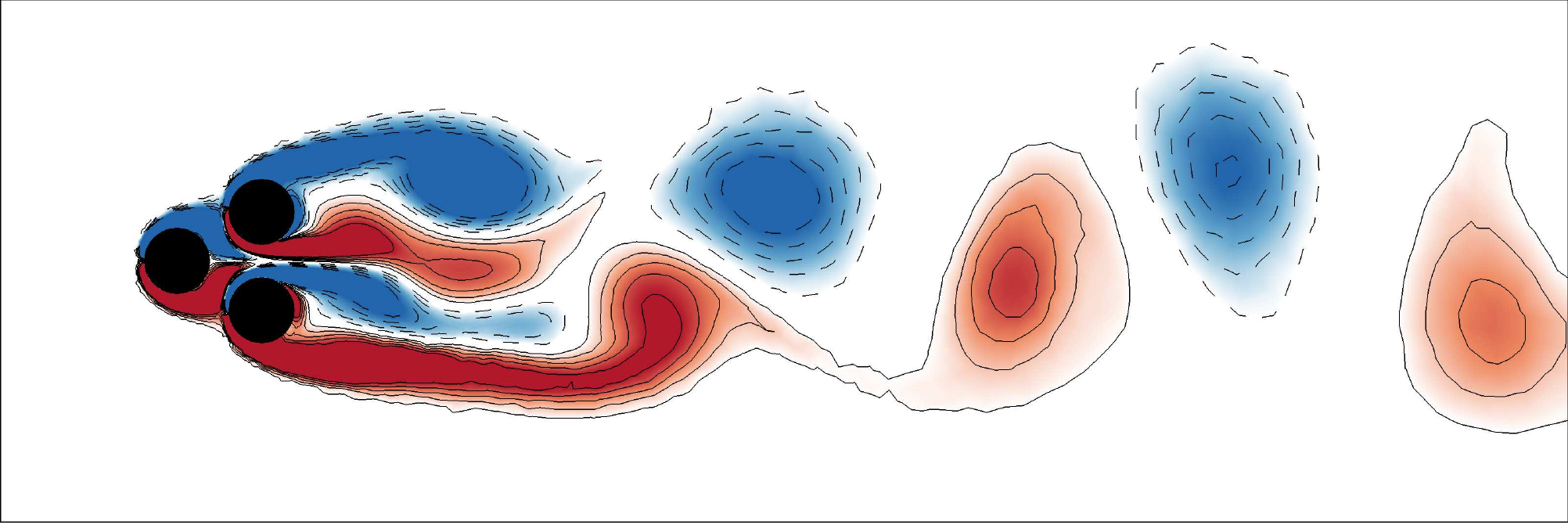}
\caption{\label{fig:Param_T7}$t+7T_1/8$}
\end{subfigure}%
\hspace{0.5cm}
\begin{subfigure}{.45\textwidth}
  \centering
\includegraphics[width=1.\textwidth]{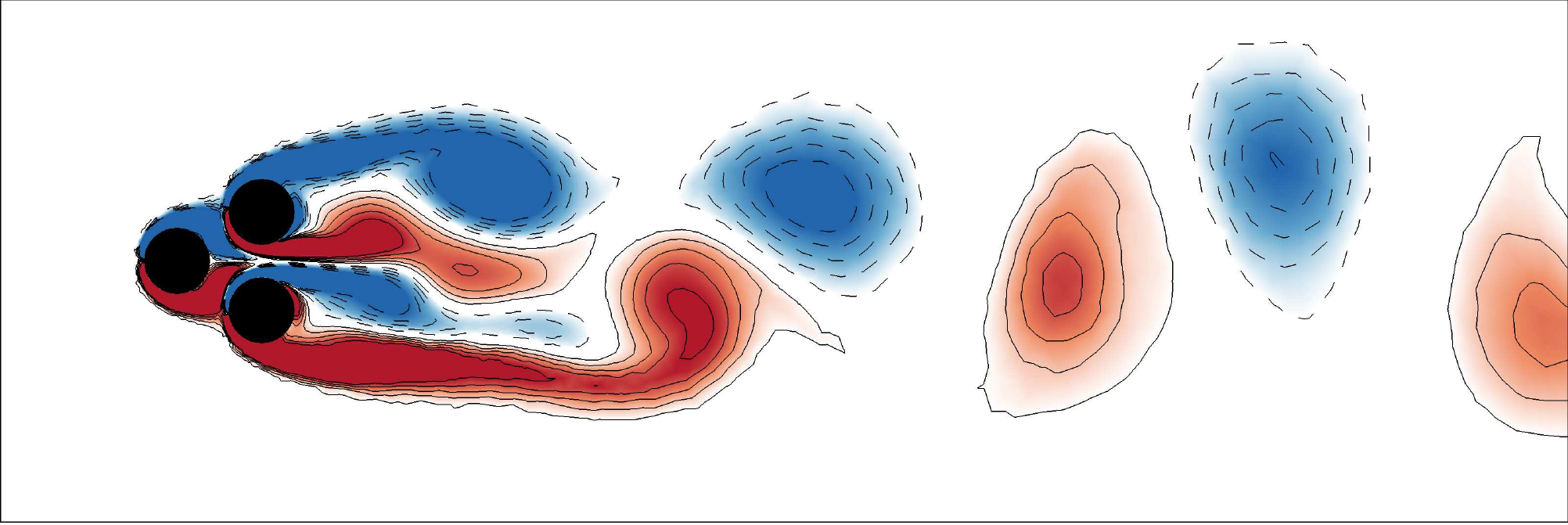}
\caption{\label{fig:Param_T8}$t+T_1$}
\end{subfigure}

\begin{subfigure}{.45\textwidth}
  \centering
\includegraphics[width=1.\textwidth]{Figures/Snapshots/SteadySolution}
\caption{Symmetric steady solution}
\end{subfigure}%
\hspace{0.5cm}
\begin{subfigure}{.45\textwidth}
  \centering
\includegraphics[width=1.\textwidth]{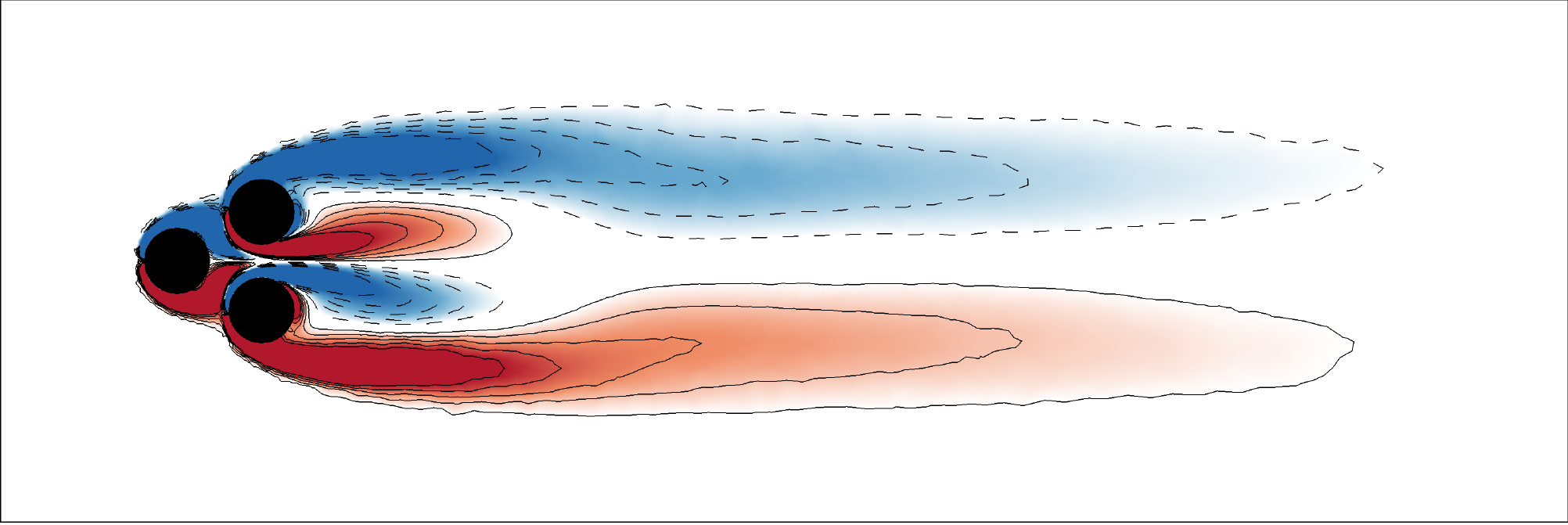}
\caption{\label{fig:Mean_Param}Mean field}
\end{subfigure}
\caption{\label{fig:Param_snap}Vorticity fields of the best base bleeding solution. (a)-(f) Time evolution of the vorticity field throughout the last period of the 1400 convective time units, (i) the objective symmetric steady solution and (j) the mean field of the forced flow.
The color code is the same as figure~\ref{fig:unforced_flow}.
$T_1$ is the period associated to the main frequency $f_1$ of the forced flow.
The mean field has been computed by averaging 100 periods.
}
\end{figure}

The characteristics of the best base bleeding solution 
leading closest to the symmetric steady solution are depicted in figure~\ref{fig:BB_description}.
In figure~\ref{fig:CL_BB}, 
the lift coefficient is displayed for the unforced transient (blue curve)
and the forced flow (red curve).
The unforced flow terminates in an asymmetric shedding with positive lift values.
After the start of forcing,  
 the lift coefficient oscillates vigorously around its vanishing mean value.
This forced statistical symmetry is corroborated 
by the oscillating jet in figures~\ref{fig:Param_T1}-\ref{fig:Param_T8}.
Base bleed increases the velocity of the rearward jet 
compared to the unforced flow. 
This jet instability mitigates the Coanda effect on the bottom and top cylinder,
i.e., the jet neither stays long at either side.

%% Figures : Base-bleeding flow characteristics ------------------------
% Base-bleeding flow snapshots -----------------------------------------
The vortex shedding persists similar to the unforced flow.
However,  the dominant frequency is increase from $f_0=0.116$ to $f_1=0.132$.
The instantaneous cost function $j_a$ in figure~\ref{fig:DSS_BB} 
shows an unsteady non-periodic behavior, 
reaching  intermittently low levels.
The broad spectral peak in figure~\ref{fig:PSD_BB} 
is a characteristic of a chaotic regime.
The phase portrait  in figure~\ref{fig:PP_BB} corroborates the non-periodic oscillatory behaviour.  
The mean field in figure~\ref{fig:Mean_Param}, 
shows that actuated mean jet is symmetric unlike the mean field of the unforced flow.
Moreover, the shear-layer  on the upper and lower sides extend further downstream as compared to the unforced state.

This parametric study reveals that base bleeding is the best symmetric steady forcing strategy to bring the flow close to the symmetric steady solution.
However, even though the cost $J_a/J_0$ is almost halved, 
the best base bleeding control fails to stabilize the flow.

% -------------------------------------------------------------------------------------------------------
%% 				Explorative Gradient Method results
%% -------------------------------------------------------------------------------------------------------
\subsection{General non-symmetric steady actuation---Explorative gradient method} \label{sec:results_EGM}
\begin{figure}
  \centerline{\includegraphics[width=0.95\linewidth]{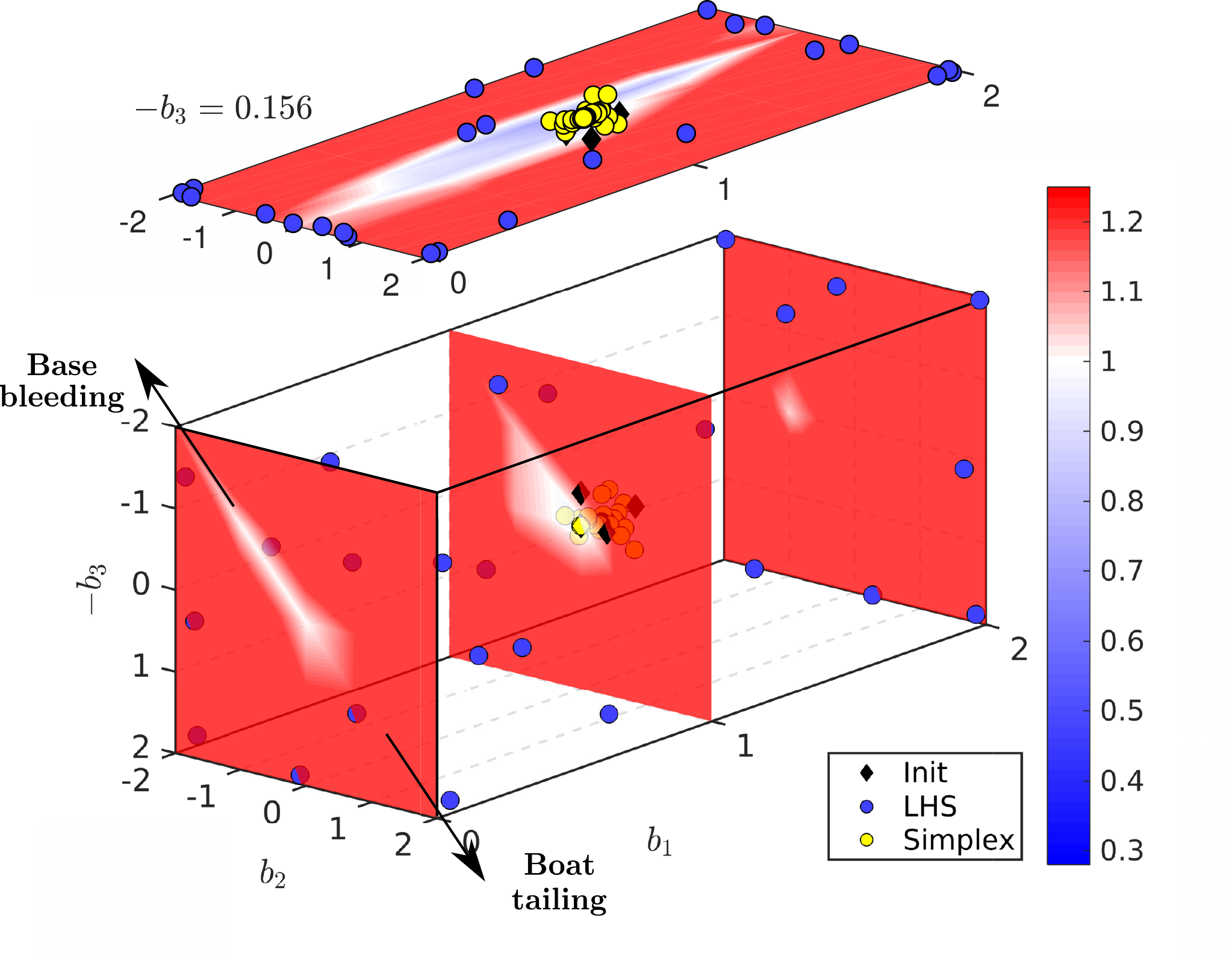}}
  \caption{Contour map of $J_a/J_0$ at the optimal plane $b_3=b^{\rm EGM}=-0.156$ found with EGM (top) and at different levels of $b_1$: $b_1=0$, $b_1=1$, $b_1=2$ (bottom).
The color code denotes white for $J_a/J_0=1$, blue for better performances and red for worse performances.
The planes are shown with $75\%$ transparency.
The four initial conditions $[1,0,0]^\intercal$, $[1.2,0,0]^\intercal$, $[1,0.4,0]^\intercal$ and $[0,0,0.4]^\intercal$ are represented by black diamonds.
Blue dots are the control laws build with the exploration phases and yellow dots are the individuals build with the exploitation phases.
All the individuals have been projected on the plane $b_3=-0.156$.
The arrows, on plane $b_1=0$, depict the base bleeding/boat tailing diagonal studied in section \S~\ref{sec:results_paramstudy}.
A parametric study shows that the minimum is close to $[b_1,b_2,b_3]^\intercal=[1,0,0]^\intercal$ whose cost  is $J_a/J_0$=0.93.}
\label{fig:Slice_Ja}
\end{figure}

In this section we aim to stabilize the symmetric steady solution 
by commanding the three cylinders with constant actuation without symmetry constraint.
This three-dimensional parameter space is explored 
with the explorative gradient method presented in section \S~\ref{sec:EGM}.
The symmetry along the $x$-axis of the fluidic pinball allows us to reduce our search space and to explore only positive values of $b_1$.
A coarse initial parametric study carried on $b_1$, $b_2$ and $b_3$ 
by steps of unity indicates that the global minimum of $J_a/J_0$ should be near  $[b_1,b_2,b_3]^\intercal = [1,0,0]^\intercal$.
Thus, we limit our research to the actuation domain $\mathcal{B} = [0,2]\times[-2,2]\times[-2,2]$.
The limitation of $b_1$ to positive values exploits the mirror symmetry of the configuration.
Figure~\ref{fig:Slice_Ja} (bottom) depicts the cost function 
in the actuation domain $\mathcal{B}$.
Three planes ($b_1=\hbox{const}$) are computed by interpolating parameters on a coarse grid.
% The planes are not computed by interpolating the EGM individuals.
The individuals computed with EGM are all shown in the 3D space.
%An additional pla
%An additional plane at $ b_3=-0.156$, 
%White regions denoting performances equivalent to the unforced case, red regions for worse performances and blue regions for better performances.
%A small blue region is located close to $(b_1,b_2,b_3)=(1,0,0)$.
% of the parametric study  <<< I don's see the individuals from the parametric study.
The four initial control laws for EGM are the center of the box 
and shifted points from this center.
The shift is $10\%$ of the box size in positive coordinate direction.
Thus, the four initial control laws are: $[1,0,0]^\intercal$, $[1.2,0,0]^\intercal$, $[1,0.4,0]^\intercal$ and $[1,0,0.4]^\intercal$.
The exploration phase is then performed in $\mathcal{B}$.
For algorithmic reasons, the explorative points 
are chosen from 1 million points obtained from a space filling LHS.
In the following, $N_i$ denotes the number of evaluations. 
The optimization processes stops after $N_i=100$ evaluations.
This corresponds to 25 iterations of the exploration/exploitation process.
Convergence is already reached around $N_i \approx 50$.
On one hand, we notice that the exploration phases (LHS in blue) focus on the boundary of the search space.
This is consistent with the goal of LHS, 
as the furthest initial individuals are on the boundary of the box.
On the other hand, the exploitation phases (simplex in yellow) stay in the same neighborhood near the initial individuals, crawling along the local gradient to find the minimum.

% Figure : EGM Constant3------------------------------------------------

% % Figure : EGM Constant3 Jb
% \begin{figure}
%   \centerline{\includegraphics[width=0.95\linewidth]{Figures/Constant3/EGM_Slice_Jb}}
%   \caption{same as figure \ref{fig:Slice_Ja} but for $J_b$. }
% \label{fig:Slice_Jb}
% \end{figure}

%%

\begin{figure}
\centering
\begin{subfigure}{.45\textwidth}
  \centering
\includegraphics[width=1.\textwidth]{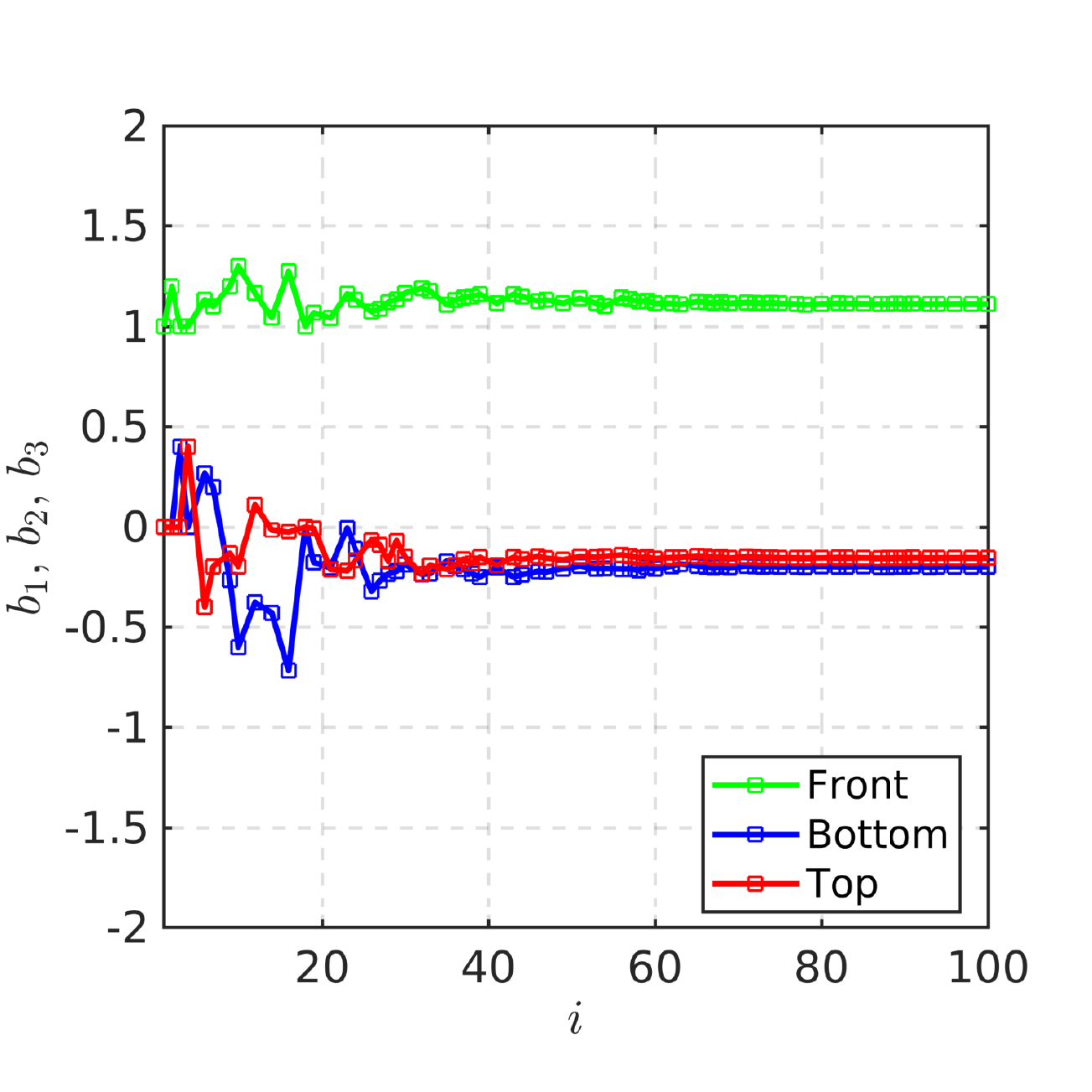}
\caption{\label{fig:b1prog} }
\end{subfigure}%
\hfil
\begin{subfigure}{.45\textwidth}
  \centering
\includegraphics[width=1.\textwidth]{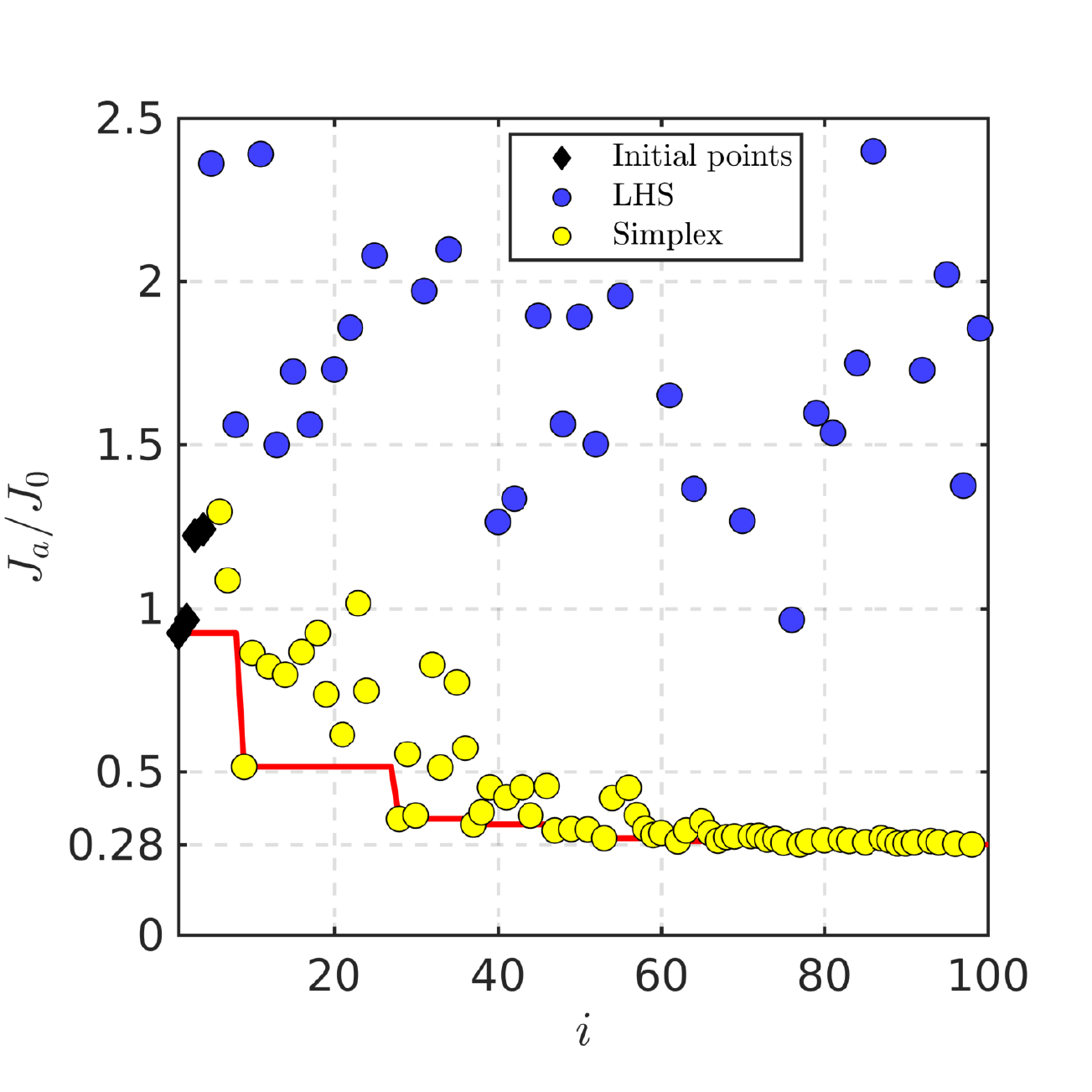}
\caption{\label{fig:Jprog} }
\end{subfigure}%
\caption{\label{fig:prog}Evolution of $b_1$, $b_2$ and $b_3$ (left) for each new simplex step indicated by the scattered squares and $J_a/J_0$ (right) according to the number of evaluations $i$ for the EGM optimization process.
The red line on (b) shows the evolution of the best cost.
The color code of the dots on (b) is the same as figure~\ref{fig:Slice_Ja}.
The best control law  is $[b_1^{\rm EGM},b_2^{\rm EGM},b_3^{\rm EGM}]^\intercal = [1.11207,-0.20025,-0.15588]^\intercal$ with $J_a/J_0=0.28$.}
\end{figure}

Figure \ref{fig:prog} shows the progression of the best control laws 
throughout the evaluations after 25 iterations of the exploration/exploitation process.
The progression is plotted according to the number of cost function evaluations
counted with the dummy index $i$.
%For EGM, the exploration step gives only one new individual at each step, whereas the exploitation step may give more individuals following the simplex step performed.
Figure~\ref{fig:b1prog} depicts the progression of the best control law after each downhill simplex step.
We notice that a plateau is reached after 50 evaluations and there are only small variations afterwards.
The final control law after 100 evaluations reads  
\begin{equation}
\label{Eqn:bEGM}
\left [ b^{\rm EGM}_1,b^{\rm EGM}_2,b^{\rm EGM}_3
\right ]^\intercal = [1.11207,-0.20025,-0.15588]^\intercal \quad \hbox{with} \quad 
 J_a=10.85
 \end{equation}

From visualizations of the control landscape of $J_a$ in figure~\ref{fig:Slice_Ja}, 
we can safely infer that \eqref{Eqn:bEGM} describes the global minimum of our search space.
Figure~\ref{fig:Jprog} shows convergence after 70 evaluations.
Thereafter,  the downhill simplex iterations show negligible improvements. 
In the whole EGM optimization, the exploration appears to be ineffective 
as the initial individuals are close to the minimum.
An EGM run with different initial individuals ( $[1,0,0]^\intercal$, $[1.5,0,0]^\intercal$, $[1,1,0]^\intercal$ and $[1,0,1]^\intercal$, corresponding to a $25\%$ of the box size shift) have been tested.
After a few iterations, this new run started sliding down towards the same minimum.
This can be explained by the fact that the neighborhood around the minimum 
is smooth enough for a downhill slide of the exploitation individuals.

% Figure : b1 and J-----------------------------------------------------

\begin{figure}
\centering
\begin{subfigure}{.45\textwidth}
  \centering
\includegraphics[width=1.\textwidth]{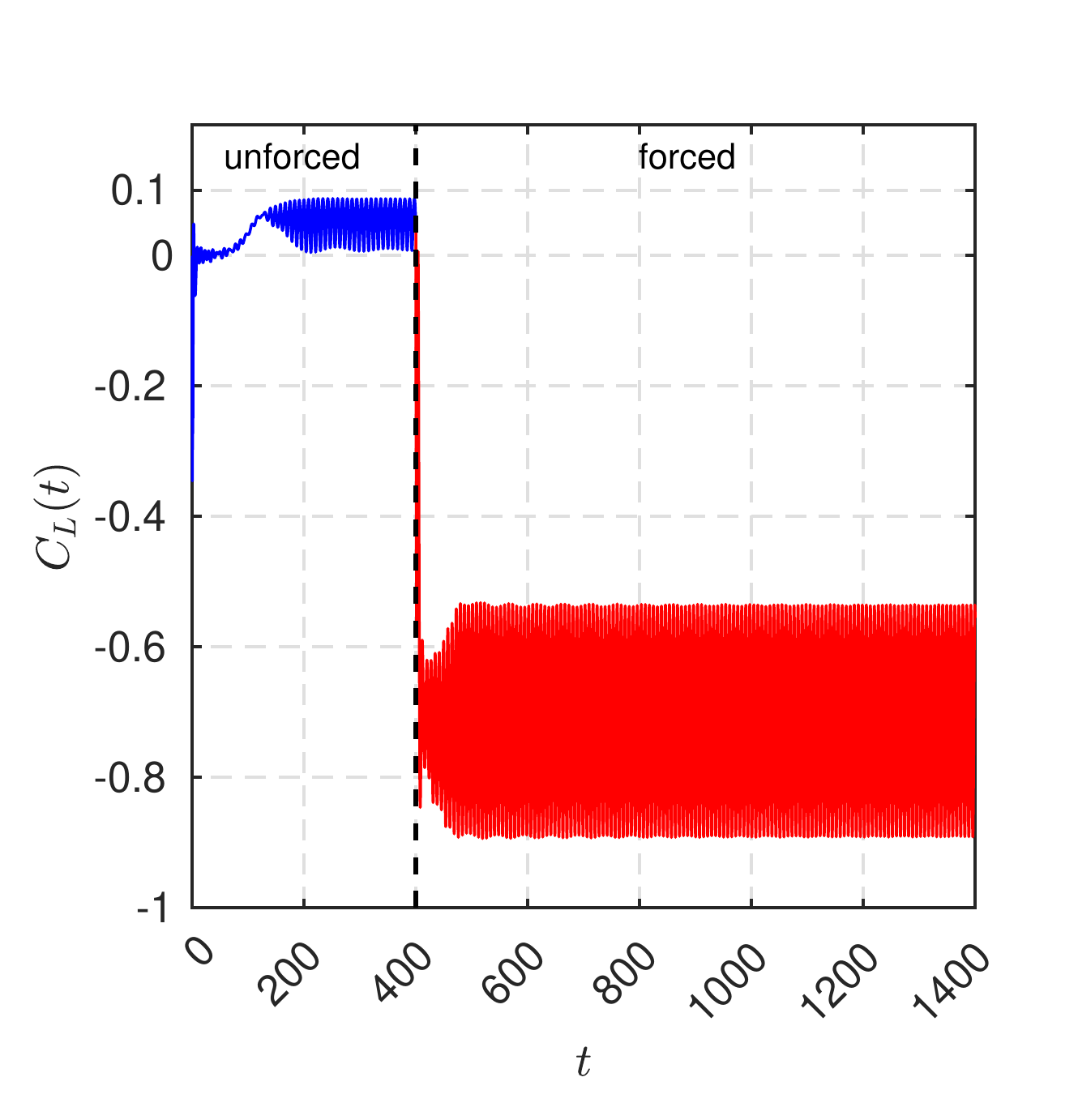}
\caption{\label{fig:CL_Constant3}}
\end{subfigure}%
\hfil
\begin{subfigure}{.45\textwidth}
  \centering
\includegraphics[width=1.\textwidth]{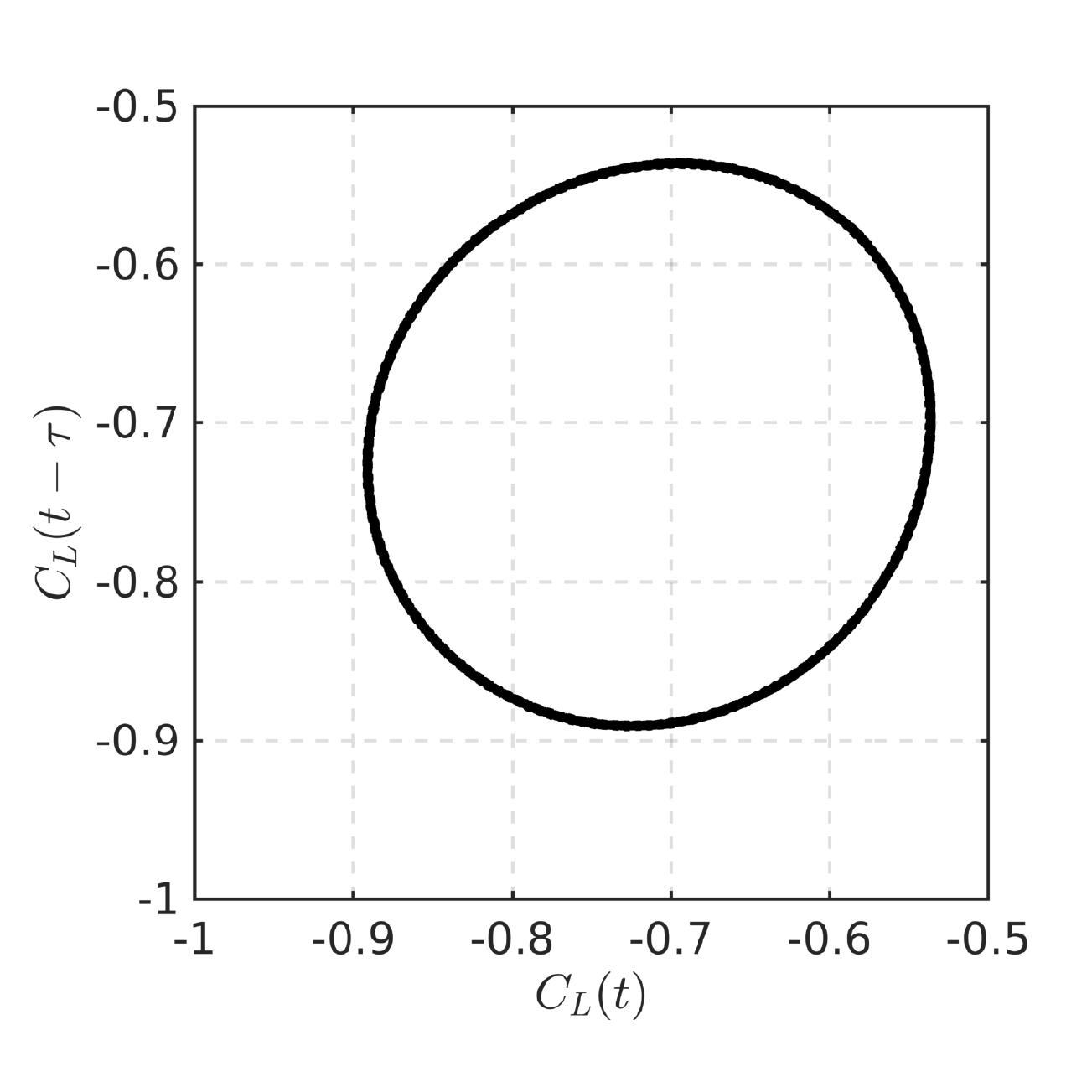}
\caption{\label{fig:PP_Constant3}}
\end{subfigure}

\begin{subfigure}{.45\textwidth}
  \centering
\includegraphics[width=1.\textwidth]{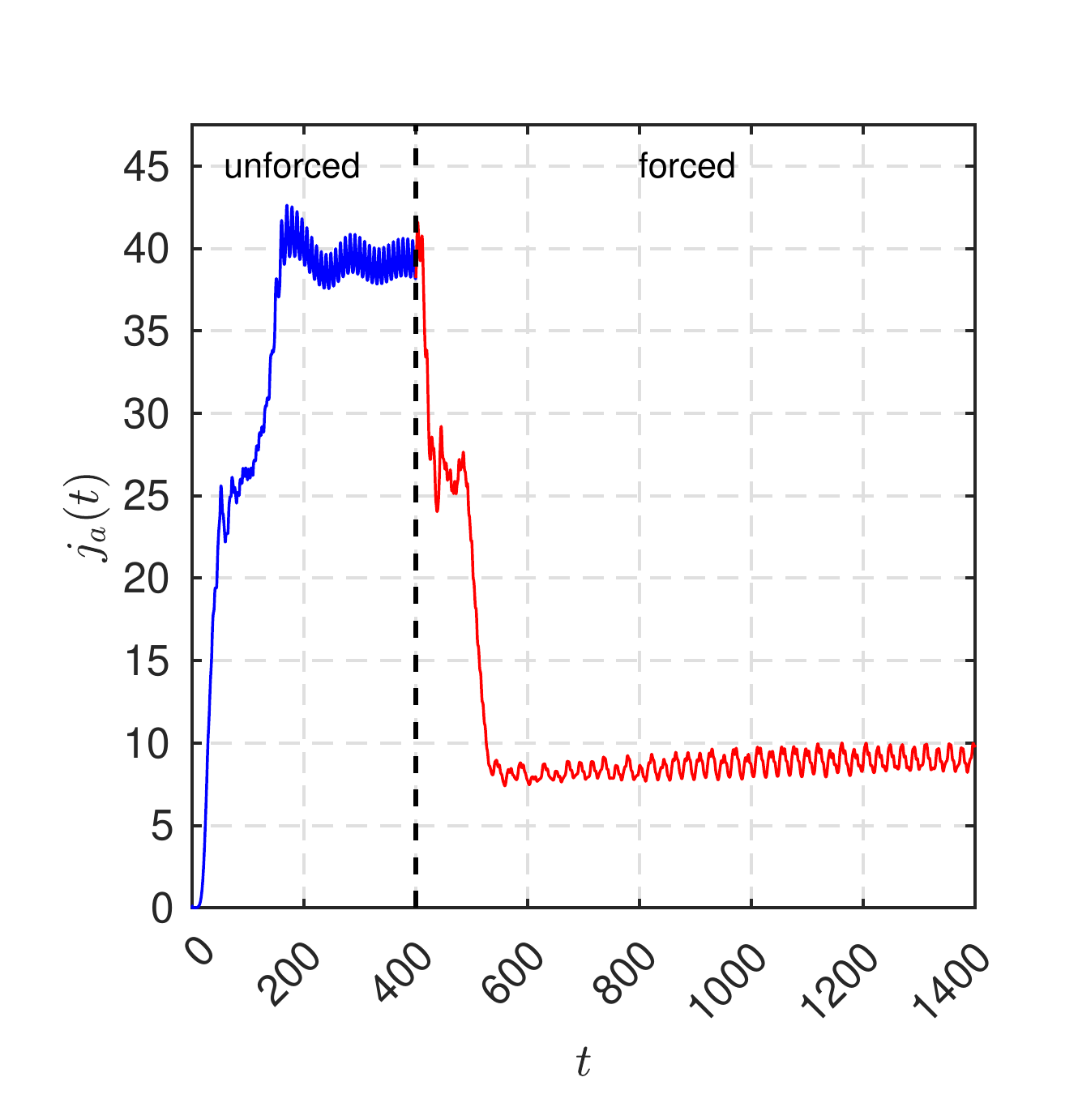}
\caption{\label{fig:DSS_Constant3}}
\end{subfigure}%
\hfil
\begin{subfigure}{.45\textwidth}
  \centering
\includegraphics[width=1.\textwidth]{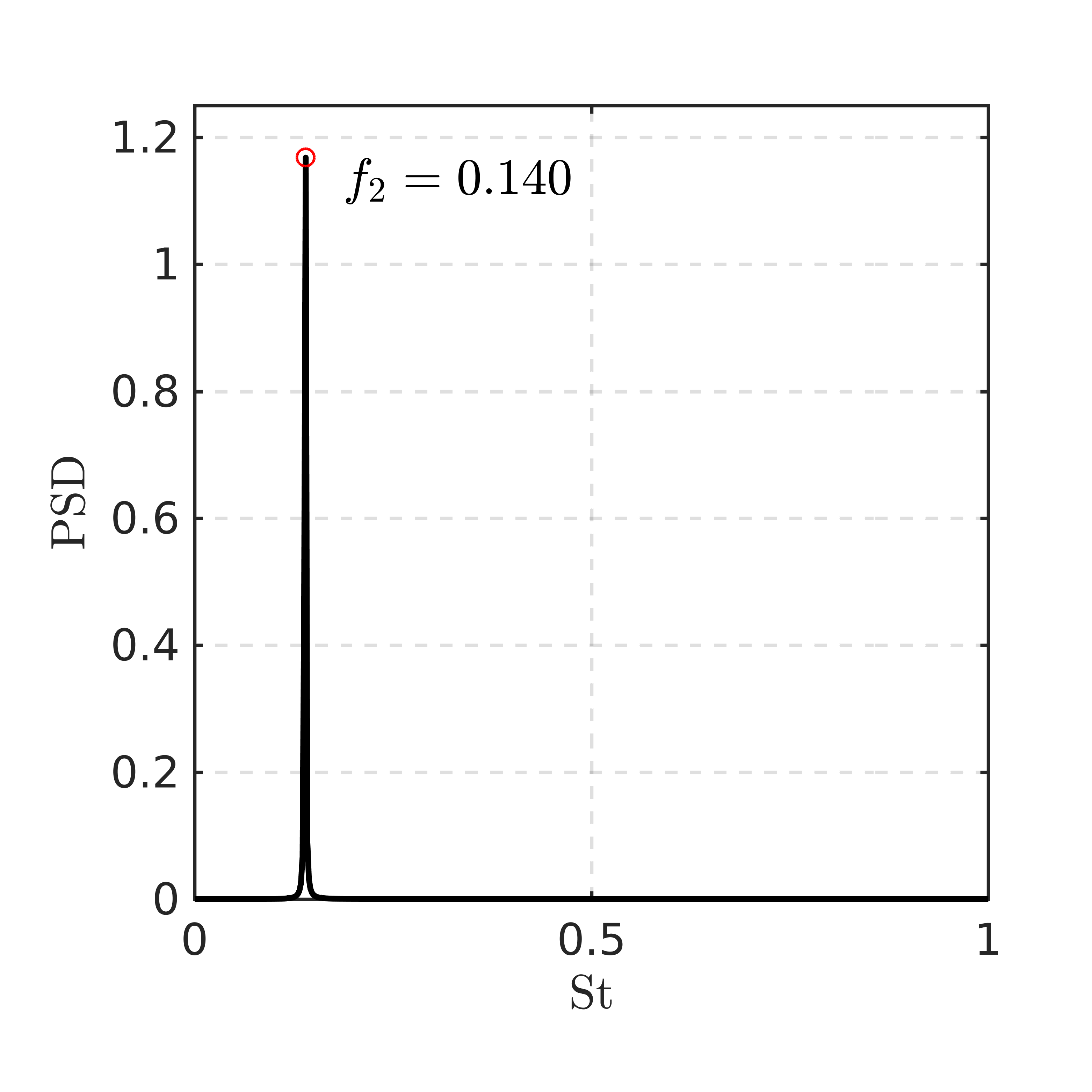}
\caption{\label{fig:PSD_Constant3}}
\end{subfigure}
\caption{\label{fig:EGM_characteristics}Characteristics of the best steady actuation found by EGM. (a) Time evolution of the lift coefficient $C_L$, (b) phase portrait (c) time evolution of instantaneous cost function $j_a$ and (d) Power Spectral Density (PSD) showing the only frequency $f_2=0.140$ of the forced flow.
The control starts at $t=400$.
The unforced phase is depicted in blue and the forced one in red.
The phase portrait and the PSD are computed over $t \in [900,1400]$ the post-transient regime.}
\end{figure}

\begin{figure}
\centering
\begin{subfigure}{.45\textwidth}
  \centering
\includegraphics[width=1.\textwidth]{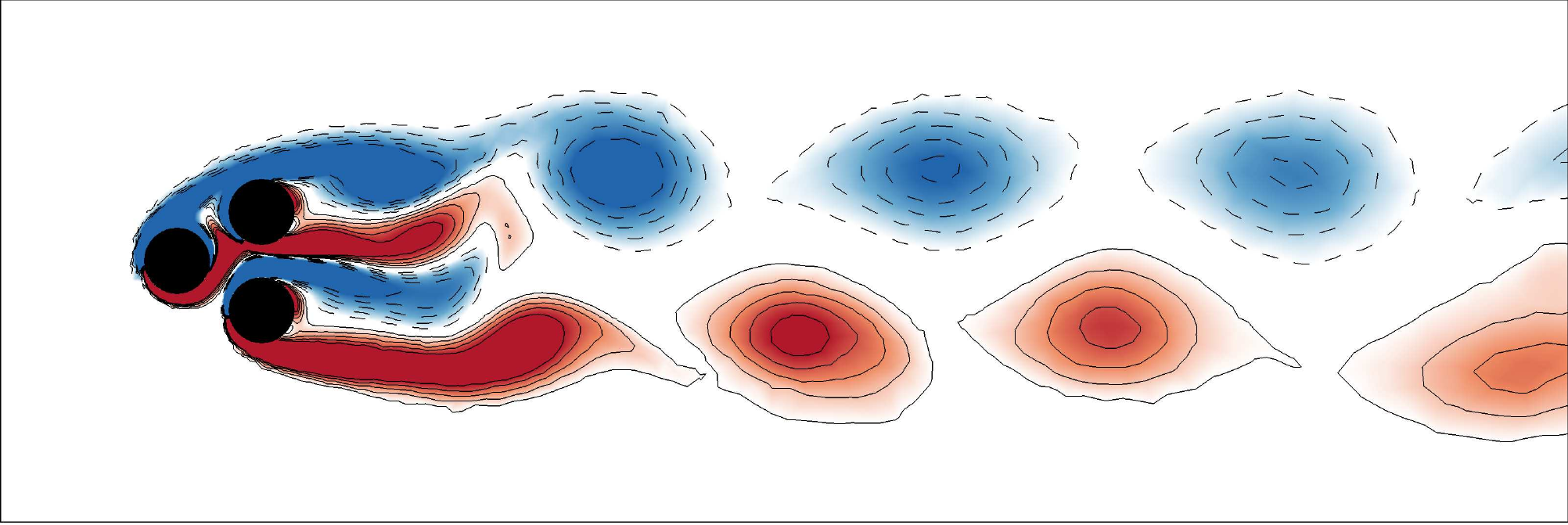}
\caption{\label{fig:EGM_T1}$t+T_2/8$}
\end{subfigure}%
\hspace{0.5cm}
\begin{subfigure}{.45\textwidth}
  \centering
\includegraphics[width=1.\textwidth]{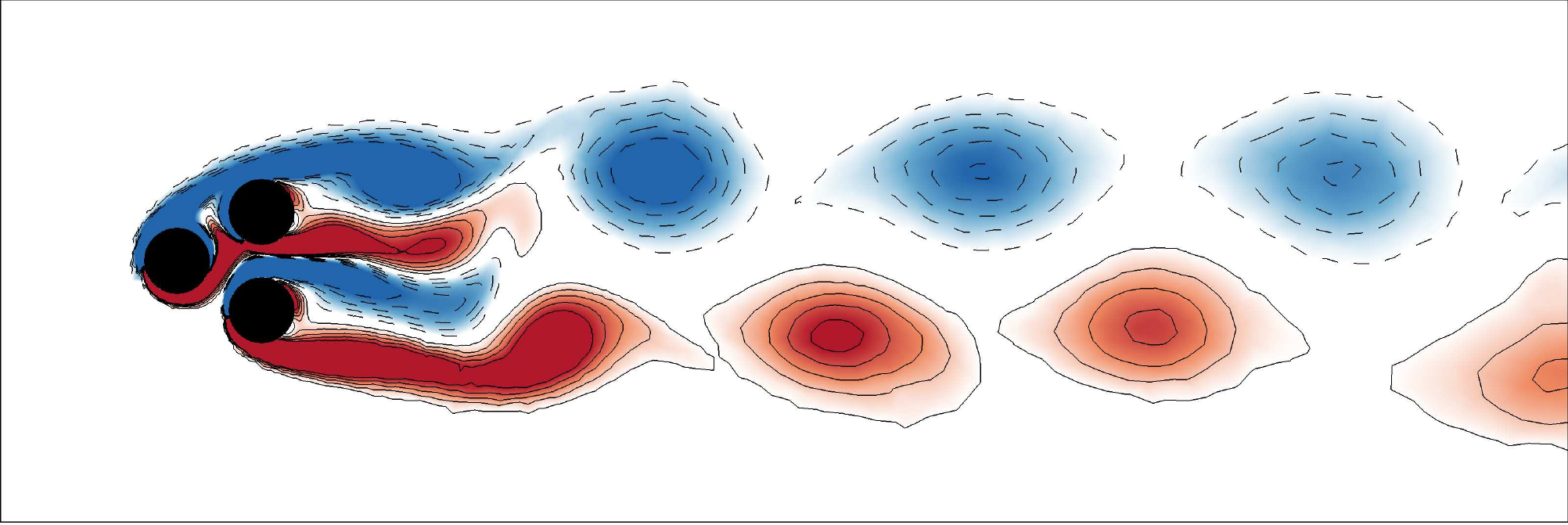}
\caption{\label{fig:EGM_T2}$t+2T_2/8$}
\end{subfigure}

\begin{subfigure}{.45\textwidth}
  \centering
\includegraphics[width=1.\textwidth]{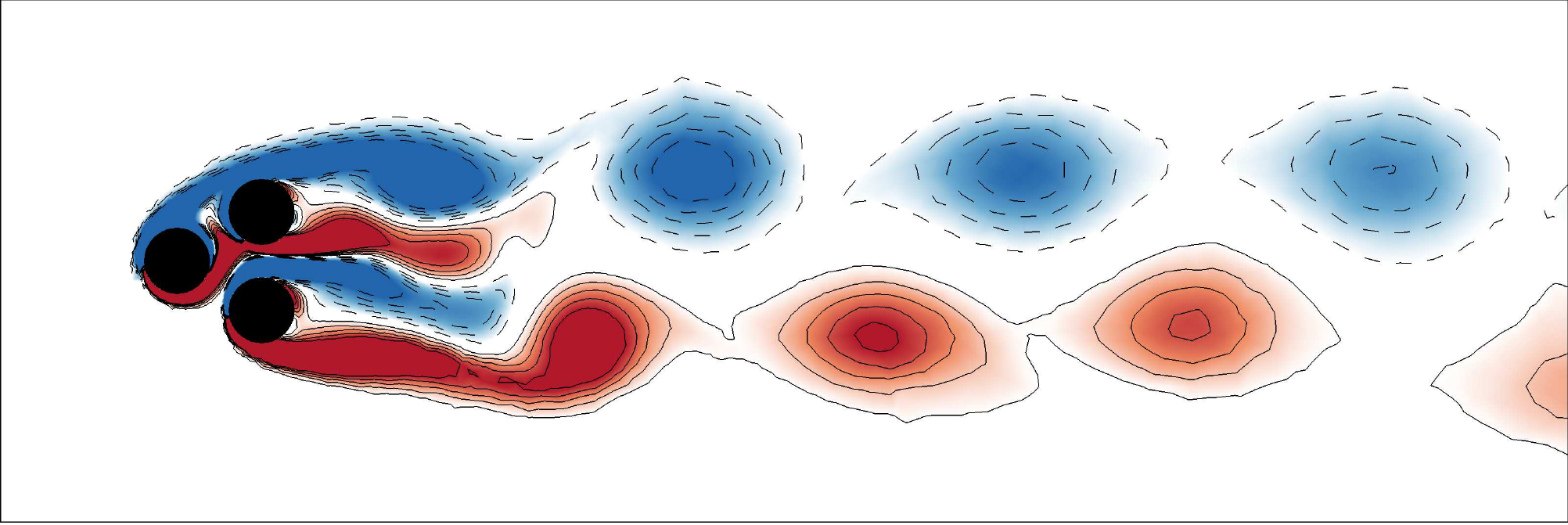}
\caption{\label{fig:EGM_T3}$t+3T_2/8$}
\end{subfigure}%
\hspace{0.5cm}
\begin{subfigure}{.45\textwidth}
  \centering
\includegraphics[width=1.\textwidth]{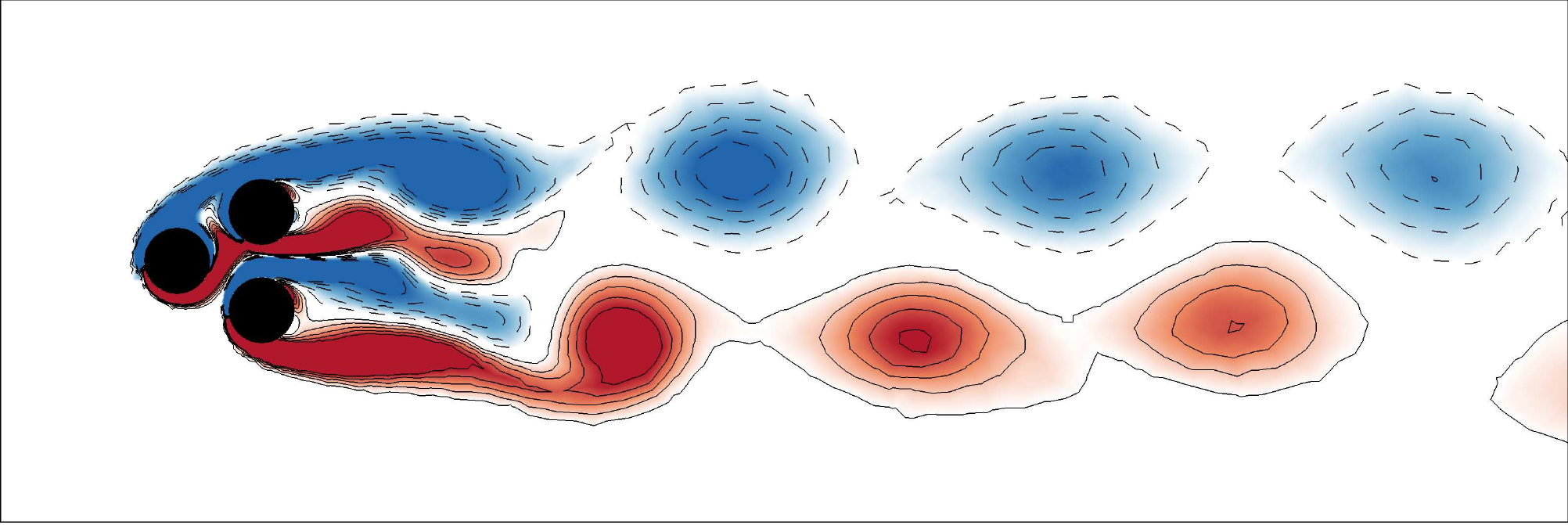}
\caption{\label{fig:EGM_T4}$t+4T_2/8$}
\end{subfigure}

\begin{subfigure}{.45\textwidth}
  \centering
\includegraphics[width=1.\textwidth]{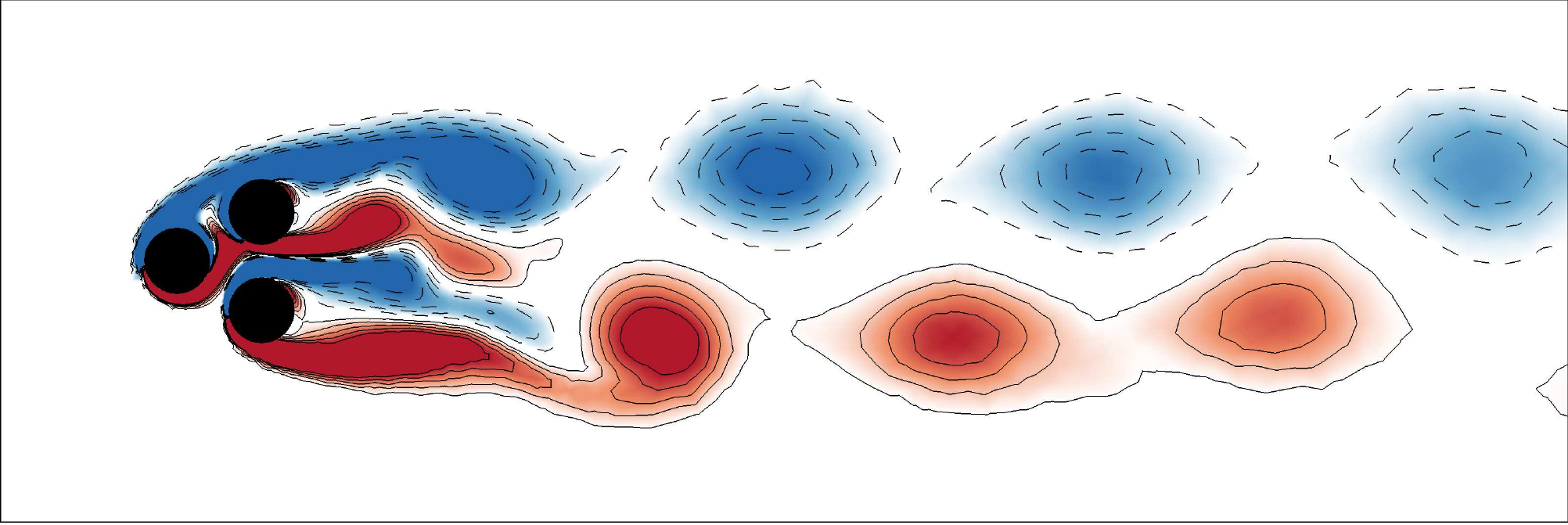}
\caption{\label{fig:EGM_T5}$t+5T_2/8$}
\end{subfigure}%
\hspace{0.5cm}
\begin{subfigure}{.45\textwidth}
  \centering
\includegraphics[width=1.\textwidth]{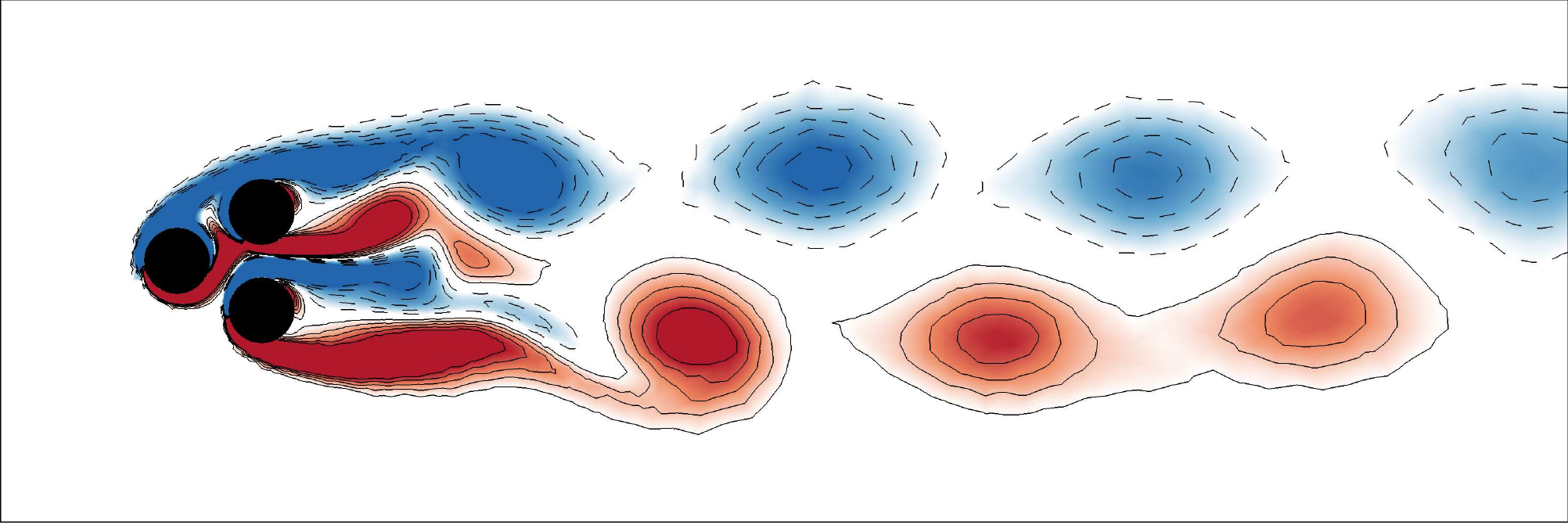}
\caption{\label{fig:EGM_T6}$t+6T_2/8$}
\end{subfigure}

\begin{subfigure}{.45\textwidth}
  \centering
\includegraphics[width=1.\textwidth]{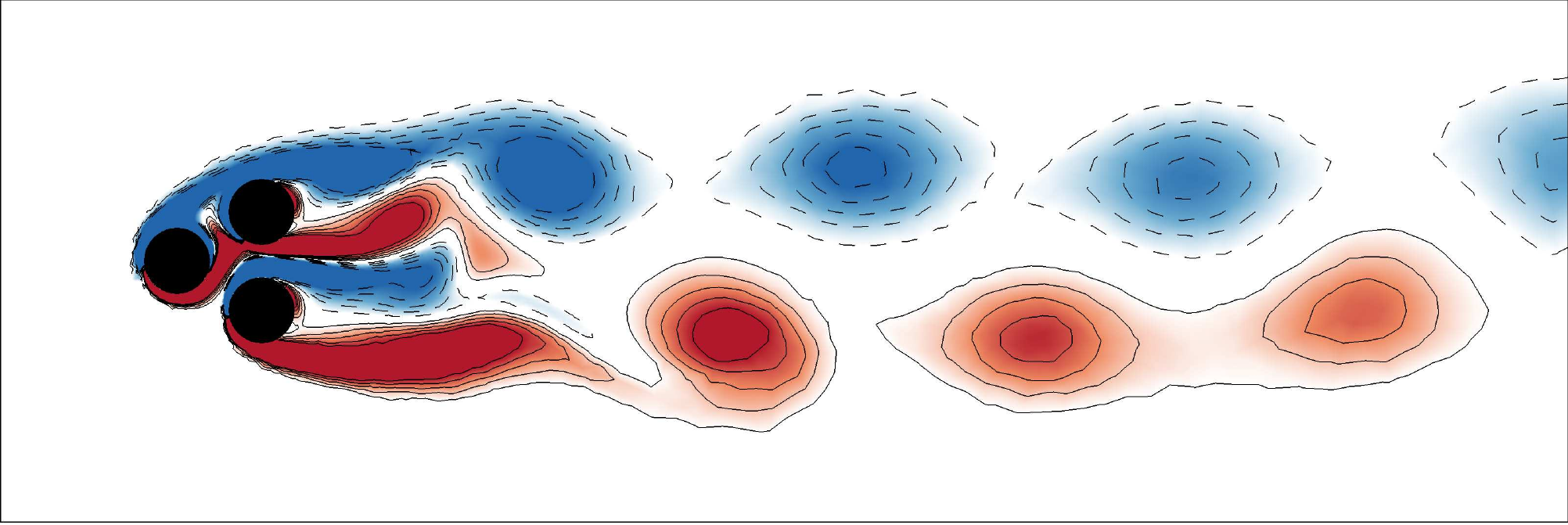}
\caption{\label{fig:EGM_T7}$t+7T_2/8$}
\end{subfigure}%
\hspace{0.5cm}
\begin{subfigure}{.45\textwidth}
  \centering
\includegraphics[width=1.\textwidth]{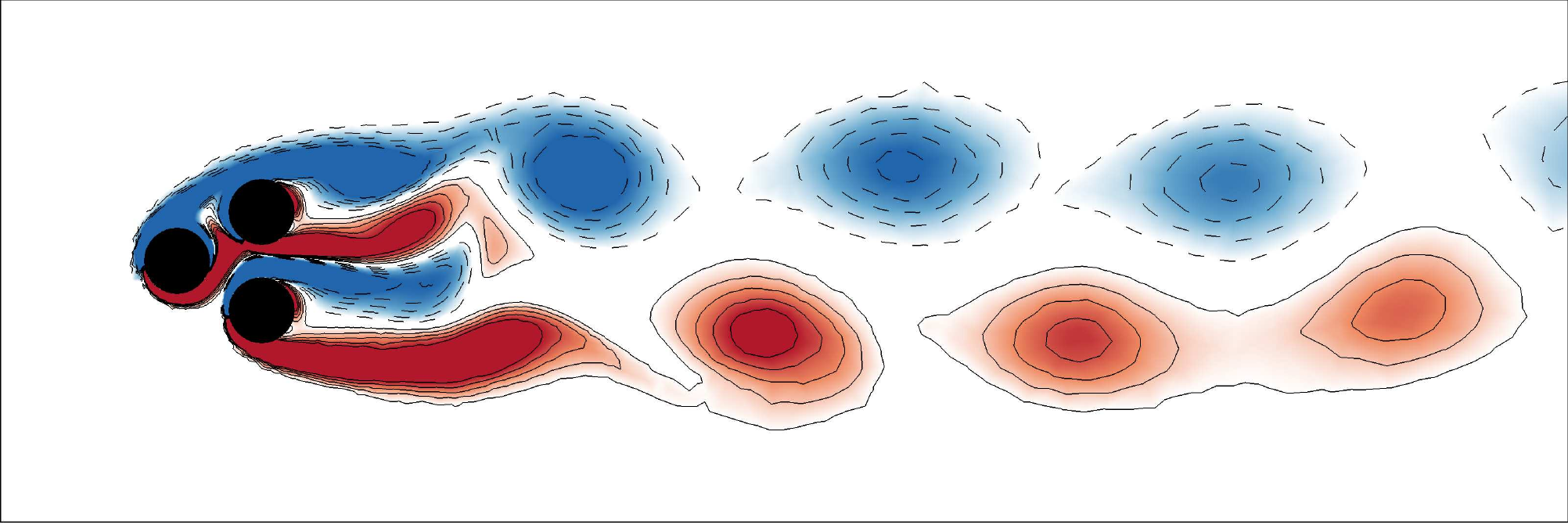}
\caption{\label{fig:EGM_T8}$t+T_2$}
\end{subfigure}

\begin{subfigure}{.45\textwidth}
  \centering
\includegraphics[width=1.\textwidth]{Figures/Snapshots/SteadySolution}
\caption{Symmetric steady solution}
\end{subfigure}%
\hspace{0.5cm}
\begin{subfigure}{.45\textwidth}
  \centering
\includegraphics[width=1.\textwidth]{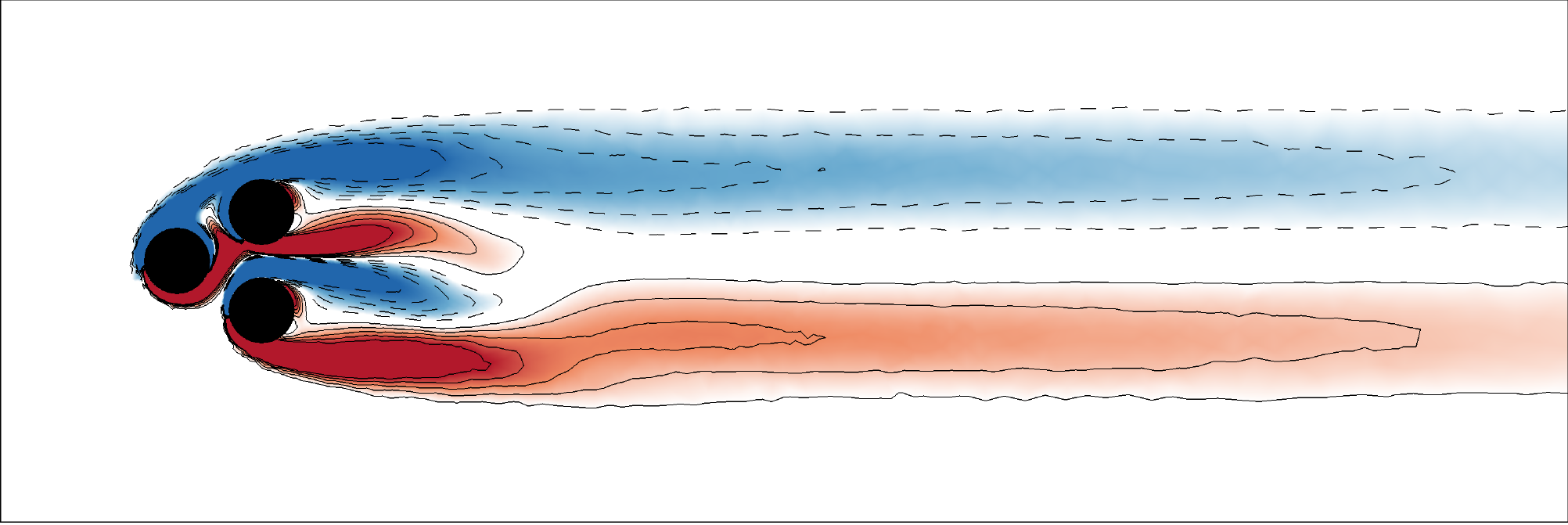}
\caption{\label{fig:Mean_EGM}Mean Field}
\end{subfigure}
\caption{\label{fig:EGM_snap}Vorticity fields of the best steady actuation found with EGM. (a)-(f) Time evolution of the vorticity field throughout the last period of the 1400 convective time units, (i) the objective symmetric steady solution and (j) the mean field of the forced flow.
The color code is the same as figure~\ref{fig:unforced_flow}.
$T_2$ is the period associated to the frequency $f_2$ of the forced flow.
The mean field has computed by averaging over 100 periods.
}
\end{figure}

The control law \eqref{Eqn:bEGM} shows that 
the front cylinder rotates almost five times faster than the two other cylinders and in opposite directions.
The asymmetry in the control law corresponds to the asymmetry in the lift coefficient in figure~\ref{fig:CL_Constant3}, where the mean value is close to -0.7.
The flow asymmetry can be visualized in the mean field (figure~\ref{fig:Mean_EGM}).
The vorticity in the vicinity of  the cylinder is directly related to the actuation; 
thus the upward deflection near the front cylinder is explained by its fast rotation, 
around 1.1 times the incoming velocity.
In addition, the tip of the positive vorticity lobe in the jet is slightly deflected downwards.
Figure~\ref{fig:EGM_T1}-\ref{fig:EGM_T8} show that EGM control
\eqref{Eqn:bEGM}  enables  a jet fluctuation around vanishing mean, like the best base bleeding solution. 
Moreover, the phase portrait and the PSD in figure~\ref{fig:EGM_characteristics} 
reveal that the flow is purely harmonic.
The main frequency $f_2=0.140$ is close to the main frequency $f_1= 0.132$ of the base bleeding solution.
Contrary to the best base bleeding solution, 
the instantaneous cost function $j_a$ stays at low levels 
with a mean value around 9.
The associated normalized cost is $J_a/J_0= 0.28$.
It is worth noting that, even though the control law $[b_1,b_2,b_3]^\intercal=[1,0,0]^\intercal$ is close to the best one found with EGM, its cost, $J_a/J_0 =0.93$, is much higher.
Moreover, the coarse description of the optimal plane $b_3=b^{\rm EGM}=-0.15588$, in figure~\ref{fig:Slice_Ja} (top), does not show any minimum a priori.
This reveals large gradients in the control landscape, near the EGM solution,
where  a small  change in the control amplitude can drastically change the associated cost $J_a/J_0$.

%% Figures : Constant3 flow characteristics-----------------------------
% EGM flow snapshots ---------------------------------------------------
In addition to the less deflected jet, 
we notice in figure~\ref{fig:EGM_snap}
 that the vortex shedding differs from the previous solution leading to a more symmetric flow.
There are now two vortex shedding of the shear layers,
one on the upper side and one on the lower side of the flow.
These shear layer dynamics hardly interact in the whole domain.
Indeed, we notice that the distance between two consecutive vortices increases significantly only before leaving the computational domain which goes along with a slightly upward deflection of the wake.
This results in extended vorticity branches in the mean field (figure~\ref{fig:Mean_EGM}) but with a lower vorticity level compared to the symmetric steady solution.
%However, because of the late interaction of the upper and lower shear-layer vortices, 
%the vorticity branches do not reach the limit of the computational domain like the symmetric steady solution.
%BRN20201007: I don't agree with what I have understood.
%GYCM20201008: You are right, now that I have modified plots of the vorticity field, the vorticity branches reach the end of the computational domain however the vorticity level at the outgoing boundary is lower.

As expected, exploring a richer search space improved the stabilization of the flow.
However, surprisingly, an asymmetric forcing managed to bring partial symmetry to the flow and reduces the cost function even further compared to the best base bleeding solution.
Experimentally, the optimization of the steady fluid pinball actuation 
also lead to asymmetric forcing \citet{Raibaudo2020pf}.
The explorative gradient method managed to converge to the global minimum in less than $N_i=100$ evaluations.
The exploration phases had a lesser impact during the optimization process as we initiated the algorithm close to the global minimum.
We can expect the exploration phases to play a  major role for more complex search space, 
comprising several minima.

%***********************************************************************
\subsection{Feedback control optimization---Gradient-enriched machine learning control} \label{sec:results_gMLC}
In this section, we optimize a feedback control law again to stabilize the unforced symmetric steady solution.
The feedback is provided by 9 velocity signals in the wake as discussed in \S~\ref{sec:control_objective}.
Several function optimizers can be used to solve the regression problem of equation~\ref{eq:control_problem}.
However, a comparison between classical MLC \citep{Duriez2016book} and gMLC has been carried out, 
showing that gradient-enriched MLC not only converges faster than MLC but also towards a better solution.
The comparison between MLC and gMLC is detailed in appendix~\ref{appB}.

In the case of the fluidic pinball, the three cylinders are our three controllers thus $Y \subset \mathbb{R}^3$.
For the  control input space $X$, we choose a grid of nine sensor downstream measuring either $x$ or $y$ velocity component.
The coordinates of the sensors are $x =5, \>6.5, \> 8$ and $y=1.25,\> 0,\> -1.25$.
The downstream position of the sensors have been chosen 
so that good  performance 
of stabilizing feedback control can be expected \citep{Roussopoulos1993jfm}:
The position is far enough  for pronounced vortex shedding 
but close enough to avoid phase decorrelation between actuation and sensing.
Moreover, sensors at different $x$ locations allow to exploit phase differences between the sensors.
The six exterior sensors are $u$ sensors while $v$ sensors are chosen for the ones on the symmetry line $y=0$, so that the signals vanish when the symmetric steady solution is reached.
Experimental realizations are typically based on one or few sensor positions.
The large number of 9 positions has the advantage that gMLC may indicate not only the 
near-optimal control law but also the best sensor location.
The information of sensors is summarized in table \ref{tab:sensor}.
%-----------------------------------------------------------------------
\begin{table}
  \begin{center}
\def~{\hphantom{0}}
\begin{tabular}{cccc}
sensor & $x$-coordinate & $y$-coordinate & velocity component \\[3pt]
\midrule
$s_1$ & 5~ & ~1.25 & $u$ \\
$s_2$ & 6.5 & ~1.25 & $u$ \\
$s_3$ & 8~ & ~1.25 & $u$ \\
$s_4$ & 5~ & ~0~~ & $v$ \\
$s_5$ & 6.5 & ~0~~ & $v$ \\
$s_6$ & 8~ & ~0~~ & $v$ \\
$s_7$ & 5~ & -1.25 & $u$ \\
$s_8$ & 6.5 & -1.25 & $u$ \\
$s_9$ & 8~ & -1.25 & $u$ \\
\end{tabular}
\caption{\label{tab:sensor}Summary of sensor information.}
\end{center}
\end{table}
We introduce time-delayed sensor signals as inputs to enrich the search space and allow ARMAX-based controllers \citep{Herve2012jfm}.
%Moreover, in order to take into account the convective nature of the flow, 
%we add time-delayed sensors as inputs of the control laws.
The delays are a quarter, half and three-quarters of the natural shedding period, 
yielding following additional lifted sensor signals and allowing to reconstruct the phase of the flow:
\begin{displaymath} 
s_{i+9}(t) = s_i(t-T_0/4), \quad s_{i+18}(t)=s_i(t-T_0/2), \quad s_{i+27}(t) = s_i(t-3T_0/4).
\end{displaymath}
For oscillatory signals, the chosen time delay 
corresponds to the first zero of the auto-correlation function
which is a common practice for construction of phase spaces.
The four time-delay coordinates is the minimum information
to determine the mean value, the amplitude and the phase of each signal at every time step.

Summarizing, 
the dimension of the sensor vector $\bm{s}$ is  $9 \times 4 =36$ and $X \subset \mathbb{R}^{36}$.
We do not include time dependent functions in the input space as we aim to stabilize the flow towards the steady solution so an open-loop strategy is not pursued.
In appendix~\ref{appA}, we detail an open-loop optimization including periodic functions.
We show that a symmetric periodic forcing at $\approx$ 3.5 times the natural frequency, manages to stabilize the flow but at the expense of high actuation power. 
So periodic functions are not included as inputs in order to avoid costly solution.
Thus, $N_b=3$, $N_s=36$ and $N_h=0$.
The control laws are then built from 9 basic operations ($+$, $-$, $\times$, $\div$, $\cos$, $\sin$, $\tanh$, $\exp$ and $\log$), 36 sensors signals $\bm{s}_{i=1..36}$ and 10 constants.
The control laws are restricted to the range $[-5,5]$ to avoid excessive actuation.
The basic operations $\div$ and $\log$ are protected in order to be defined on $\mathbb{R}$ in its entirety.
The cost function has been computed over 1000 convective time units, so that the post-transient regime is fully established and the transient phase has a lesser weight.

For the implementation of the gMLC algorithm on the fluidic pinball, we start with a Monte Carlo step of $N_{\rm MC}=100$ individuals, the crossover probability and mutation probability are both set at $P_c=P_m=0.5$. 
Indeed, as the evolution phase is mostly an explorative phase, the mutation probability is increased, from 0.3, in previous studies, to 0.5, to improve the exploration capability.
Moreover, even though crossover is an exploitative operator, it is likely to find new minima thanks to recombinations of radically different control laws.
That is why, the crossover and mutation probabilities are both set to 0.5.
The dimension of the subspace is set to $N_{\rm sub}=10$, so it is large enough to explore a rich subspace but not too large to avoid a slowdown in the optimization process.
Evidently with a subspace of higher dimension the control law can be more finely tuned.
To assure that the subplex step effectively goes down the local minimum, we choose to evaluate $N_G=50$ individuals during the exploitation phase.
Test runs with $N_G=5$ have been carried out and showed that the learning process was slower.
We believe one reason is that each exploration phase changes systematically the subspace, which makes it difficult for the subplex to improve effectively in only a few steps, thus, subplex has almost no benefit in the early phases.
Table~\ref{tab:gMLC_parameters} summarizes all the parameters for gMLC.
The secondary optimization problem (equation~\ref{eq:second_opti_problem}) used to build a matrix representation for the control laws, is solved with LGP.
To speed up the computation, we choose to solve the secondary optimization problem with 100 individuals evolving through 10 generations.
Finally, our implementation is enhanced by a screening of the individuals to avoid reevaluating individuals that have different mathematical expressions but are numerically equivalent, just as \citep{Cornejo2019pamm}.
%Thus, for a random sampling of sensor signals if one control law gives the sames outputs as a previously evaluated individual, then it is discarded and a new one is generated.
%In our simulations, 1000 random samples have been chosen to carry out the comparison.
This screening is used only in steps where the individuals are generated stochastically, meaning in the Monte Carlo step and in the exploration phases.
This improvement is also used in LGP to solve the secondary optimization problem.
We choose our stopping criterion to be a total number of evaluations
to mimic experimental conditions.
In this study, the limit is set to 1000
following prior experience and practical considerations. 
The authors have observed convergence within this limit
for all MLC studies with dozens of configurations.
In addition, wind tunnel experiments with 1000 evaluations 
and 5--20 seconds testing time can easily be performed in one day.

%-----------------------------------------------------------------------
\begin{table}
  \begin{center}
\def~{\hphantom{0}}
\begin{tabular}{>{\centering}p{2cm}>{\centering}p{5cm}>{\centering\arraybackslash}p{5cm}}
parameter & description & value \\[3pt]
\midrule
$N_b$ & number of actuators & 3 \\
$N_s$ & number of sensors & 9 sensors $\times$ 4 delays = 36\\
$N_h$ & number of periodic functions & 0\\
$N_{\rm MC}$ & number of Monte Carlo individuals & 100 \\
$N_{\rm sub}$ & subplex size & 10 \\
$P_c$ & crossover probability & 0.5 \\
$P_m$ & mutation probability & 0.5 \\
$N_G$ & number of individuals per phase & 50 \\
$N_c$ & number of constants & 10 \\
& constant range  & [-1,1] \\
& operations & $+$, $-$, $\times$, $\div$, $\cos$, $\sin$, $\tanh$, $\exp$, $\log$ \\
\end{tabular}
\caption{\label{tab:gMLC_parameters}gMLC parameters for the fluidic pinball.}
\end{center}
\end{table}

\begin{figure}
 \centerline{\includegraphics[width=1\linewidth]{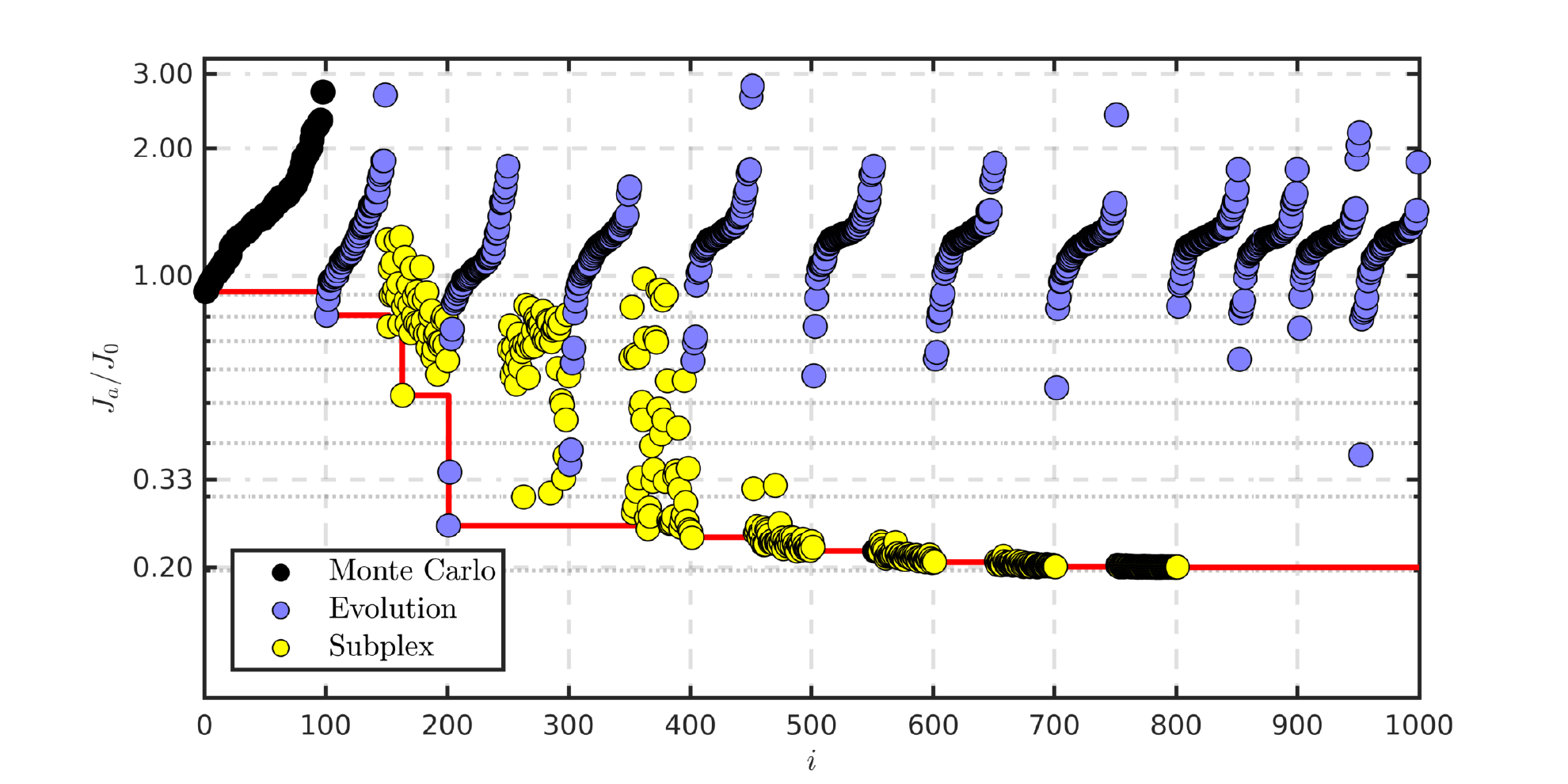}}
 \caption{Distribution of the costs during the gMLC optimization process.
 Each dot represents the cost $J_a/J_0$ of one individual.
 The color of the dots represents how the individuals have been generated.
 Black dots denote the individuals which are 
 randomly generated (Monte Carlo).
 Blue dots refer to individuals which are generated from a genetic operator (exploration). 
 And yellow dots correspond to individuals arising from the subplex method (exploitation).
 The individuals from the Monte Carlo step and the exploration phase have been sorted following their costs.
The red line shows the evolution of the best cost.
The vertical axis is in log scale.}
\label{fig:gMLC_Learning}
\end{figure}

Figure~\ref{fig:gMLC_Learning} presents the learning process of gMLC for the stabilization of the fluidic pinball.
We notice that the first exploration phase, individuals $i=101,\ldots,150$, 
already improved the best cost compared to the Monte Carlo phase.
The following exploitation, individuals $i=151,\ldots,200$, 
present a steep descent, improving  the best solution even further.
During this phase, we notice a clear trend for the cost of the new individuals.
This trend indicates that the simplex is going down towards a minimum.
But this descent is interrupted by the next exploration phase. 
Individuals $i=201,\ldots,250$, 
 greatly improve the best solution.
Particularly, two individuals have a much lower cost that the ones in the simplex, suggesting that a new minima have been found. 
The next exploitation phase with individuals $i=251,\ldots,300$ brings no improvement.
% This is because as exploration generated new individuals. % This sentence can be removed as the following paragraph explains it better.
% BRN20201007: I don't understand
% GYCM20201008: I meant that as the simplex has now a new best individual, some simplex steps are required to "bring" the other individuals in the neighborhood of the new minimum.
The high values of cost in the exploitation steps following the exploration phases is explained by the fact that as we are exploring new minima, shrink steps must be performed to bring the simplex towards the new minima; and the shrink steps replaces all individuals in the simplex except the best one.
As we are leaving one minimum for another one, 
the intermediate values can be arbitrarily high until the simplex reached the neighborhood of the new minimum.
The next exploration phase with individuals $i=301,\ldots,350$ 
also give good individuals that have been included in the simplex.
After 350 evaluations, the only improvements are performed by exploitation phases.
Even if the best cost keeps decreasing slowly, the improvements are small, 
indicating that we are close to the  minimum.
Once we reach a plateau, further improvement can only be performed if an exploration phase finds an individual close to a better minimum.
That is why after 800 individuals, we performed only exploration phases.
%A new minimum may have been found during the last exploration step.
%In order to explore its neighborhood a new simplex step needs to be performed, but with a different simplex.
The final control law build with gMLC reads
\begin{equation}
\label{Eqn:bgMLC}
\begin{array}{lll}
b^{\rm gMLC}_1 &=&
  = -0.0004\sin(\cos(s_{30})) -0.0034(s_{6} + s_{22}) -0.0033(\log(s_{11})) -0.0305( s_{3}) \\
& & -0.0098(s_{16} + s_{15}) + 0.0055 s_{35}(s_{16} + 0.31016) -0.0091( s_{3} - s_{23}) \\
& & + 0.9206 \tanh(s_{16}) -0.1238\cos(s_{31})+0.1907,\\
b^{\rm gMLC}_2  & =& -0.0459( \log(\log(s_{31})))-0.1946,\\
b^{\rm gMLC}_3  & = & -0.0004(  0.841471s_{34} - s_{36}  ) -0.0043\log(s_{9} ) -0.0022(s_{25} - s_{16})\\
& & -0.0098 (\cos(s_{3}) - s_{16}) + 0.9206\log(\tanh(\exp(s_{2}))) -0.0295 \\  
J_a &=& 7.82.
\end{array}
 \end{equation}

%% Figure: Control law description--------------------------------------

%% Figures : gMLC flow characteristics----------------------------------
\begin{figure}
\centering
\begin{subfigure}{.45\textwidth}
  \centering
\includegraphics[width=1.\textwidth]{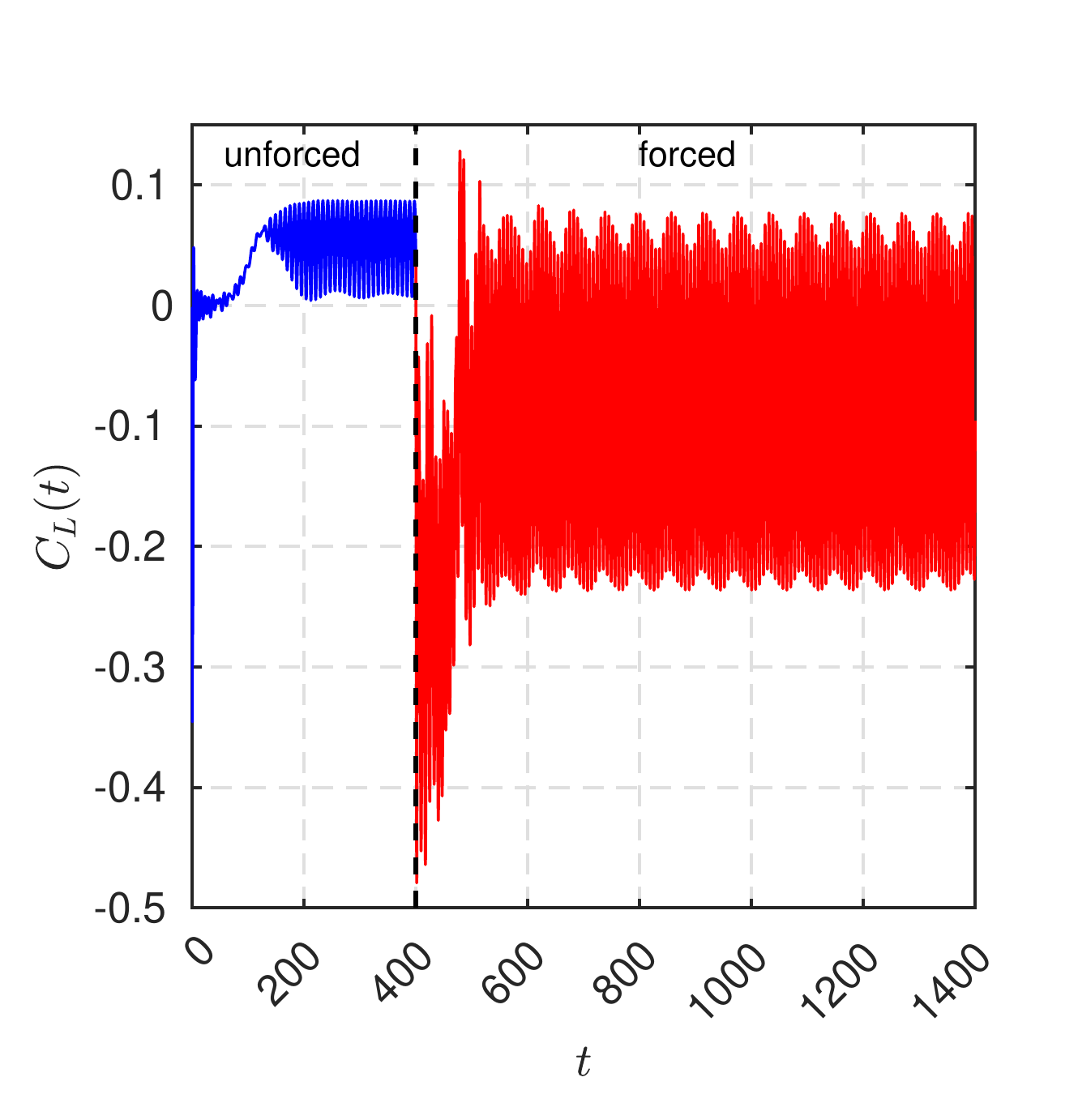}
\caption{\label{fig:CL_gMLC}}
\end{subfigure}%
\hfil
\begin{subfigure}{.45\textwidth}
  \centering
\includegraphics[width=1.\textwidth]{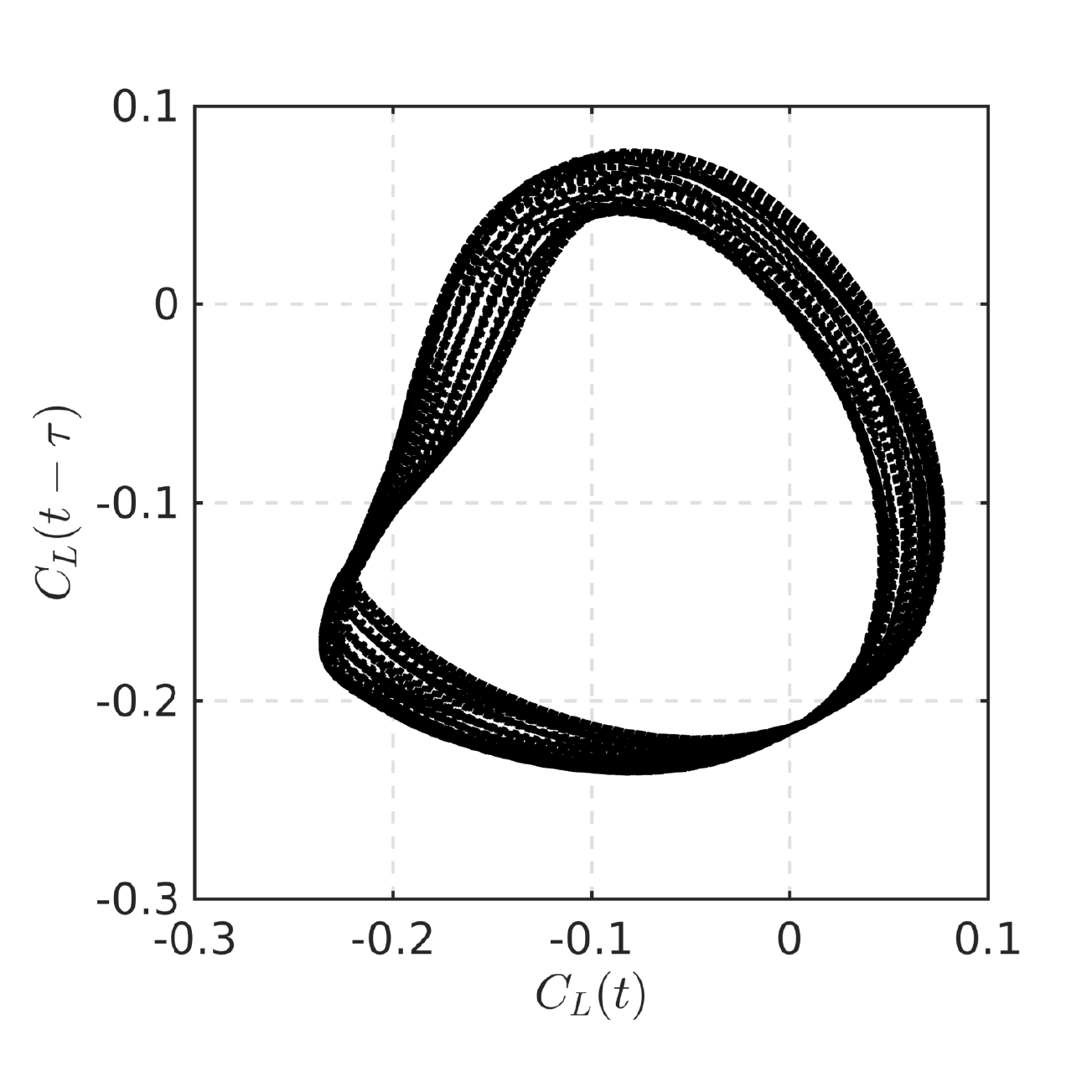}
\caption{\label{fig:PP_gMLC}}
\end{subfigure}

\begin{subfigure}{.45\textwidth}
  \centering
\includegraphics[width=1.\textwidth]{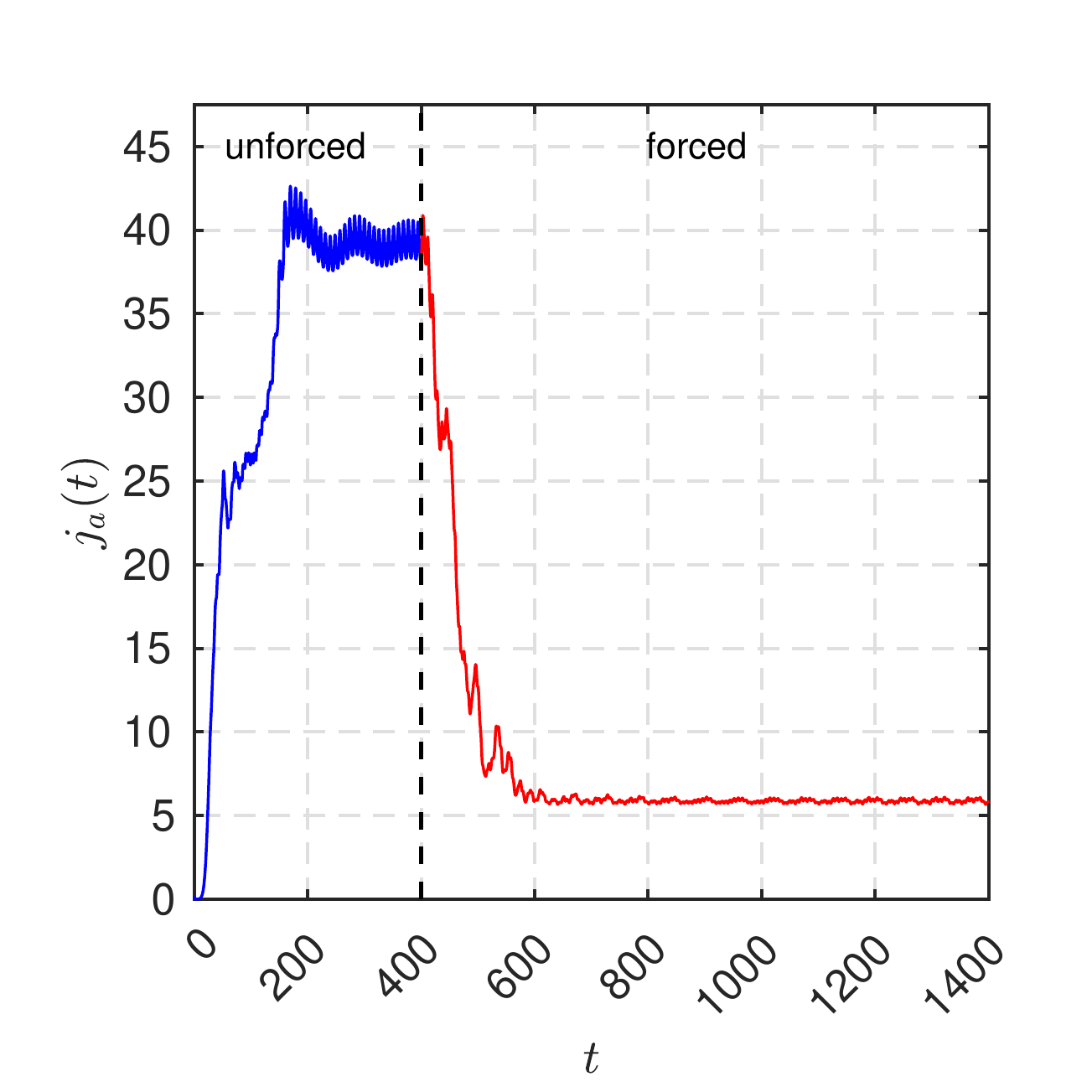}
\caption{\label{fig:DSS_gMLC}}
\end{subfigure}%
\hfil
\begin{subfigure}{.45\textwidth}
  \centering
\includegraphics[width=1.\textwidth]{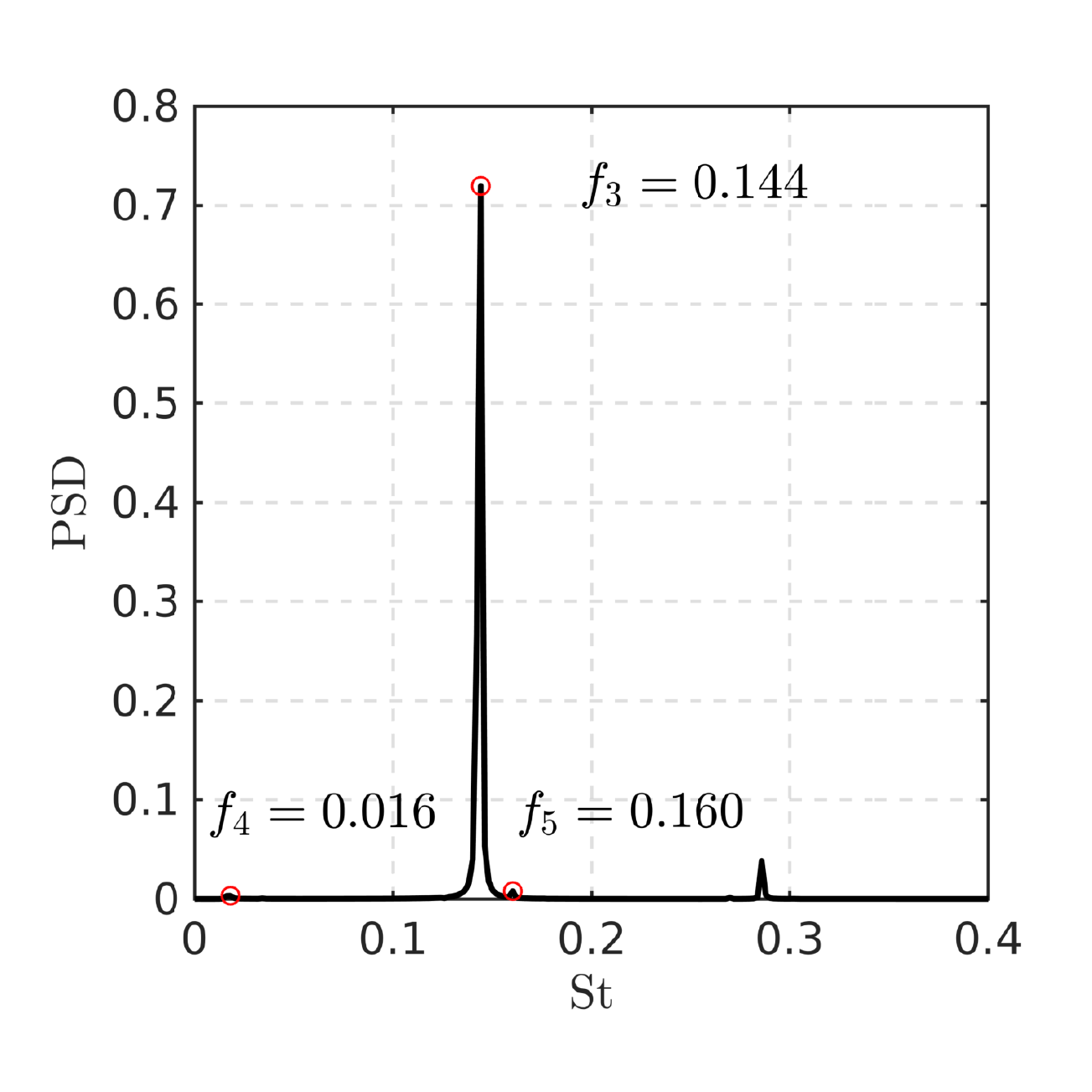}
\caption{\label{fig:PSD_gMLC}}
\end{subfigure}
\caption{\label{fig:gMLC_characteristics}Characteristics of the flow controlled by the best feedback control law found with gMLC. (a) Time evolution of the lift coefficient $C_L$, (b) phase portrait, (c) time evolution of instantaneous cost function $j_a$ and (d) Power Spectral Density (PSD) showing the frequency $f_3=0.144$ of the forced flow, one of its harmonics and two low-power frequencies $f_4=0.016$ and $f_5=0.160$.
The control starts at $t=400$.
The unforced phase is depicted in blue and the forced one in red.
The phase portrait and the PSD are computed over $t \in [900,1400]$, during the post-transient regime.}
\end{figure}

Figure~\ref{fig:gMLC_characteristics}
 presents the characteristics of the flow controlled by the best control law $\bm{K}^{\mathrm{gMLC}}$ built with gMLC.
 %BRN20201007: Don't we want to write down that control law?
 This control law is detailed later specially in table~\ref{tab:control_laws}.
 %GYCM20201008: We could write it down but I feel it won't be useful for interpretation. I believe table 3 brings a better description of the control law.
In figure~\ref{fig:CL_gMLC}, we can see that even if the resulting lift coefficient is still asymmetric, the mean value (around $-0.1$) is closer to 0 as compared to the EGM solution.
The PSD in figure~\ref{fig:PSD_gMLC} shows a dominant frequency at $f_3=0.144$ and one of its higher harmonics.
A small peak can be seen for $f_4 \approx 0.016$.
The nonlinear interaction between the frequencies $f_3=0.144$ and $f_4=0.016$ gives rise to another small peak at $f_5=0.160$.
The phase portrait in figure~\ref{fig:PP_gMLC} reveals drifts in pronounced oscillations due to the low frequency modulation.
The presence of the dominant frequency $f_3=0.144$ and its harmonic in the spectrum is consistent with the periodic behavior of the flow.
The $f_4=0.016$ peak is responsible for the width of a predominant limit-cycle dynamics in the phase portrait.

The evolution of the instantaneous cost function $j_a$  in figure~\ref{fig:DSS_gMLC} shows a plateau after $200$ convective time units, reaching an even lower level (around 6), compared to the EGM solution (around 9).
The associated cost $J_a/J_0=0.20$ is better than the EGM solution at $J_a/J_0=0.28$.

%-----------------------------------------------------------------------
%\begin{figure}
%  \centerline{\includegraphics[width=1\linewidth]{Figures/gMLC/gMLC_Control}}
% \caption{Time series of the actuation command for the best feedback control law found with gMLC.}
%\label{fig:gMLC_Control}
%\end{figure}

 %% gMLC low snapshots--------------------------------------------------
 \begin{figure}
 \centering
 \begin{subfigure}{.45\textwidth}
   \centering
 \includegraphics[width=1.\textwidth]{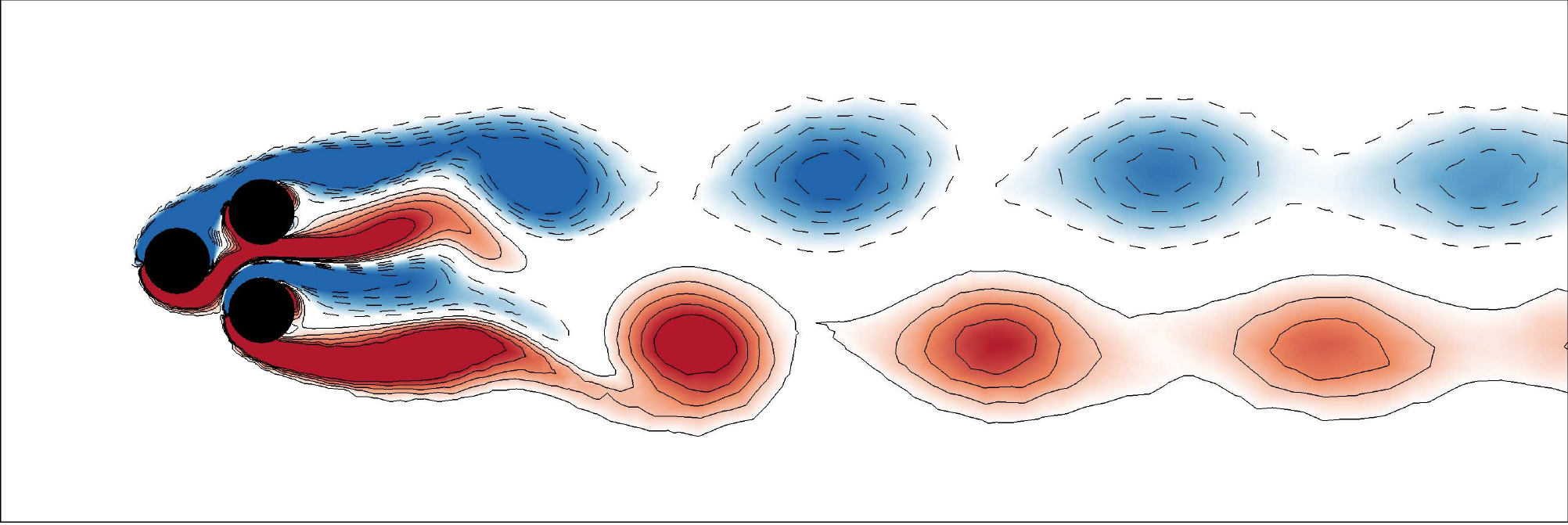}
 \caption{\label{fig:gMLC_T1}$t+T_3/8$}
 \end{subfigure}%
 \hspace{0.5cm}
 \begin{subfigure}{.45\textwidth}
   \centering
 \includegraphics[width=1.\textwidth]{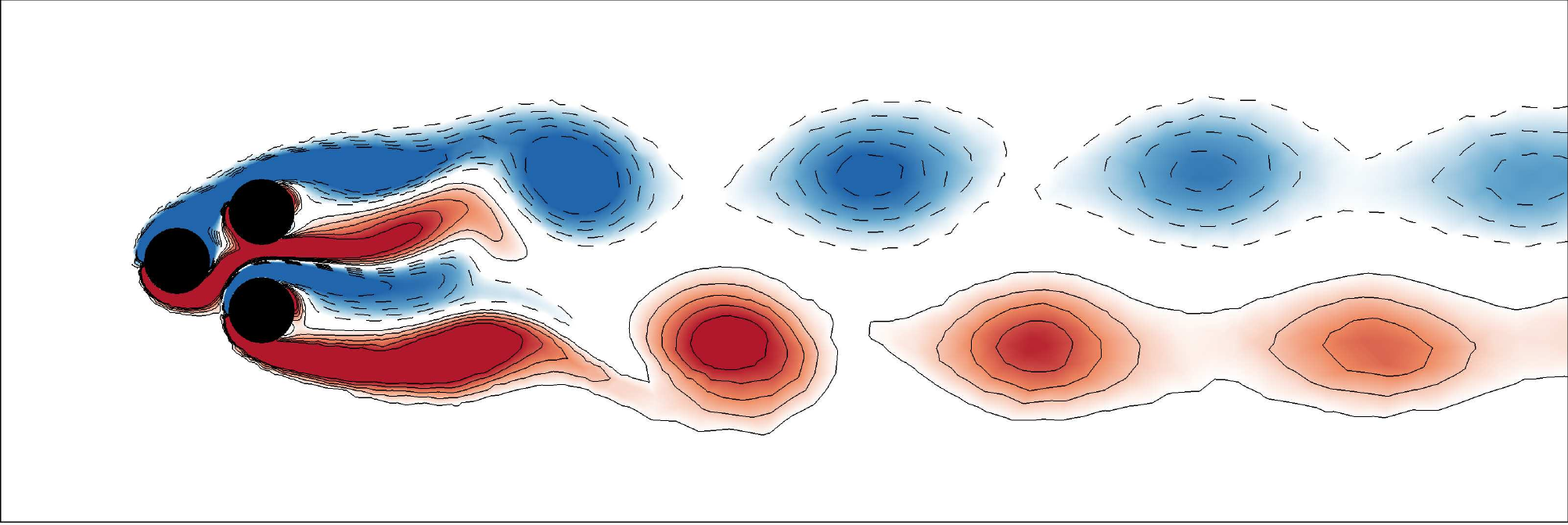}
 \caption{\label{fig:gMLC_T2}$t+2T_3/8$}
 \end{subfigure}

 \begin{subfigure}{.45\textwidth}
   \centering
 \includegraphics[width=1.\textwidth]{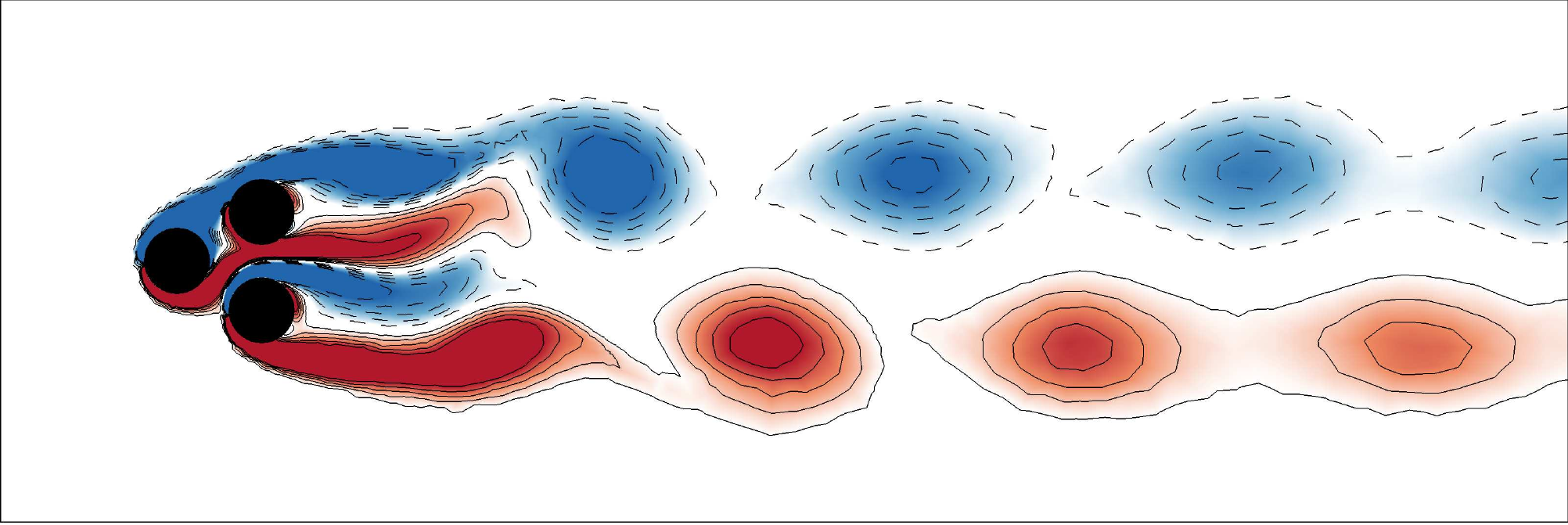}
 \caption{\label{fig:gMLC_T3}$t+3T_3/8$}
 \end{subfigure}%
 \hspace{0.5cm}
 \begin{subfigure}{.45\textwidth}
   \centering
 \includegraphics[width=1.\textwidth]{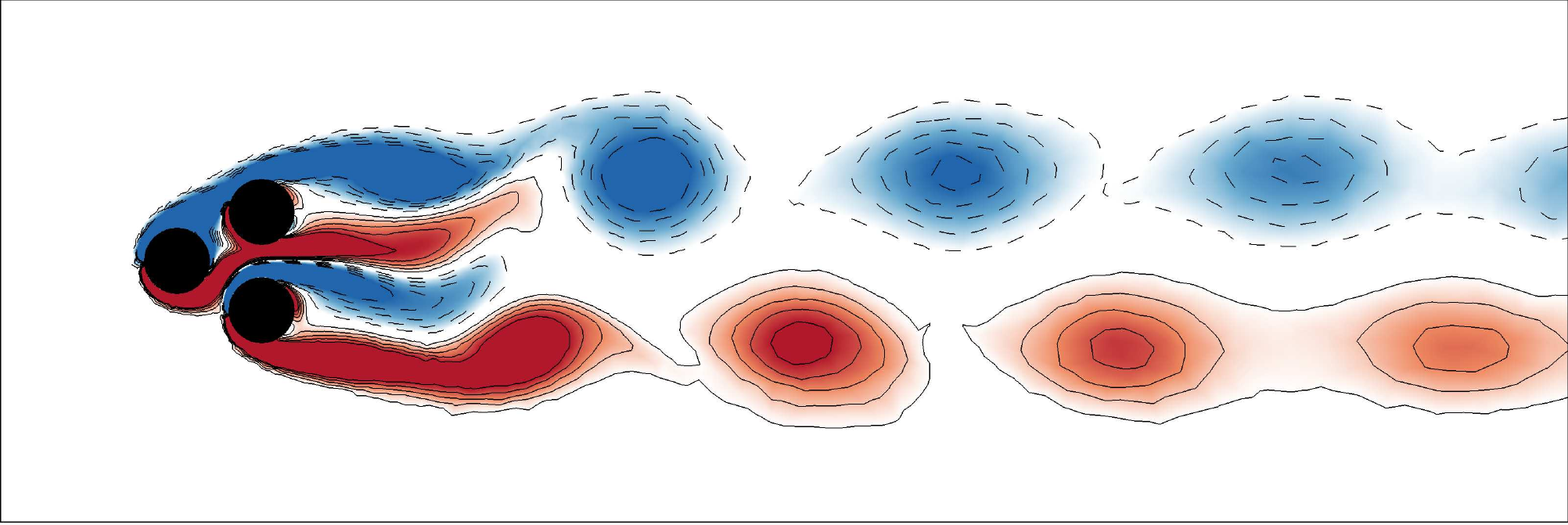}
 \caption{\label{fig:gMLC_T4}$t+4T_3/8$}
 \end{subfigure}

 \begin{subfigure}{.45\textwidth}
   \centering
 \includegraphics[width=1.\textwidth]{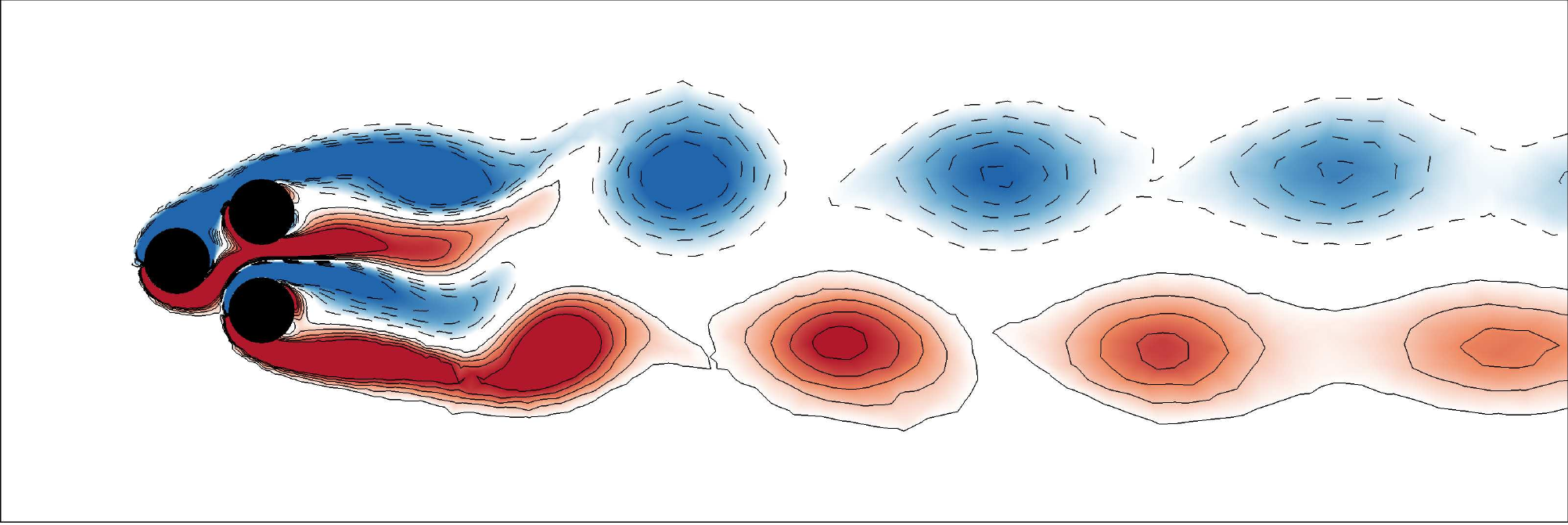}
 \caption{\label{fig:gMLC_T5}$t+5T_3/8$}
 \end{subfigure}%
 \hspace{0.5cm}
 \begin{subfigure}{.45\textwidth}
   \centering
 \includegraphics[width=1.\textwidth]{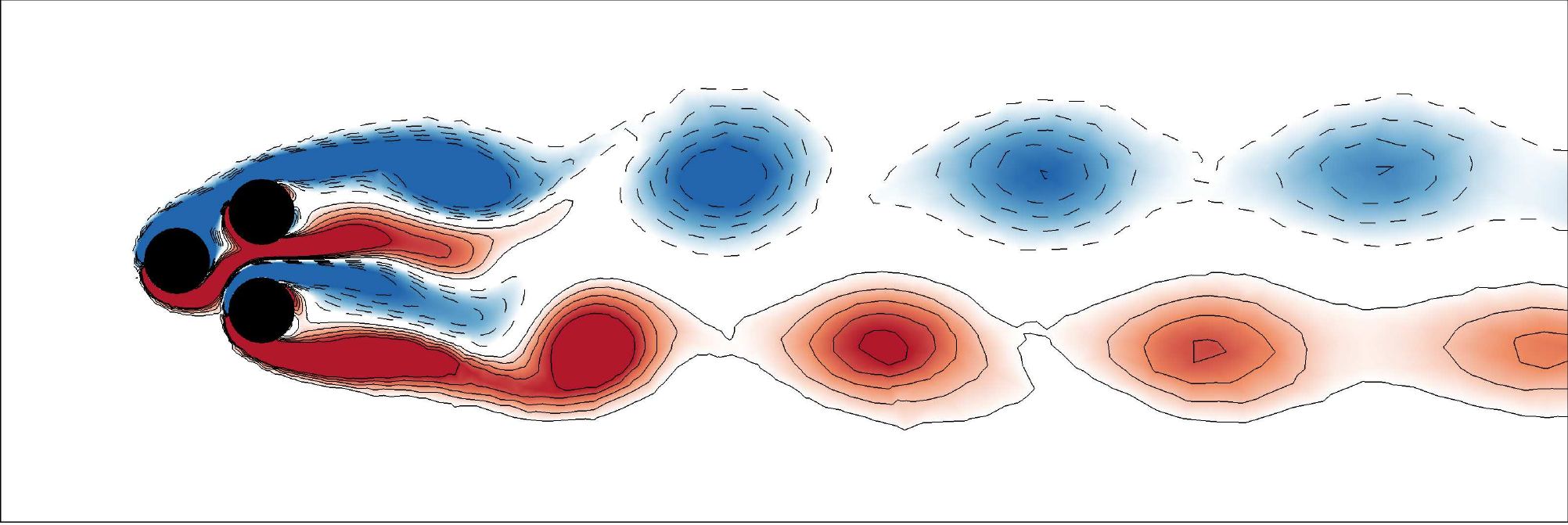}
 \caption{\label{fig:gMLC_T6}$t+6T_3/8$}
 \end{subfigure}

 \begin{subfigure}{.45\textwidth}
   \centering
 \includegraphics[width=1.\textwidth]{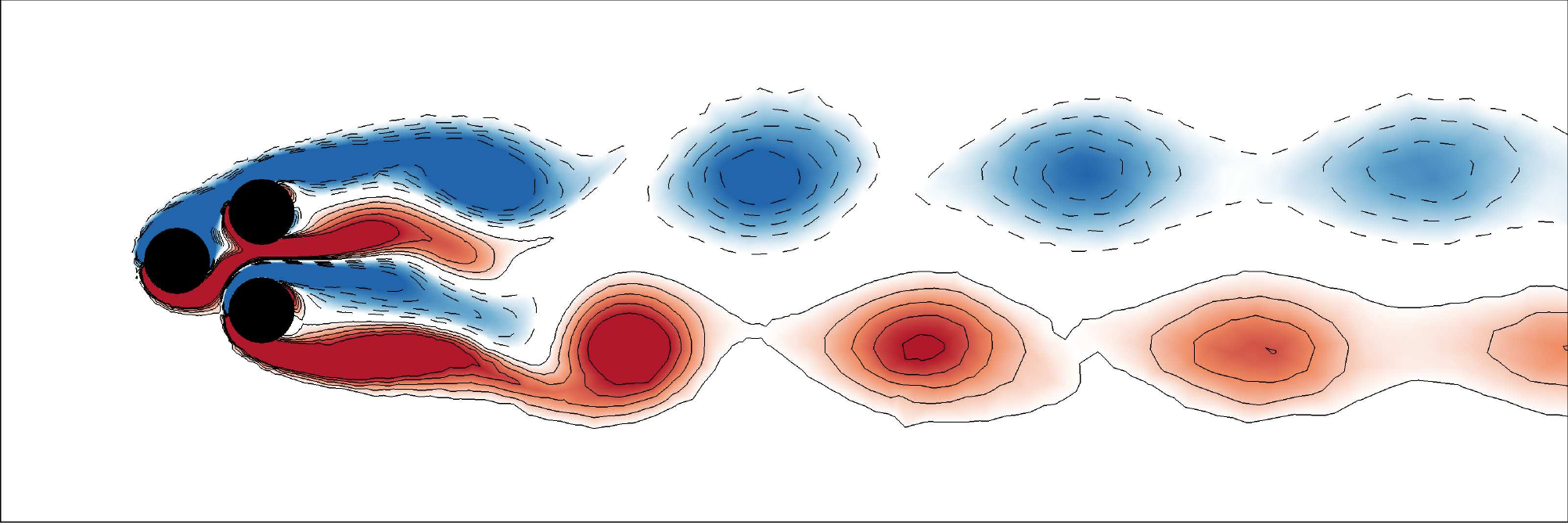}
 \caption{\label{fig:gMLC_T7}$t+7T_3/8$}
 \end{subfigure}%
 \hspace{0.5cm}
 \begin{subfigure}{.45\textwidth}
   \centering
 \includegraphics[width=1.\textwidth]{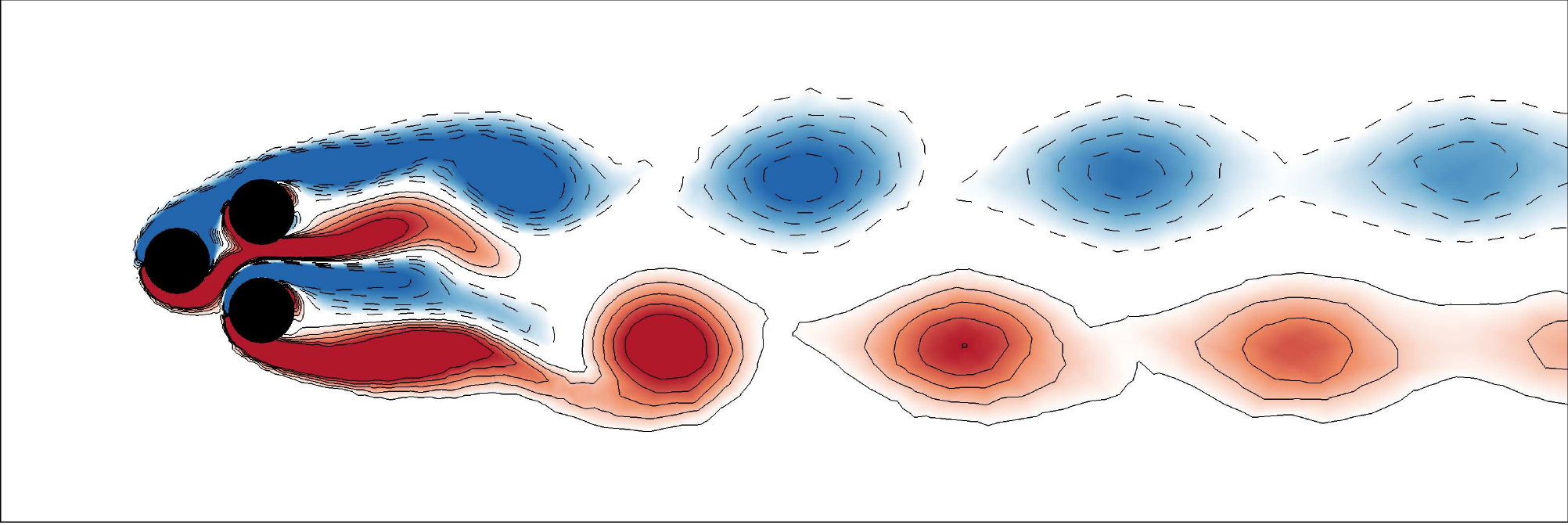}
 \caption{\label{fig:gMLC_T8}$t+T_3$}
 \end{subfigure}
 
  \begin{subfigure}{.45\textwidth}
   \centering
 \includegraphics[width=1.\textwidth]{Figures/Snapshots/SteadySolution}
 \caption{Symmetric steady solution}
 \end{subfigure}%
 \hspace{0.5cm}
 \begin{subfigure}{.45\textwidth}
   \centering
 \includegraphics[width=1.\textwidth]{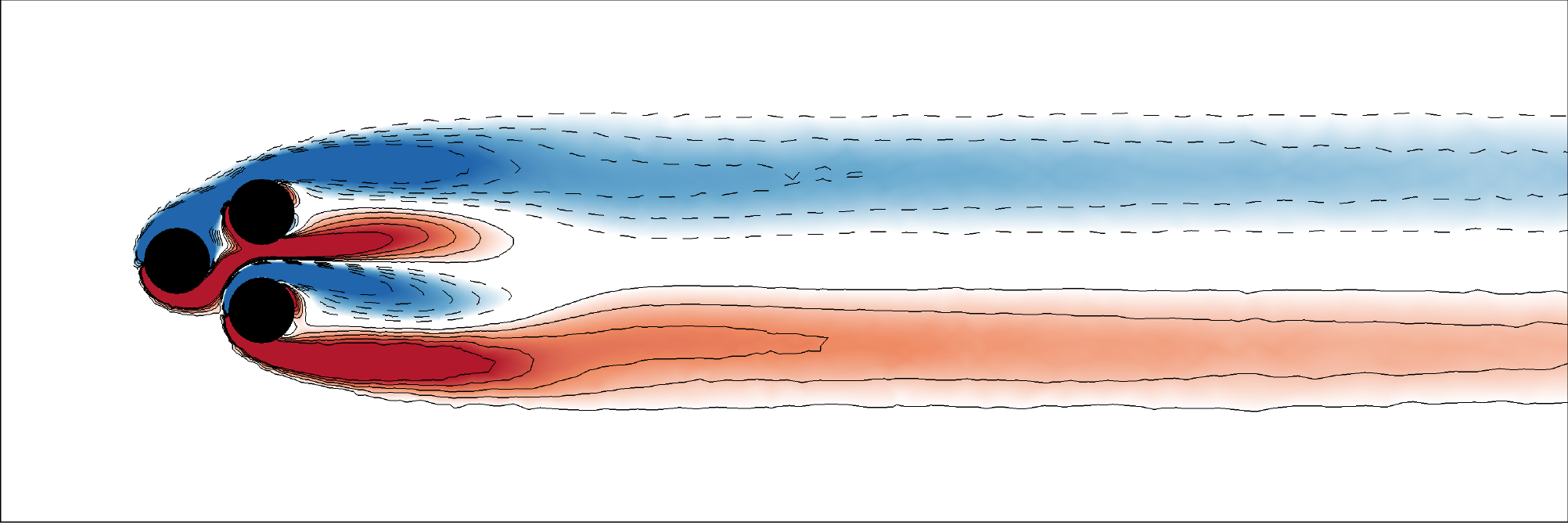}
 \caption{\label{fig:Mean_gMLC}Mean field}
 \end{subfigure}
 \caption{\label{fig:gMLC_snap}Vorticity fields of the best feedback control found with gMLC. (a)-(f) Time evolution of the vorticity field throughout the last period of the 1400 convective time units, (i) the objective symmetric steady solution and (j) the mean field of the forced flow.
The color code is the same as figure~\ref{fig:unforced_flow}.
 $T_3$ is the period associated to the frequency $f_3$.
 The mean field has been computed by averaging 100 periods.}
 \end{figure}
\begin{table}
  \begin{center}
\def~{\hphantom{0}}
\sisetup{scientific-notation = true}
\begin{math}
\begin{array}{cccccc}
\# & b_1 & b_2 & b_3 & \text{weight} & J_a/J_0\\[3pt]
\midrule
1 & \sin(\cos(s_{30})) & 0 & 0.841471s_{34} - s_{36} & \num{-0.0004} & 0.91 \\
2 & s_{6} + s_{22} & 0 & 0 & \num{-0.0034} & 0.94 \\
3 & 0 & 0 & \log(s_{9}) & \num{-0.0043} & 0.97 \\ 
4 & 0 & 0 & s_{25} - s_{16} & \num{-0.0022} & 0.97 \\
5 & \log(s_{11}) & 0 & 0 & \num{-0.0033} & 0.95 \\ 
6 & s_{3} & 0 & 0 & \num{-0.0305} & 0.92 \\ 
7 & s_{16} + s_{15} & 0 & \cos(s_{3}) - s_{16} & \num{-0.0098} & 0.97 \\
8 & s_{35}(s_{16} + 0.31016) & 0 & 0 & \num{0.0055} & 0.80 \\ 
9 & s_{3} - s_{23} & 0 & 0 & \num{-0.0091} & 0.88\\ 
10 & 1 & \log(\log(s_{31})) & 0 & \num{-0.0459} & 0.93 \\ 
\textbf{11} & \bm{\tanh(s_{16})} & \bm{-0.187071} & \bm{\log(\tanh(\exp(s_{2})))} & \bm{9.206\times 10^{-1}} & \bm{0.26} \\ 
12 & 0.540302 & -0.144304 & -0.0144074 & \num{0.0687} & 0.34 \\ 
\textbf{13} & \bm{\cos(s_{31})} & \bm{-0.144304} & \bm{-0.0144074} & \bm{-1.238\times 10^{-1}} & \bm{0.36} \\ 
\textbf{14} & \bm{0.949948} & \bm{-0.144304} & \bm{-0.0144074} & \bm{2.100\times 10^{-1}} & \bm{0.39}
    \end{array}
    \end{math}
    \caption{\label{tab:control_laws}Summary of the 14 control laws composing $\bm{K}^{\rm gMLC}$ described in equation \eqref{Eqn:bgMLC}.
    For each control law, we present $b_1$, $b_2$, $b_3$, the associated weight and the reduced cost $J_a/J_0$.
    The three best control laws  are $\#11$, $\#13$ and $\#14$.}
    \end{center}
    \end{table}
\begin{table}
  \begin{center}
\def~{\hphantom{0}}
\begin{tabular}{lccc}
cylinder & mean value & main frequency & peak-to-peak amplitude \\[3pt]
\midrule
front ($b_1$, green) & ~0.48 & 2$f_3$ & ~0.12 \\
bottom ($b_2$, blue) & -0.19 & 2$f_3$ & ~0.03 \\
top ($b_3$, red) & -0.02 & ~$f_3$ & $<0.01$ \\
\end{tabular}
\caption{\label{tab:control}Summary of control law information.
The frequencies and peak-to-peak amplitude have been computed on the post-transient regime.
}
\end{center}
\end{table}

Table~\ref{tab:control_laws}, \ref{tab:control} and figure~\ref{fig:gMLC_robustness}a give more details on the control law $\bm{K}^{\rm gMLC}$ built with gMLC.
Firstly, we can see that even though the simplex comprises $N_{\rm sub}=10$ individuals, subplex build the control law $\bm{K}^{\rm gMLC}$ by linearly combining 14 control laws.
Indeed after a few iterations of simplex, 
all the individuals are eventually a linear combination of the initial individuals forming simplex.
When a new individual is introduced in the basis thanks to the exploration phase, the exploitation phase will combine the remaining individuals with the new one, making the next individual a linear combination of more than 10 individuals.
It is important to note that even after the introduction of new individuals with the exploration phase, the subspace to explore changes but the dimension remains.
In this case, with $N_{\rm sub}=10$, the dimension of the subspace is 9.
The repetition of this process builds each time more complex control laws.
Thus, in table~\ref{tab:control_laws}, individuals $i=11,12,13,14$ 
have been introduced thanks to exploration phases.
The control laws with the strongest weights are $i=11, 13$ and $14$, 
whereas the weight associated with the other control laws are at least one order of magnitude lower.
Control law $i=11$ is also the one with the lowest cost $J_a/J_0=0.26$.
$\bm{K}^{\mathrm{gMLC}}$ is then mainly based on $i=11$ and corrected by the remaining control laws.
This indicates that on the last phase of the learning, it is the minimum in the neighborhood of $i=11$ that has been found.

Moreover, table~\ref{tab:control_laws} shows that all three control components
$b_1^{\rm gMLC}$, $b_2^{\rm gMLC}$ and $b_3^{\rm gMLC}$ of the gMLC control law include sensor information.
%%% BRN20201007: Check the sentence above.
% GYCM20201008: Modified
However, figure~\ref{fig:gMLC_robustness}a shows that the actuation command associated with $\bm{K}^{\rm gMLC}$ for the two rearward cylinders ($b_2$ and $b_3$) are nearly constant.
This is partially due to the low weights associated to the control laws with sensor signals.
We can also assume that the sensor signals cancel each other, leading to such low peak-to-peak amplitudes.
Table~\ref{tab:control} shows the characteristics of the actuation command during the post-transient regime.
A spectral analysis shows that the main frequency of the actuation command for the front and bottom cylinder are twice the main frequency of the flow $f_3$, revealing that the actuation is a function of the flow.
% especially for the front cylinder whose amplitude peak-to-peak is 0.12. % not needed
%%%% BRN20201007: I do not understand the previous sentence.
% GYCM20201008: I wanted to highlight the fact that the main frequency of the actuation command was two times the ffor the front and bottom cylinder. I corrected it
Thus,
gMLC managed to build a combination between asymmetric steady forcing and feedback control.
Finally, like EGM, the best solution found is asymmetric 
but with lower amplitudes.
Consequently, 
 the associated actuation power is lower compared to general steady actuation found with EGM: $J_b=0.2018$ for the general steady actuation and $J_b=0.0391$ for the feedback control law found with gMLC.

The controlled flow is depicted over one period in figure~\ref{fig:gMLC_T1}-\ref{fig:gMLC_T8}.
First, we notice that the jet fluctuates around a vanishing mean, as for the EGM actuation.
Also, the vortex shedding of the upper and lower shear layers hardly interact.
Compared to the EGM solution, the stability of the wake is improved 
as the two Kelvin-Helmholtz vortices keep their transverse distance to the symmetry line 
until the very end of the computational domain.
This is explained by the re-energization of the shear layers thanks 
to the vigorous rotation of the front cylinder at twice the main frequency $f_3$ of the controlled flow,
like \citet{Protas2004pf}.
%%%% What is f_3?
The mean field, in figure~\ref{fig:Mean_gMLC}, is similar to the symmetric steady solution.
Indeed, we notice that the vorticity regions extend to the end of the computation domain, 
like the symmetric steady solution.
Also, like for the best general steady actuation, the region near the cylinders is non-symmetrical due to the action.
However, contrary to the symmetric steady solution, the mean field of the feedback control has a narrower region between the vorticity regions upstream and a wider region downstream.

As expected, gMLC manages to find a new solution that surpasses the best general steady actuation found with EGM.
Surprisingly, gMLC built a non-trivial solution, 
combining asymmetric steady forcing and feedback control for the front cylinder, controlling the flow with a direct feedback of the phase of the flow, i.e., phasor control \citep{Brunton2015amr}.
Interestingly, gMLC composed a control law that forces the flow at twice the main frequency.
In addition, compared to the best general steady actuation, the actuation power is significantly reduced.
Lastly, the learning process of gMLC exploited both the evolution phases and the simplex steps to rapidly build better solutions.
Thanks to the evolution phases, new minima have been successfully found and thanks to the simplex steps, the solutions have been improved even more.
The progress of the subplex steps show that local gradient information can be exploited in a subspace of an infinite dimension space to minimize a cost function.
Building on this success, we believe that gradient-enriched MLC 
will greatly accelerate the optimization of control laws for MIMO control
as compared to the linear genetic programming  control.

\section{Discussion}
\label{Sec:Discussion}

%FUTURE VALUE AND LIMITATION
This section discusses design aspects of the proposed  methodology
which are of relevance to this and other configurations.
In \S~\ref{ToC:S5:Feedback}, the role of feedback is assessed.
\S~\ref{ToC:S5:MIMO} discusses the role of the number of actuators and sensors for the the learning process.
In \S~\ref{ToC:S5:Noise}, 
the effect of the dynamics complexity  and noise on learning speed is discussed.
Finally, robustness for other operating conditions 
is elaborated in \S~\ref{ToC:S5:Robustness}.

%% Averaged control --------------------------------------------------------------------------------------

\subsection{The need for feedback}
\label{ToC:S5:Feedback} \begin{figure}
  \centerline{\includegraphics[width=0.8\linewidth]{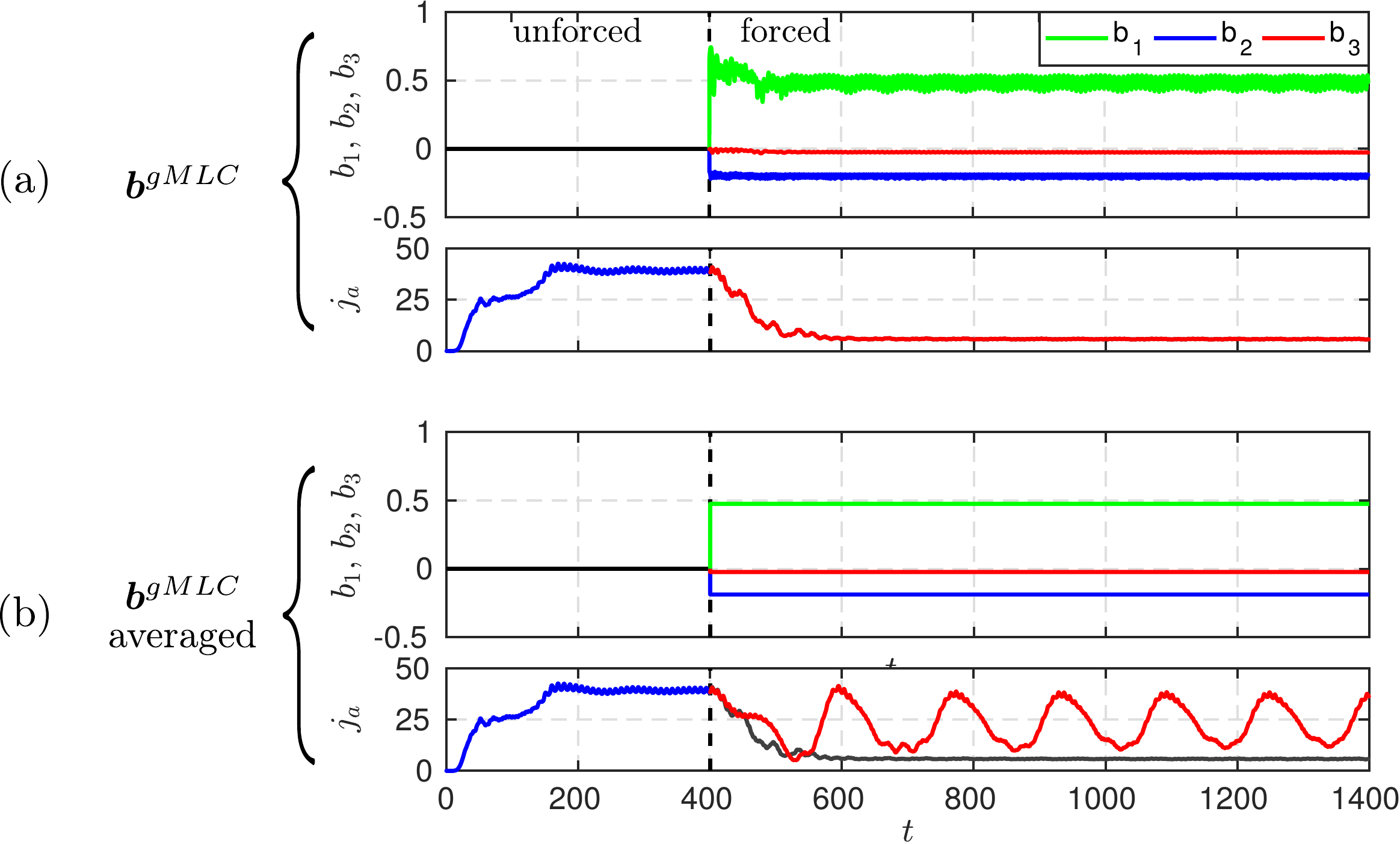}}
 \caption{Time series of the actuation command $\bm{b}=[b_1,b_2,b_3]^\intercal$  and the instantaneous cost $j_a$ for the best feedback control law found with gMLC (a) and (b) for the flow controlled by a steady control law whose values are the averaged gMLC control during the post-transient regime.%;
 }
\label{fig:gMLC_robustness}
\end{figure}
Feedback plays an important role in the gMLC control.
Figure~\ref{fig:gMLC_robustness}a shows the corresponding evolution of the actuation commands
and instantaneous cost function $j_a$.
The actuation commands lead to constant cylinder rotation 
with small fluctuation from the sensor signal.
The cost function converges to a steady value after some 200 non-dimensional time units.
In figure~\ref{fig:gMLC_robustness}b,
the actuation commands are replaced by their respective post-transient averaged value
of the last 500 time units.
Now, the cost function fluctuates periodically 
between the optimal and the unforced value.
The associated averaged cost is $J_a/J_0=0.59$ 
and about three times the  optimal gMLC value $J_a/J_0=0.20$.
The important role of feedback is corroborated with another test.
The actuation commands of the gMLC control are recorded 
and applied in an open-loop manner to the flow with a random initial condition.
Again, the performance $j_a$ largely fluctuates.
Evidently, the small feedback fluctuations play an important role in the stabilization.
Intriguingly, similar observations are made by the authors 
for stabilizing experimental cavity fluctuations 
and will be described in the dissertation of the first author \citep{cornejomacedaPhD}.

%% --------------------------------------------------------------------------------------
%% Large number of sensors and actuators

\subsection{Number of sensors and actuators}
\label{ToC:S5:MIMO}
The control performance is found to increase
as the search space is generalized from single parameter steady base bleeding forcing
to three parameter steady actuation to feedback with 9 sensors.
Generally, increasing the number of actuators and sensors
can be expected to enhance the maximum control performance
albeit with eventually diminishing returns.
On the other hand, 
the learning time will also increase with the number of actuators and sensors
and with the complexity of control laws, e.g., inclusion of time-delay coordinates.
Evidently, there is a trade-off between performance gains 
from increasing the search space
and the limitations of the testing time.
Like in model identification \citep[see, e.g.,][]{AbuMostafa2012book}, 
one can expect an optimal level of complexity for given testing time.
From MLC with dozen's of configurations \citep[see, e.g.,][]{Noack2019springer},
our observation is that the learning time is weakly affected by the number of control law inputs 
but increases with the number of uncorrelated actuation commands.

The subplex iteration of gMLC is found to 
significantly accelerate the learning process.
Evolutionary methods are known to underperform for convergence of identified minima,
i.e., a strength of gradient-based approaches.
Gradient-based methods have another advantage
of staying in well-performing subspaces.
In contrast, genetic operations, like mutation and crossover,
tend to bread new individuals leaving these subspaces.
These observations are particularly relevant 
when a symmetry or invariant of the control law is performance critical.
The inclusion of known symmetries or invariants in gMLC 
might be achieved by pre-testing and excluding individuals 
which strongly depart from these constraints.
An example of self-discovery of such symmetries and invariants 
is reported in \citet{Belus2019} for deep reinforcement learning.

%% Influence of the complexity of the flow --------------------------------------------------------------------------------------
\subsection{Complexity of dynamics and noise}
\label{ToC:S5:Noise}
The applicability of gMLC to turbulent flows will be addressed in future works
starting with \citet{cornejomacedaPhD}.
Already MLC has been successfully applied to learning distributed actuation 
for mixing optimization of a turbulent jet \citep{Zhou2020jfm}.
Recent experimental applications of gMLC include 
mitigation of cavity oscillations, 
drag reduction of a generic truck model under yaw
and lift increase of airfoil under angle of attack at a Reynolds number near one million.
Performance and reproducibility of gMLC control are encouraging 
and outperform other methods, including MLC.
Hence, the very optimization principle 
of iterating between exploration (for discovering new minima) 
and exploitation (for a fast descent towards the minimum) seems sound.
Yet, numerical studies of multi-frequency forcing of the fluidic pinball
foreshadow challenges for asymptotic regimes.
When the actuation space has many `idle' direction with near constant performance,
the gradient-based descent may be trapped in local minima.
One cure is a subplex method on `active subspaces' 
aligned with direction of performance gradients.

Genetic programming is a powerful regression solver 
which is successfully validated in 
dozens of experiments, Navier-Stokes simulations, 
and dynamical systems \citep{Ren2020jh,Noack2019springer}. 
It can learn complex laws for $O(10)$ signals and $O(10)$ actuation commands
by testing $O(1000)$ individuals over $O(1000)$ characteristic times each,
i.e., $O(1,000,000)$ characteristic times in total \citep{Wu2018ef,Zhou2020jfm}.
Yet, it may not be the most effective choice under several conditions:
(1) the total testing time is restricted to much smaller budgets
typical for three-dimensional simulations;
(2) the control law is smooth or can be expected 
to depend linearly or affinely to the sensor signals,
(3) the flow performance responds immediately to good or bad actuation.
Smoothness is well exploited in cluster-based control \citep{Nair2019jfm}.
Deep reinforcement learning may learn quickly the optimal actuation 
in case of fast performance response  \citep{Fan2020pnas,Rabault2019jfm}.
A combination of techniques may also benefit a quick learning such as the merging of genetic algorithm and downhill simplex in \citet{Maehara2013}.
Future work will give more indications about good choices or combinations of machine learning algorithms.

%%--------------------------------------------------------------------------------------
%% Robustness
\subsection{Robustness of the control}
\label{ToC:S5:Robustness}
The current study optimizes control for a single Reynolds number.
Its robustness will be addressed in future work.
We can distill few rules of thumb for robustness 
from past experience with experiments.
First, if the actuation mechanism relies 
on changing large-scale coherent structures,
like synchronizing vortex formation \citep{Parezanovic2016jfm},
the control learned for one condition 
is likely to be robust for a range of conditions.
Second, the control law should be learned in a non-dimensional form.
For instance, the Strouhal number of an actuation 
can be expected to be more relevant for different velocities than the value in Hertz
unrelated to the velocity change \citep{Gautier2015jfm}.
Third, in an ideal scenario, the intended range of operating conditions
is already included in the cost function.
For instance, a control law may be evaluated 
at different operating conditions
or in a slow transient between them \citep{Asai2019aiaa,Ren2019pof}.
This will, however, dramatically  increase the testing time.
The learning time saved by smarter algorithms, like gMLC,
may be invested in assuring robustness for multiple operating conditions.
\citet{Tang2020} provide an inspiring example for deep reinforcement learning.

\section{Conclusions}
\label{Sec:Conclusions}

%\begin{figure}
%  \centerline{\includegraphics[width=0.85\linewidth]{Figures/Summary_Costs}}
%  \caption{Summary of the performances for the best solutions for each method/search space couple.
%  The relative distance to the symmetric steady solution $J_a/J_0$ and the actuation power $J_b$ is represented. The costs are computed over 1000 convective time units.
%  The last column shows the results for a periodic forcing optimization performed with EGM, see appendix \ref{appA}.}
%\label{fig:Summary_costs}
%\end{figure}

    \setul{0.5ex}{0.3ex}
    \definecolor{Yellow}{rgb}{1,1,0.0}
    \definecolor{Blue}{rgb}{0,0.0,1}
    \setulcolor{Blue}

\begin{figure}
% table caption is above the table       % Give a unique label
% For LaTeX tables use
\begin{center}
\def~{\hphantom{0}}
\begin{tabular}{>{\centering}p{2.75cm}>{\centering}p{2cm}>{\centering}p{2cm}>{\centering}p{3.5cm}p{1.8cm}}
  search space & dimension & method & $N_i$ & \ul{$J_a/J_0$}, 
      \setulcolor{Yellow}
  \ul{$J_b$}
 \\
\midrule
unforced natural & - & - & - & \includegraphics{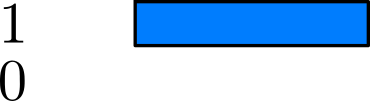}
\\ \rule{0pt}{5ex}
symmetric steady & 1 &param. study & - & \includegraphics{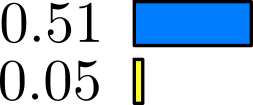}
\\ \rule{0pt}{5ex}
general steady & 3 & EGM & 77 & \includegraphics{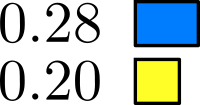}
\\ \rule{0pt}{5ex}
feedback control & $\infty$ & gMLC & 800   & \includegraphics{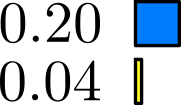} 
\\ \multicolumn{5}{r}{250 for $J_a/J_0<0.26$\rule{11mm}{0pt}}
\\ \rule{0pt}{5ex}
feedback control & $\infty$ & MLC & 900 & \includegraphics{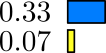}
\\ \rule{0pt}{5ex}
periodic forcing & 2 & EGM & 74 & \includegraphics{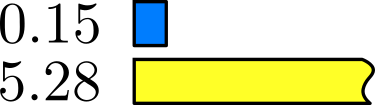}
\end{tabular}
\end{center}
\caption{\label{tab:summary_methods}
    Summary of the performances for the best solutions of each search space.
    The first column describes the search space.
The second column contains the respective dimension of the studied search space.
 The method and the number of evaluations needed to arrive 
 at the presented solution are listed in the third and fourth column, respectively.
  The fifth column shows the relative distance to the symmetric steady solution $J_a/J_0$ (in blue) and the actuation power $J_b$ (in yellow).
For the gMLC feedback control optimization, 
the best control law studied has been found after 800 evaluations 
but the cost $J_a/J_0$ was already under 0.26 already after 250 individuals.
The fifth row corresponds to the solution found with standard MLC after 1000 evaluations
as elaborated in appendix \ref{appB}. 
The last row shows the results for a periodic forcing optimization performed with EGM, see appendix \ref{appA}.}
\end{figure}

We have stabilized the wake behind a fluidic pinball
with three independent cylinder rotations 
in successively larger search spaces for control laws.
Table~\ref{tab:summary_methods} summarizes the corresponding performances
quantified by the average distance between the controlled flow and the steady symmetric solution.
First, steady symmetric forcing is employed.
A base bleed solution with a cylinder rotation of 28\% of the oncoming velocity
leads to a flow which is 49\% closer to the symmetric solution
than the unforced attractor.
Other studies also report about a stabilizing effect of base bleed on bluff body wakes \citep{Wood1964jras,Bearman1967aq}.
In contrast, Coanda forcing, i.e., two symmetric cylinder rotations
which accelerate the outer flow, may completely stabilize the flow.
Yet,  this new wake has no long vortex bubble 
and is further away from the symmetric steady solution than the unforced vortex shedding.

Second, a general non-symmetric actuation 
is optimized with the explorative gradient method.
Surprisingly, an asymmetric actuation reduces the average distance
between the flow and the steady target solution further to 28\%
of the unforced value.
This asymmetric actuation leads to shear layer vortices 
which do not interact and thus
do not form von K\'arm\'an vortices.
The mean flow is slightly asymmetric, 
but largely mimics the elongated steady symmetric solution.
The price for the better performance is a larger actuation power
(see table~\ref{tab:summary_methods}).
Intriguingly, 
machine learning control also
leads to distinctly asymmetric actuation 
in experiments  \citep{Raibaudo2020pf} and simulations \citep{Cornejo2019pamm}
for other cost functions.

\begin{figure}
  \centerline{\includegraphics[width=1\linewidth]{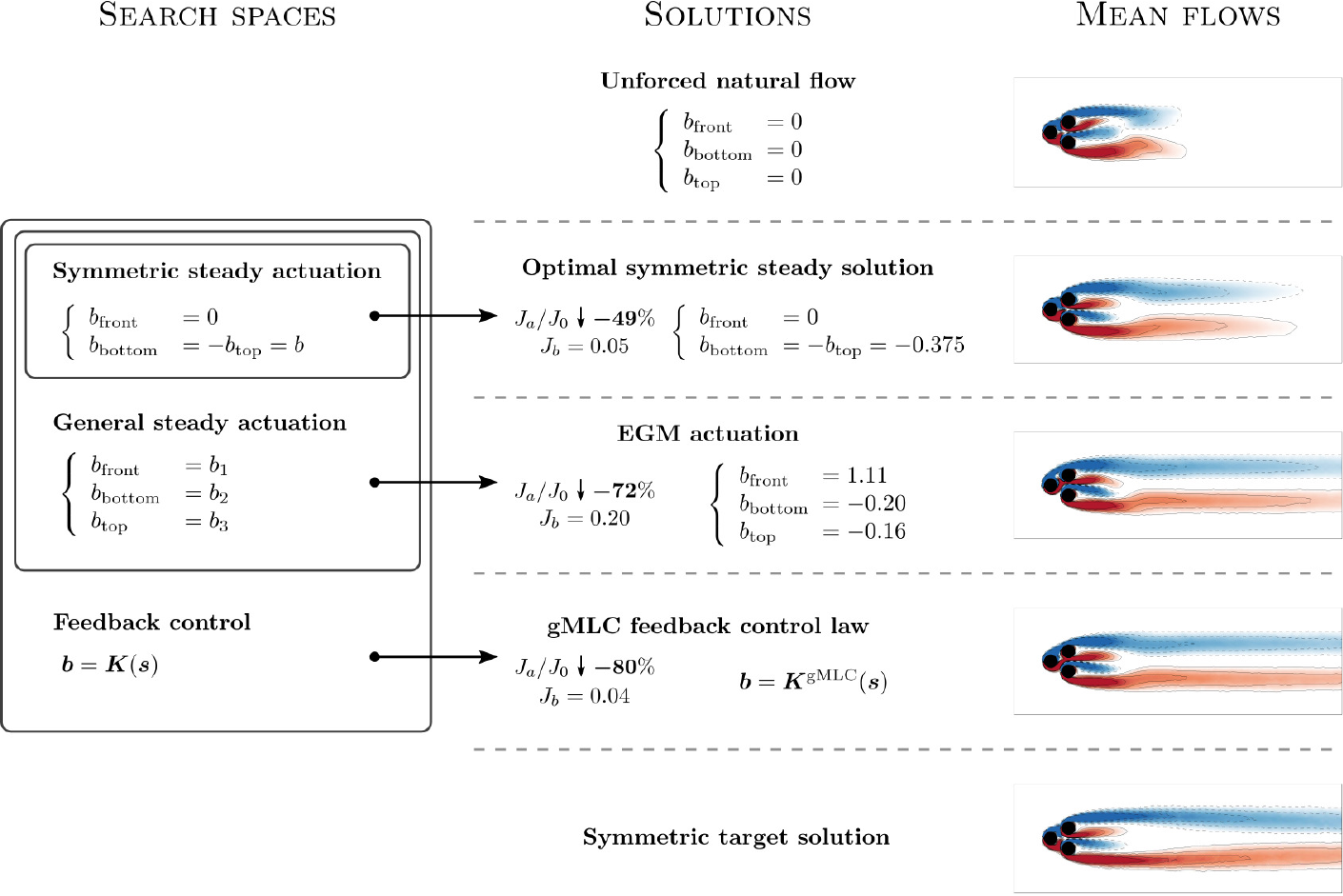}}
 \caption{Summary of the optimized stabilization solutions obtained for each search space.
The Venn diagram (left) depicts the hierarchy of the control law spaces.
The corresponding \emph{optimal} solutions are presented along with their  performances, control laws (center) and mean fields (right).
The mean field of the statistically asymmetric unforced flow is depicted in the top row
and the symmetric target solution in the bottom row.
 }
\label{fig:summary_stabilization}
\end{figure}

Third, a feedback actuation obtained from gradient-enriched machine learning control
brings the flow even closer to the steady target solution.
The associated actuation power is smaller 
than the previous optimized steady actuations
(see table~\ref{tab:summary_methods}).
The actuation is a combination of asymmetric steady forcing
and phasor control.
The resulting flow looks similar to the optimal asymmetric steady forcing.
Figure~\ref{fig:summary_stabilization} 
summarizes the results for the hierarchy of control search spaces.

%We note also that for all three actuation power associated with the steady symmetric control ($J_b=0.05$) , the generalized steady control ($J_b=0.20$) and the gMLC control ($J_b=0.04$) are all inferior to the first component Ja/J0, especially for the gMLC control.
%Indeed, following the topolgy of the control landscape of the general steady control it is unlikely to have a solution that reduces $J_a$ as much as the best solution with less actuation power.
%As for the gMLC solution, it has the lowest actuation power so far.
%So, it is more than likely that these solutions are the more power-efficient ones, making acuation power penalization unnecessary.

The feedback control does not seem to have the authority
to completely stabilize the symmetric target solution,
like for the cylinder wake controlled by a volume force \citep{Gerhard2003aiaa}.
The wake can  be `almost' stabilized for short periods of time,
starting from the unforced flow.
Then, new coherent structures emerge and lead to residual shear-layer shedding.
This lack of complete authority for stabilization
may be explained by the complexity of the dynamics.
The fluidic pinball has a primary instability
associated with von K\'arm\'an vortex shedding,
a secondary pitchfork instability associated with the centerline jet, and
two Kelvin-Helmholtz instabilities of the top and bottom shear layer.

Intriguingly, symmetric high-frequency forcing
can bring the flow even closer to the steady target solution 
but with an actuation power which is roughly two orders of magnitude
larger than the previous control laws 
(see table~\ref{tab:summary_methods}).
\citet{Protas2004pf} and \citet{Thiria2006jfm} also find a stabilizing effect
of high-frequency forcing on vortex shedding.
The thickening of the shear layers by high-frequency vortices 
reduces the gradients and thus the instability.
To summarize, machine learning control has automatically discovered
well known stabilizing mechanisms, 
like base bleed and phasor control,
but added an unexpected asymmetric forcing 
and combination of this open-loop actuation 
and phasor feedback for improved performance.

The presented stabilizations are expected 
to be independent of the employed optimizer
as different approaches lead to very similar results.
The chosen optimizers balance exploitation (downhill descend of found minima)
and exploration (search for better minima).
The optimization has been effected 
in a three-dimensional parameter space for steady forcing
and a feedback space with three actuation inputs and nine sensing outputs.
Starting point is Latin hypercube sampling as exhaustive exploration 
of the parameter space
and linear genetic programming control as effective regression solver
with explorative and exploitive features. 
The search has been  significantly accelerated
by intermittently adding gradient-based descends.
The resulting explorative gradient method and gradient-enriched
machine learning control seem efficient for both exploration and exploitation.
Future research shall focus on accelerated learning.

The control performance may be further improved 
by allowing for more general control laws 
comprising the history of the sensor signals,
like in ARMAX-based control \citep{Herve2012jfm}.
Other generalizations of machine learning control include 
multiple pre-defined operating conditions, 
adaptive control for unknown operating conditions, 
automated learning of the response model 
from the control law to performance following \citet{Fernex2020prf}
and automated learning of control-oriented modeling based on \citet{LiH2020jfm}.
The fluidic pinball represents an attractive plant of sufficient dynamic complexity
with manageable computational load for these developments.

%\begin{acknowledgements}
\section*{Acknowledgements}
The authors thank Antoine Blanchard, Nan Deng, Luc Pastur and Themis Sapsis 
for fruitful discussions and enlightening comments.
We also thank the anonymous referees for their insightful suggestions.
This work is supported by the French National Research Agency (ANR)
via FLOwCON project “Contr\^ole d’\'ecoulements turbulents en boucle ferm\'ee par apprentissage automatique” funded by the ANR-17-ASTR-0022, the German National Science Foundation (DFG grants SE 2504/1-1 and SE 2504/3-1), the iCODE Institute, research
project of the IDEX Paris-Saclay and by the Hadamard Mathematics
LabEx (LMH) through the grant number ANR-11-LABX-0056-LMH in the
“Programme des Investissements d'Avenir”. 
%\end{acknowledgements}
\section*{Declaration of interests}
The authors report no conflict of interest.

%%Print the glossary
%\glsaddall
%\setglossarystyle{list}
%\printglossaries

\appendix
 \section{Comparison between MLC and gMLC} \label{appB}
In this appendix, 
we compare the performances of a machine learning control (MLC), 
based on linear genetic programming control \citep{Li2018am,Zhou2020jfm}, 
and the proposed gradient-enriched MLC (gMLC) variant for the stabilization of the fluidic pinball.
gMLC is described in section \S~\ref{sec:results_gMLC}.
MLC differs from gMLC in two respects.
First, the evolution is from generation to generation, i.e., groups of individuals.
Second, unlike gMLC, no gradient information is employed.
The first generation of randomly generated individuals 
evolves through generations thanks to three genetic operations:
\begin{itemize}
\item \emph{Crossover}: a stochastic recombination of two individuals, giving two new individuals exploiting parts of the first two individuals;
\item \emph{Mutation}: a stochastic change in one individual, giving one new individual more or less different from the previous one; 
\item \emph{Replication}: an identical copy of one individual, assuring memory of good individuals throughout the generations.
\end{itemize}
During the evolution process, 
the better-performing individuals are selected with larger probability
to build new individuals thanks to the genetic operators.
The best individuals are selected thanks to a tournament selection method.
As in \citet{Duriez2016book}, we choose a tournament selection of size 7 for 100 individuals.
A genetic operation is chosen randomly following given probabilities: 
the crossover probability $P_c$, the mutation probability $P_m$ and the replication probability $P_r$.
The probabilities add up to unity $P_c+P_m+P_r=1$.
The set of parameters $[P_c,P_m,P_r]^\intercal=[0.6,0.3,0.1]^\intercal$ suggested in \citet{Duriez2016book} have been chosen for MLC.
A parametric study varying $P_c$, $P_m$ and $P_r$ with a 0.1 step has been carried out on the stabilization of a Landau oscillator by forcing only on one of its components.
As MLC is a stochastic process, 
we perform 100 test runs for each probability combination.
This parametric study reveals that this probability combination $[P_c,P_m,P_r]^\intercal=[0.6,0.3,0.1]^\intercal$ is one of the best.
Among all the probability combinations, this combination is one of those that converges towards better solutions in average, with one of the lowest dispersion of the final solutions over the 100 test runs, showing that this combination is also one of the more robust.
%Thus, 
%this is the probability combination that has been chosen for MLC.
This or a very similar probability combination 
has already been used in \citet{Duriez2016book} and dozens of experiments \citep{Noack2019springer}.

In addition to  \emph{crossover}, \emph{mutation} and \emph{replication}, 
we transfer the best individual of the previous generation 
to the next one via the \emph{elitism} operation.
This operation assures that the best individual is always in the latest generation
so that `the winner does not get lost'.

The architecture of the linear programming control laws are the same for MLC and gMLC, including the mathematical operations, number of constants, 
number of registers, as well as inputs and outputs (see table~\ref{tab:gMLC_parameters}).

\begin{figure}
 \centerline{\includegraphics[width=0.90\linewidth]{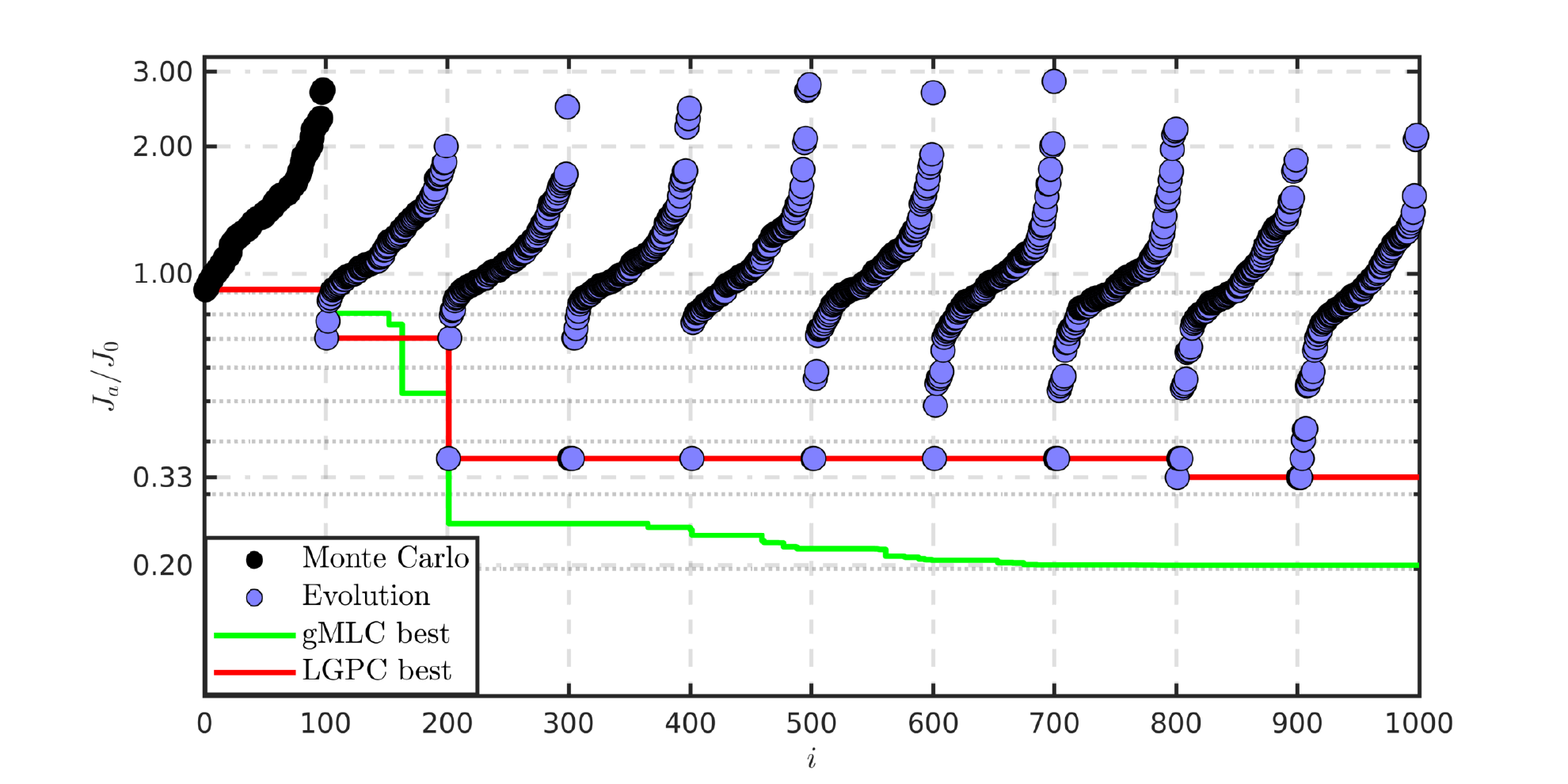}}
 \caption{Distribution of the costs during the MLC optimization process.
 Each dot represents the cost $J_a/J_0$ of one individual.
 The color of the dots represent how the individuals have been generated.
 Black dots for the individuals randomly generated by a Monte Carlo process (individuals $i=1,\ldots,100$), 
 blue dots for the individuals generated from a genetic operator (individuals $i=101,\ldots,1000$).
 For each generation the individuals have been sorted according to their cost.
 The red line shows the evolution of the best cost for the MLC optimization process.
 The green curve corresponds to the gMLC optimization process.
The vertical axis is in log scale.}
\label{fig:MLCvsgMLC}
\end{figure}

The cost function is evaluated over 1000 convective time units, both in MLC and gMLC.
The MLC and gMLC algorithms are compared over 1000 evaluations.
For MLC, a population of 100 individuals is chosen to evolve over 10 generations.
For a fair comparison, MLC and gMLC share the same initial Monte Carlo generation, 
comprising the first 100 randomly generated individuals.
Figure~\ref{fig:MLCvsgMLC} shows the distribution of the costs $J_a/J_0$ as a function of the evaluations.
We notice that for both algorithms the first exploration phase makes great improvement.
In the second generation, 
the best cost is  0.80 for gMLC and 0.70 for MLC.
Note that MLC's better performance is understandable as 100 individuals have been evaluated for
the second generation whereas only 50 individuals have been evaluated for gMLC.
After testing 200 individuals, 
gMLC surpasses MLC thanks to the subplex steps, 
reaching a cost $J_a/J_0=0.36$.
For the second evolution phase, 
both MLC and gMLC perform well reaching low levels of $J_a/J_0$: 0.36 for MLC and 0.26 for gMLC.
Then, MLC achieves only small progress after 900 evaluations, the cost improves from 0.36 to 0.33.
The series of blue dots at $J_a/J_0=0.36$ from $i=201$ to $i=900$ represents several instances of the best individual of generation 3, duplicated thanks to elitism.
For gMLC, figure \ref{fig:gMLC_Learning} shows that evolution phases do not bring any progress after 250 evaluations 
%%% BRN20201005 I do not see these 250 evaluations. Do you mean the plateau i=200...350?
and further improvement is made thanks to the subplex steps. 
As described in section \S~\ref{sec:results_gMLC}, 
the evolution phases help to enrich the simplex subspace.
The subplex steps manage to reduce the cost function from 0.26 to 0.20.
We notice that after 600 evaluations all new subplex individuals have the same cost.
Hence, gMLC surpasses MLC with a smaller number of evaluations 
and enables improvement/fine-tuning of the control laws in the final phase.
%-----------------------------------------------------------------------

%\begin{figure}
% \centerline{\includegraphics[width=0.90\linewidth]{Figures/Progression/Results}}
% \caption{Evolution of the cost function of the best control law for the three strategies(top) Monte Carlo, (middle) MLC and (bottom) gMLC runs. The three strategies sharethe same 100 first individuals generated randomly (Monte Carlo). The first row shows asnapshot of the fluidic pinball for the lowest value of the cost (in red in the second row).The last row shows the actuation command.
% The MLC solution is  $(b_1,b_2,b_3) = (0.5404,-0.4255,0.01235)$.}
%\label{fig:MLCJ}
%\end{figure}
 % Comparison between MLC and gMLC \label{appC}
\section{Optimal periodic forcing} 
\label{appA}
%% Ja, Jb maps ---------------------------------------------------------
\begin{figure}
\centering
\begin{subfigure}{.45\textwidth}
  \centering
\includegraphics[width=\linewidth]{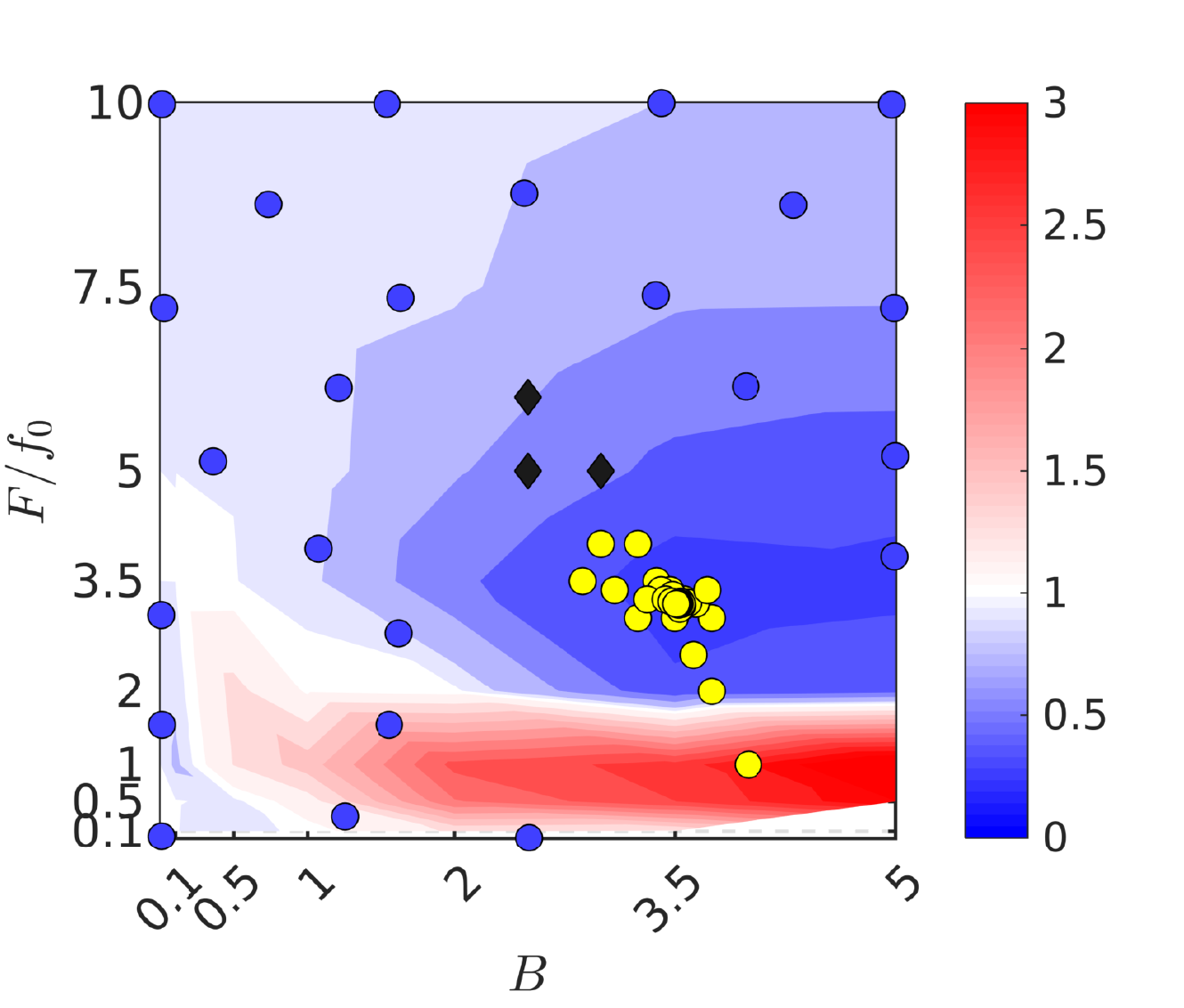}
\caption{\label{fig:EGM_periodic_ja}$J_a/J_0$.}
\end{subfigure}%
\hfil
\begin{subfigure}{.45\textwidth}
  \centering
\includegraphics[width=1.\textwidth]{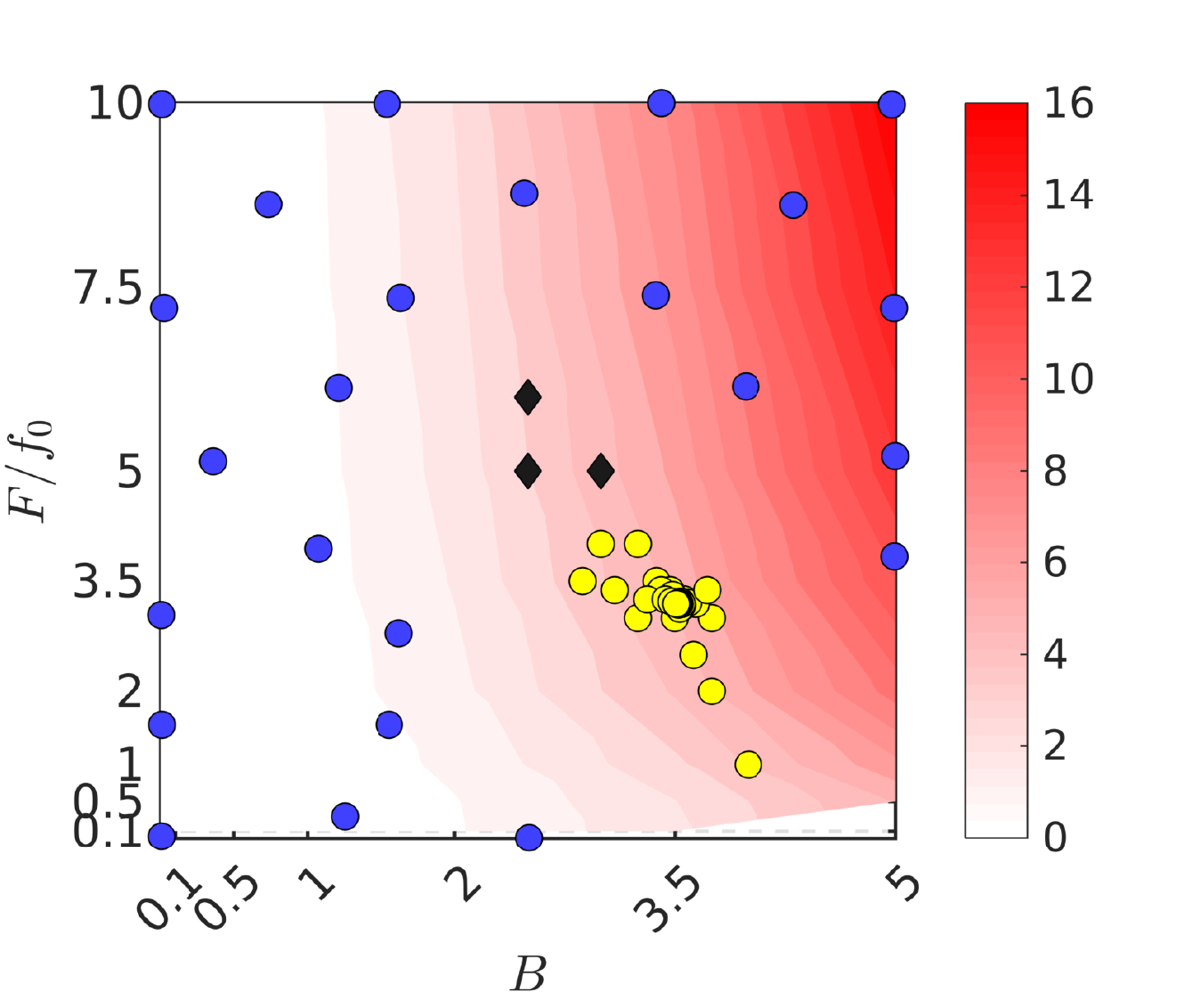}
\caption{\label{fig:EGM_periodic_jb}$J_b$.}
\end{subfigure}
  \caption{Contour plot for $J_a/J_0$ (a) and $J_b$ (b) as a function of the amplitude $B$ and the normalized frequency $F/f_0$. For (a), blue (red) regions denote good (bad) performances 
  while white regions correspond to costs that are equivalent to the natural flow. 
  For (b), the color code describes the actuation energy. 
  % The white zone on the bottom right is due to the limits of our simulation capabilities.
  The symbols represent the individuals tested with EGM: 
  black diamonds for the initial conditions, 
  blue solid circles for  exploration phases and 
  yellow solid circles for the exploitation phases.
  For the legend, refer to figure~\ref{fig:Slice_Ja}.}
\label{fig:EGMPeriodicMap}
\end{figure}
%-----------------------------------------------------------------------

In this appendix we aim to stabilize the symmetric steady solution thanks to a symmetric periodic forcing.
In this case, the two back cylinders oscillate in opposite directions whereas the front cylinder stays still.
The control ansatz is the following:
\begin{equation*}
\begin{array}{cc}
b_1 = & 0 \\
b_2 = &  ~B \cos (2\pi F t) \\
b_2 = & -B \cos (2\pi F t)
\end{array}
\end{equation*}
with $B$, the amplitude of the oscillations, and $F$, the frequency, being the two parameter to optimize.
The search space is limited to $[B,F/f_0]^\intercal \in [0,5]\times[0,10]$ as higher amplitudes and frequencies would be beyond our solver capabilities.
This two-dimensional search space is explored with EGM.
The contour plot in figure~\ref{fig:EGMPeriodicMap} depicts the search space based on $J_a/J_0$ and $J_b$.
The contour plot has been produced thanks to simulations 
for $B \in \{ 0.1,\>0.5,\>1,\>2,\>3.5,\>5 \}$ and 
$F/f_0 \in \{ 0.1,\> 0.5,\>1,\>2,\>3.5,\> 5,\>7.5,\>10 \}$.
The steps are finer for low frequencies and low amplitudes.
The individuals have been evaluated over 250 convective time units.
We notice that there is only one minimum on the plane, close to $[B,F/f_0]^\intercal = [3.51,3.19]^\intercal$.
Also, forcing at frequencies close to the natural frequency resonates with the flow and drastically increases the distance to the steady solution for high amplitudes.
For $J_b$, the contour map expectedly displays high values at high frequencies and large amplitudes.
The three initial control laws for EGM are the center of the box and increments of 1/5 of the box size in each direction: $[2.5,5]^\intercal$, $[3,5]^\intercal$, $[2.5,6]^\intercal$.
As expected, the LHS steps (in blue) spread rather evenly in the domain 
whereas the simplex steps (in yellow) quickly descend into the global minimum.

\begin{figure}
\centering
\begin{subfigure}{.45\textwidth}
  \centering
\includegraphics[width=1.\textwidth]{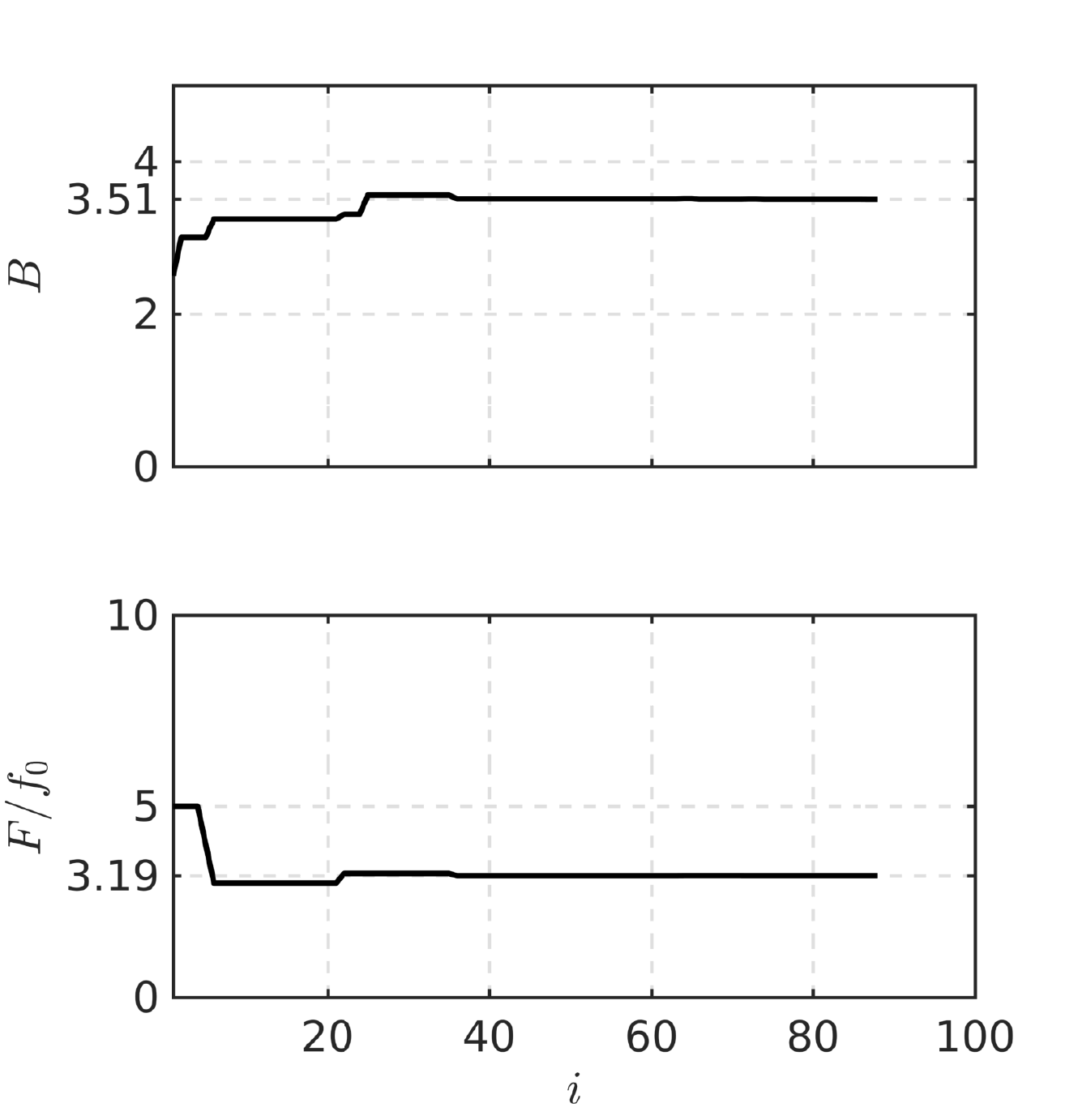}
\caption{\label{fig:AmpFreqprog}}
\end{subfigure}%
\hfil
\begin{subfigure}{.45\textwidth}
  \centering
\includegraphics[width=1.\textwidth]{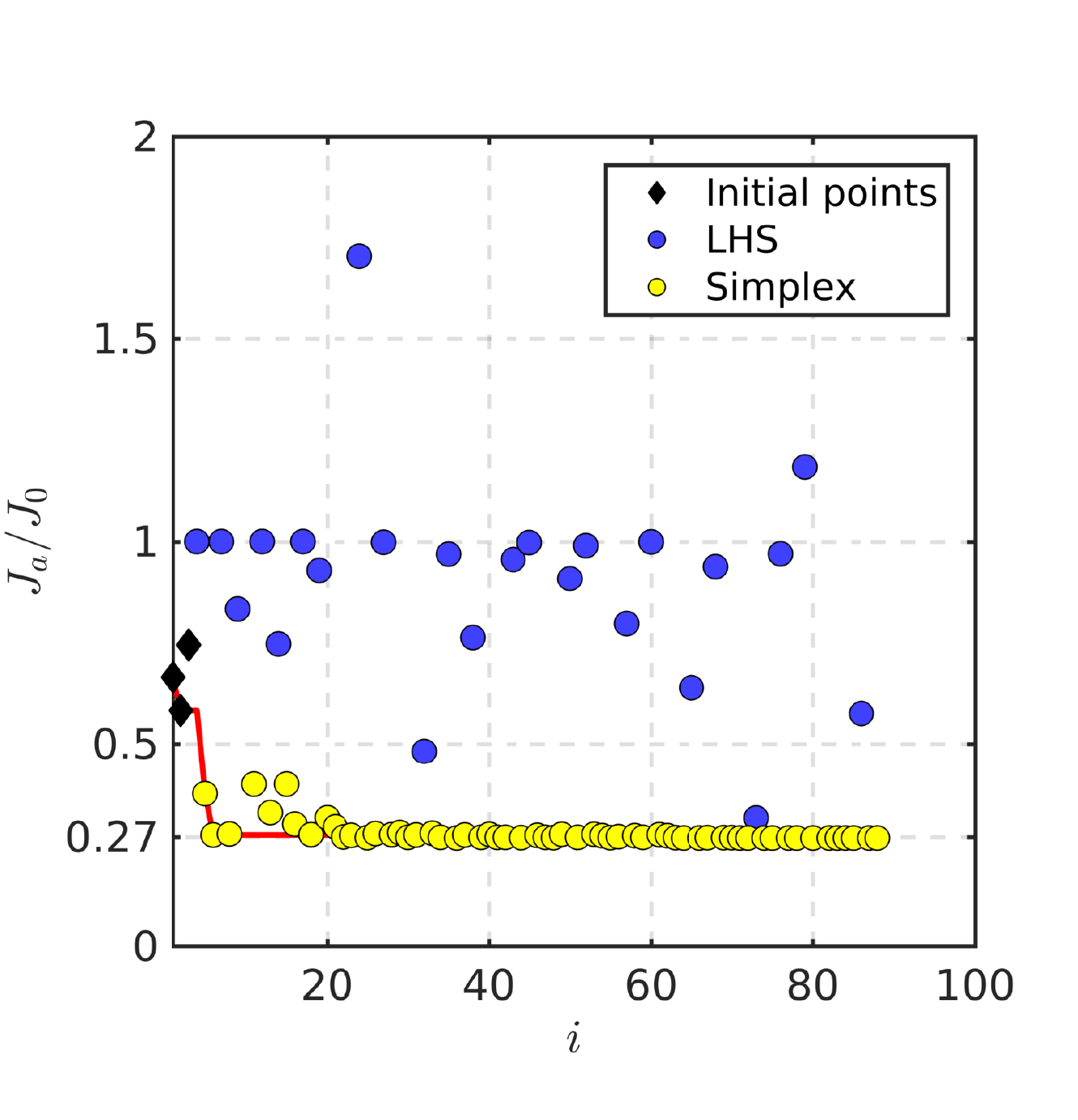}
\caption{\label{fig:JPerioprog}}
\end{subfigure}%
\caption{\label{fig:Perioprog}Evolution of (a) the amplitude $B$ 
and the normalized frequency $F/f_0$ and (b) the reduced cost $J_a/J_0$ as a function of the number of evaluations $i$, for the EGM optimization process.
The red line in (b) shows the evolution of the best cost.
The evaluation time is 250 convective time units.}
\label{fig:fig}
\end{figure}

\begin{figure}
\centering
\begin{subfigure}{.45\textwidth}
  \centering
\includegraphics[width=1.\textwidth]{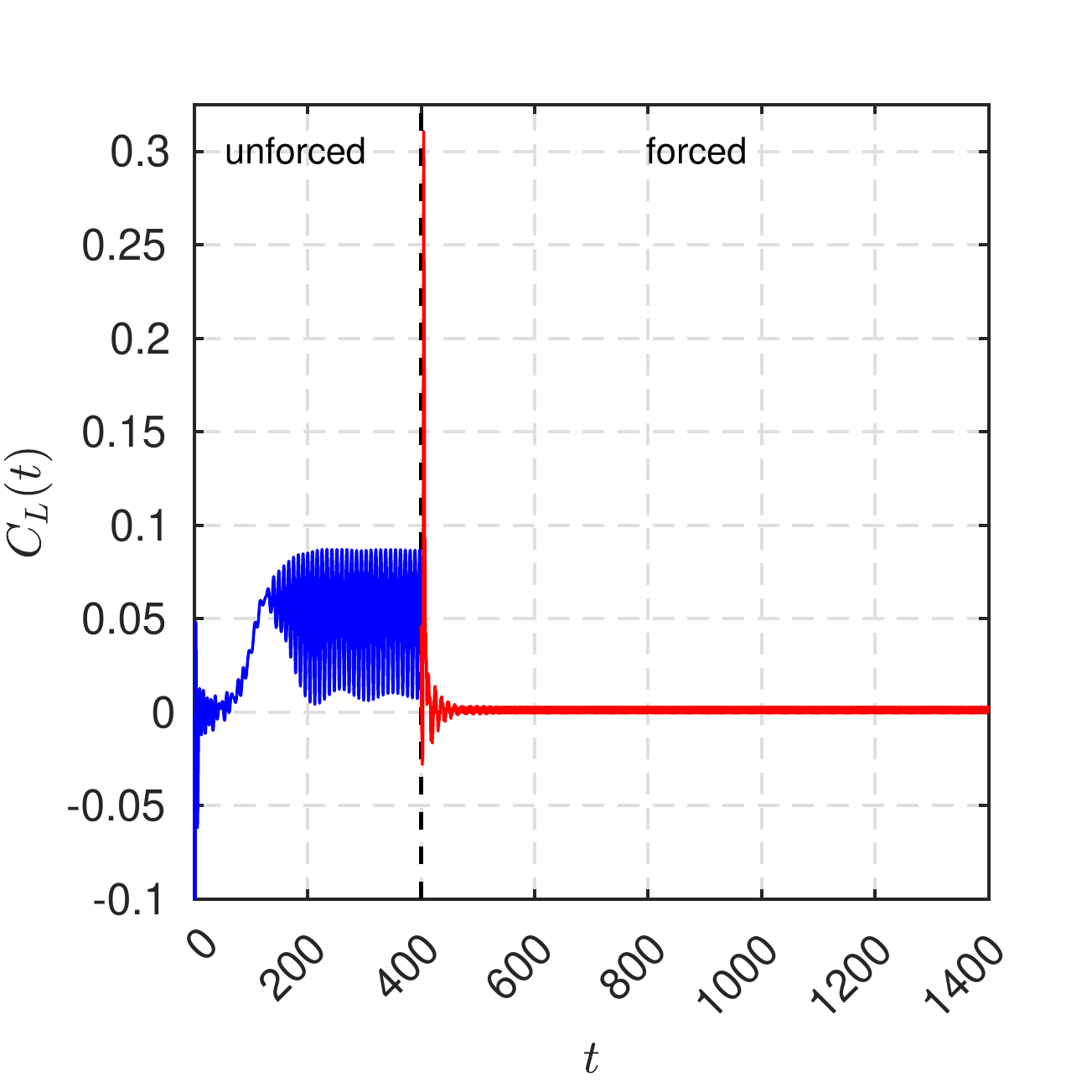}
\caption{\label{fig:CL_Perio}}
\end{subfigure}%
\hfil
\begin{subfigure}{.45\textwidth}
  \centering
\includegraphics[width=1.\textwidth]{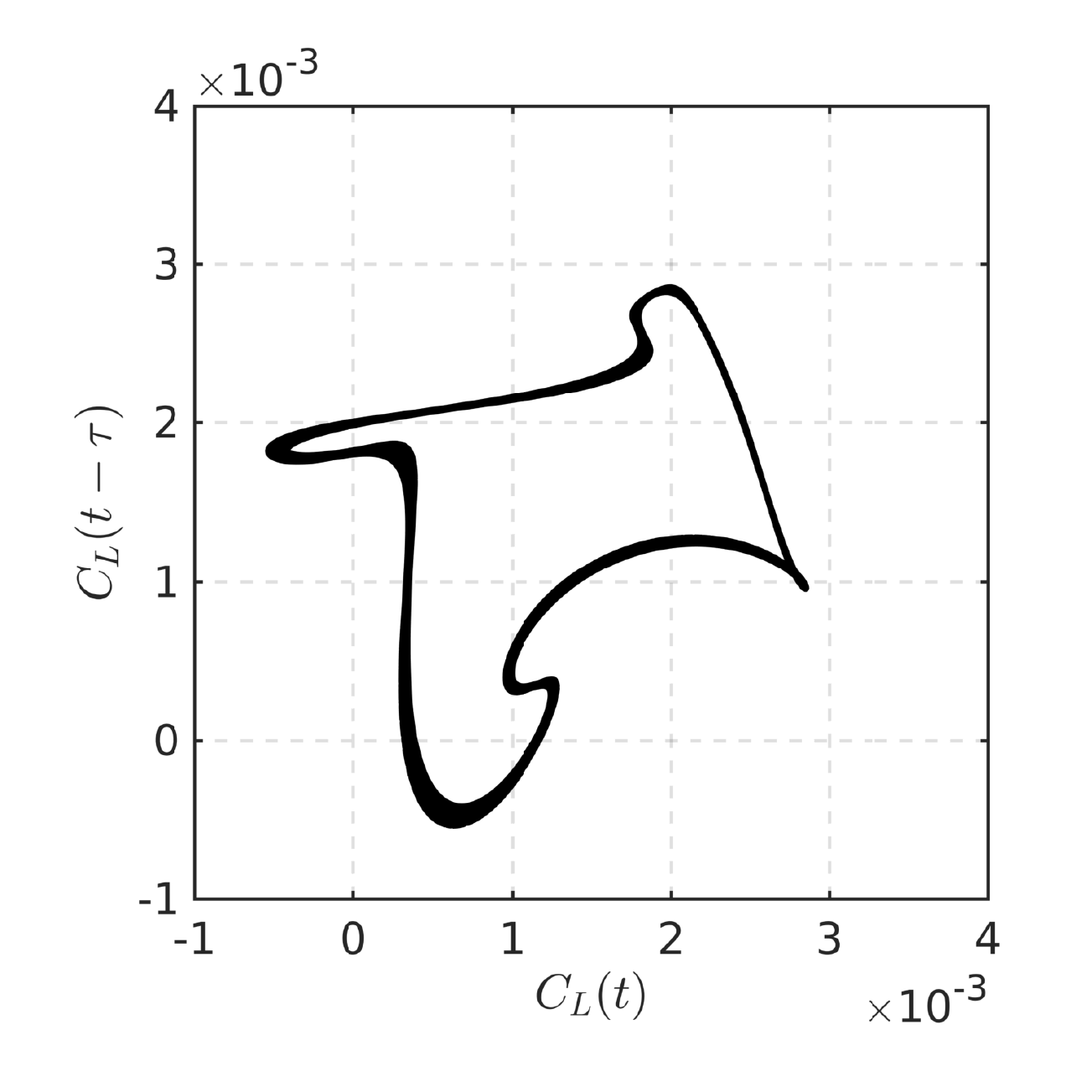}
\caption{\label{fig:PP_Perio}}
\end{subfigure}

\begin{subfigure}{.45\textwidth}
  \centering
\includegraphics[width=1.\textwidth]{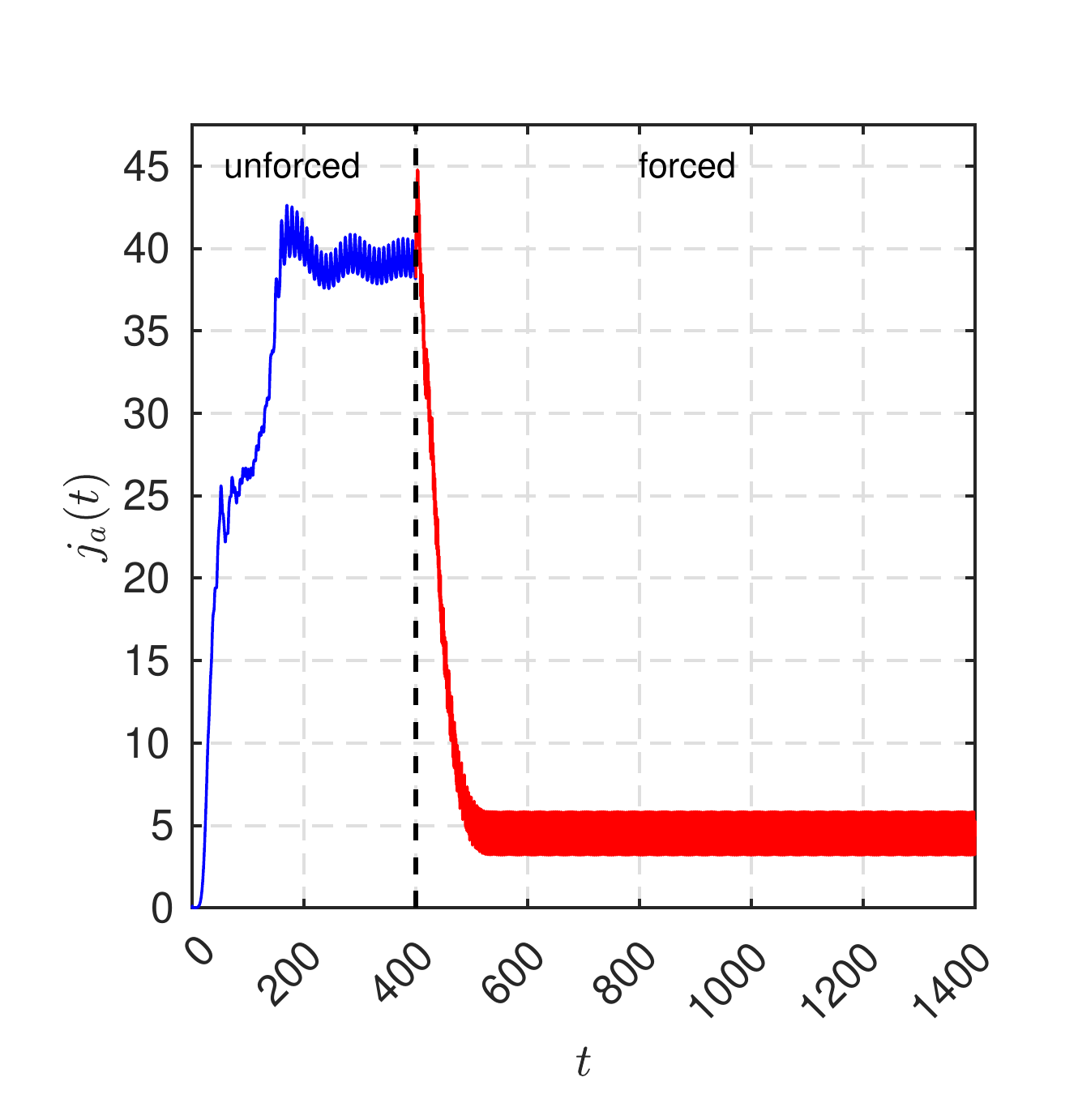}
\caption{\label{fig:DSS_Perio}}
\end{subfigure}%
\hfil
\begin{subfigure}{.45\textwidth}
  \centering
\includegraphics[width=1.\textwidth]{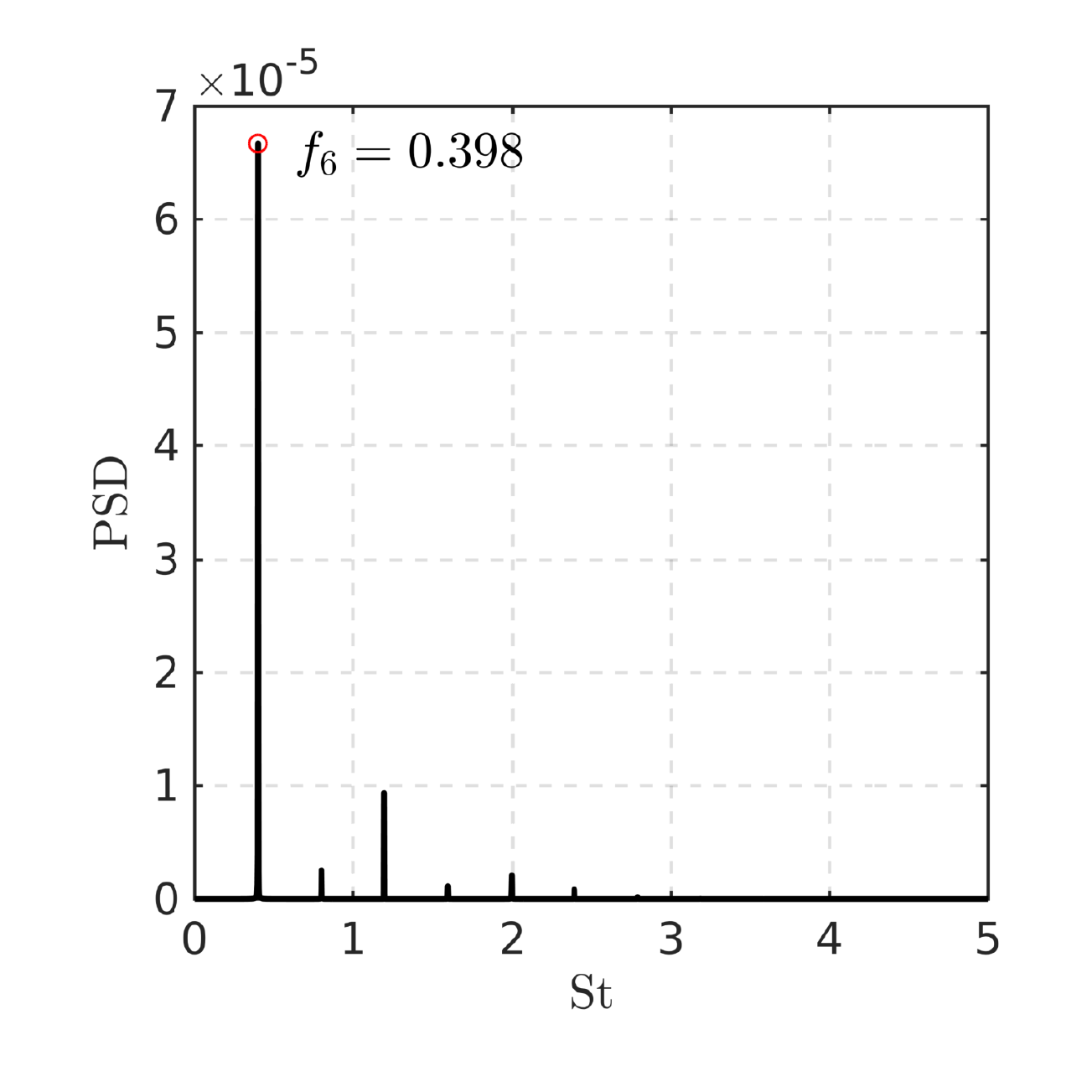}
\caption{\label{fig:PSD_Perio}}
\end{subfigure}
\caption{\label{fig:EGM_Perio_characteristics}Characteristics of the best periodic forcing found with EGM. (a) Time evolution of the lift coefficient $C_L$, (b) phase portrait (c) time evolution of instantaneous cost function $j_a$ and (d) Power Spectral Density (PSD) showing the main frequency $f_6=0.398$ of the forced flow and six harmonics.
The control starts at $t=400$.
The unforced phase is depicted in blue and the forced one in red.
The phase portrait and the PSD are computed over $t \in [900,1400]$, during the post-transient regime.}
\end{figure}

\begin{figure}
\centering
\begin{subfigure}{.45\textwidth}
  \centering
\includegraphics[width=1.\textwidth]{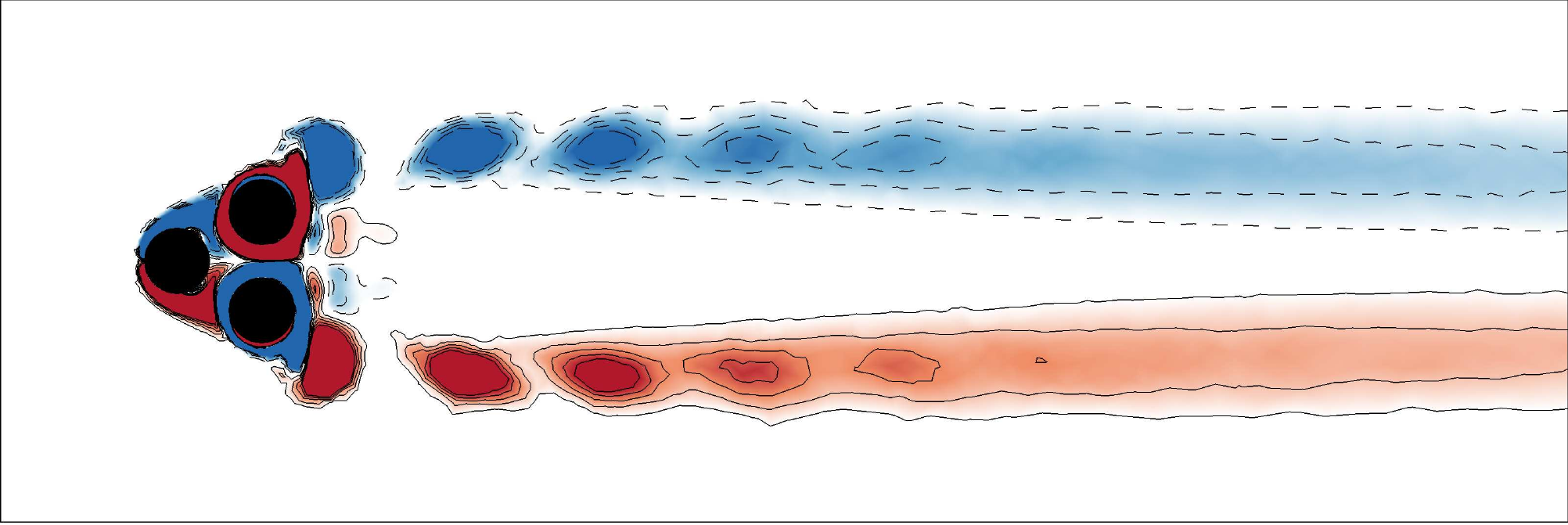}
\caption{\label{fig:EGM_Perio_T1}$t+T_6/8$}
\end{subfigure}%
\hspace{0.5cm}
\begin{subfigure}{.45\textwidth}
  \centering
\includegraphics[width=1.\textwidth]{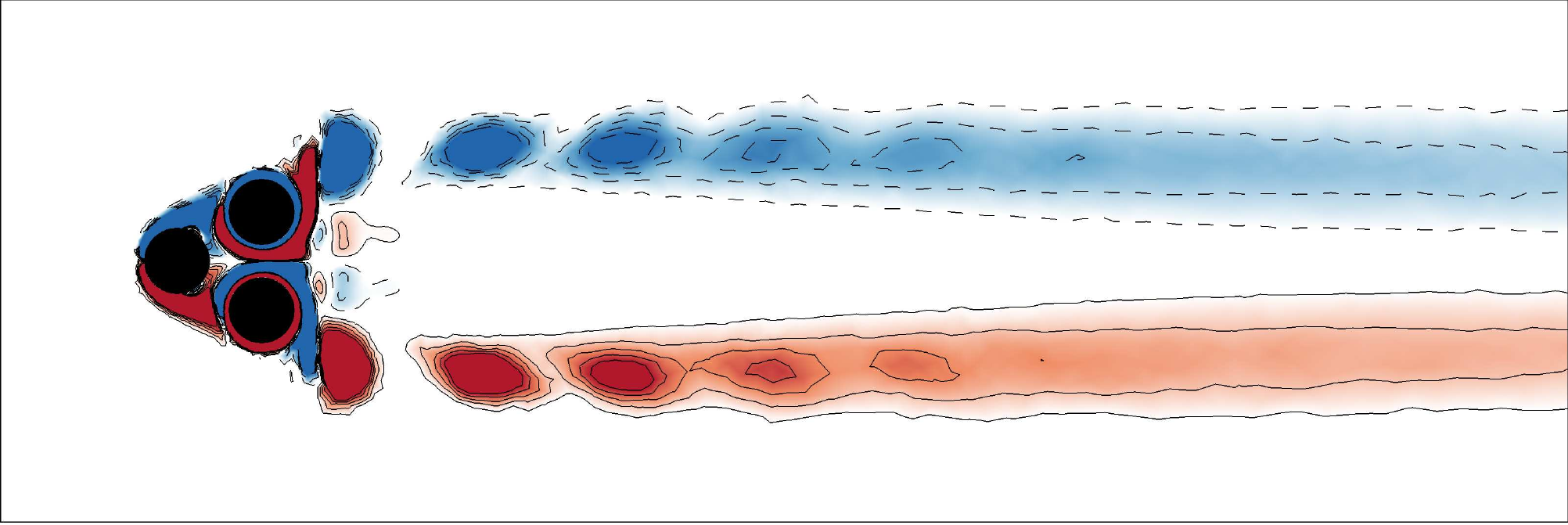}
\caption{\label{fig:EGM_Perio_T2}$t+2T_6/8$}
\end{subfigure}

\begin{subfigure}{.45\textwidth}
  \centering
\includegraphics[width=1.\textwidth]{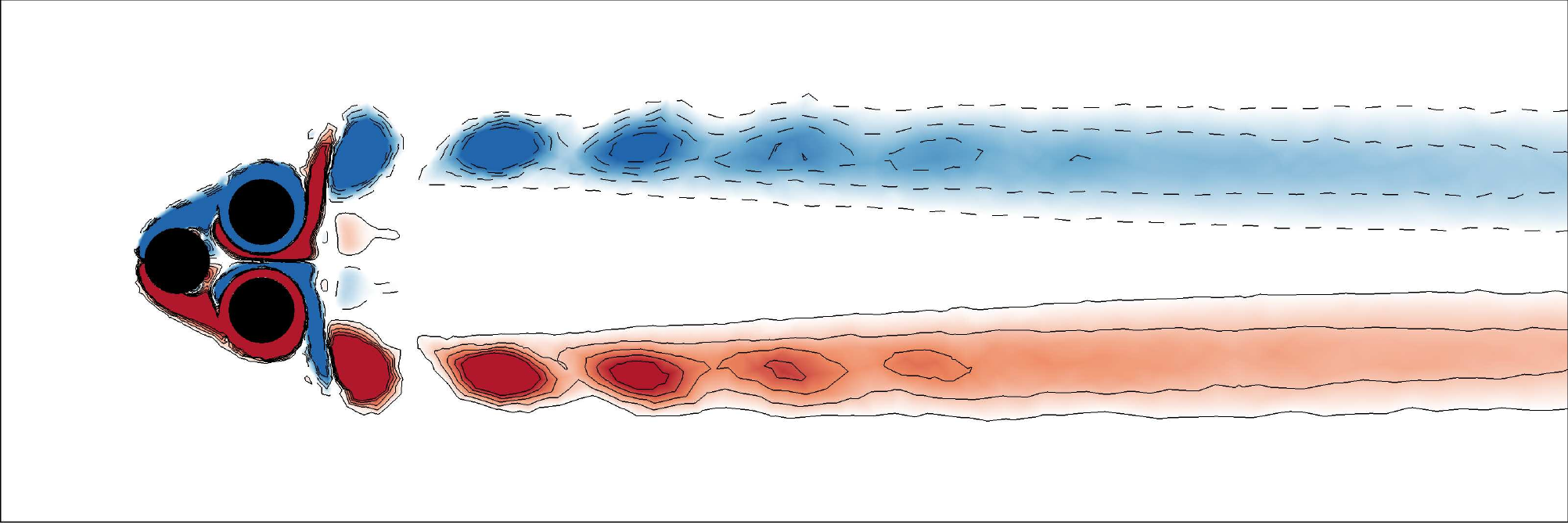}
\caption{\label{fig:EGM_Perio_T3}$t+3T_6/8$}
\end{subfigure}%
\hspace{0.5cm}
\begin{subfigure}{.45\textwidth}
  \centering
\includegraphics[width=1.\textwidth]{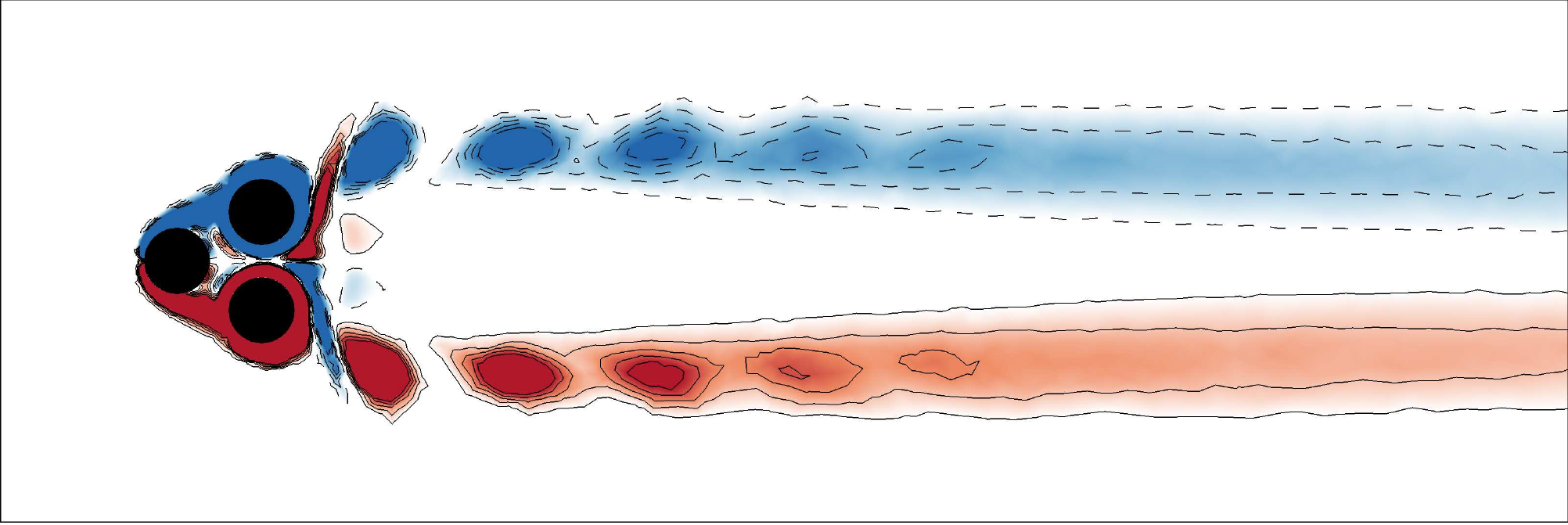}
\caption{\label{fig:EGM_Perio_T4}$t+4T_6/8$}
\end{subfigure}

\begin{subfigure}{.45\textwidth}
  \centering
\includegraphics[width=1.\textwidth]{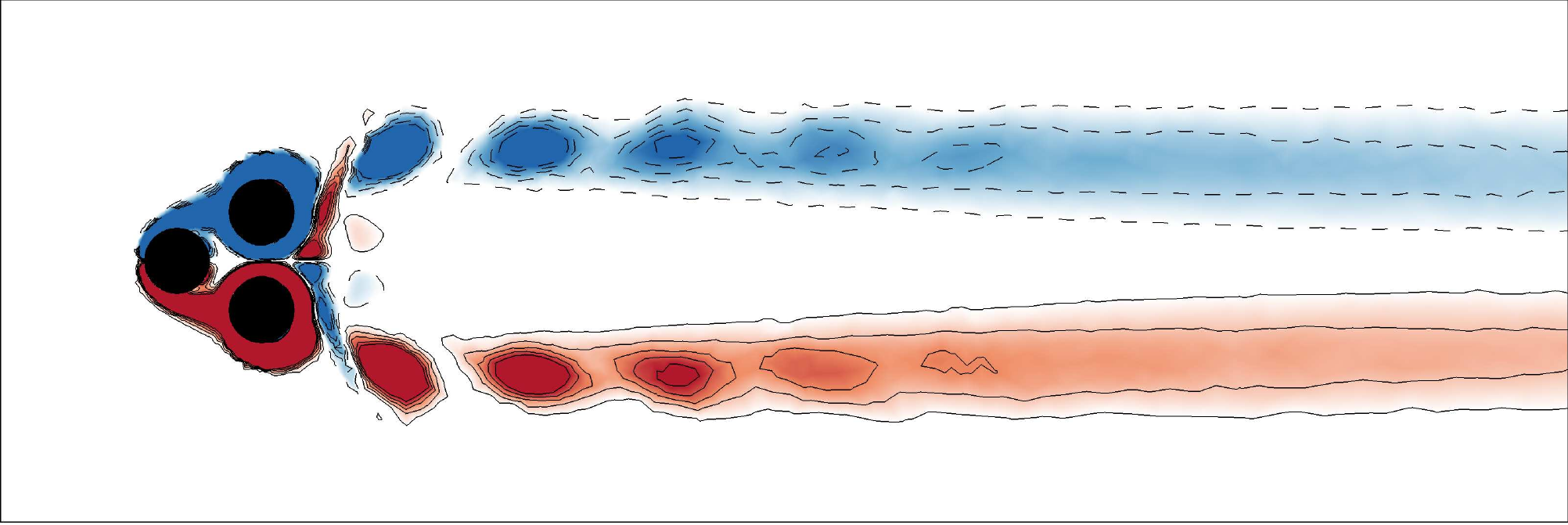}
\caption{\label{fig:EGM_Perio_T5}$t+5T_6/8$}
\end{subfigure}%
\hspace{0.5cm}
\begin{subfigure}{.45\textwidth}
  \centering
\includegraphics[width=1.\textwidth]{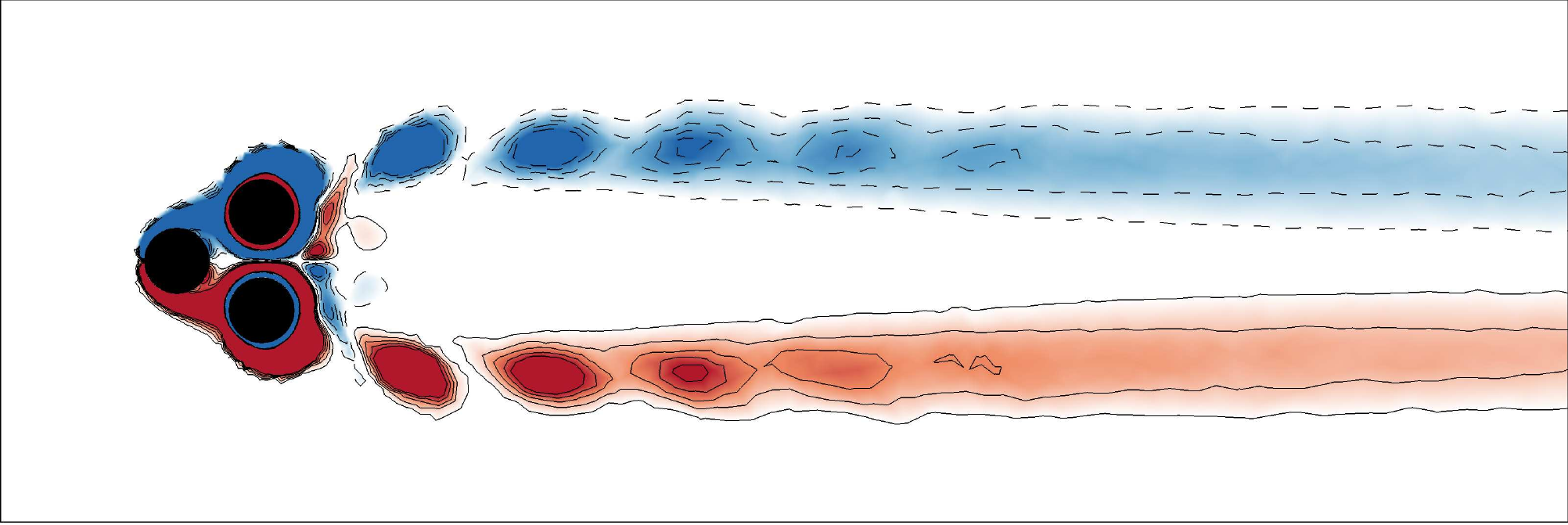}
\caption{\label{fig:EGM_Perio_T6}$t+6T_6/8$}
\end{subfigure}

\begin{subfigure}{.45\textwidth}
  \centering
\includegraphics[width=1.\textwidth]{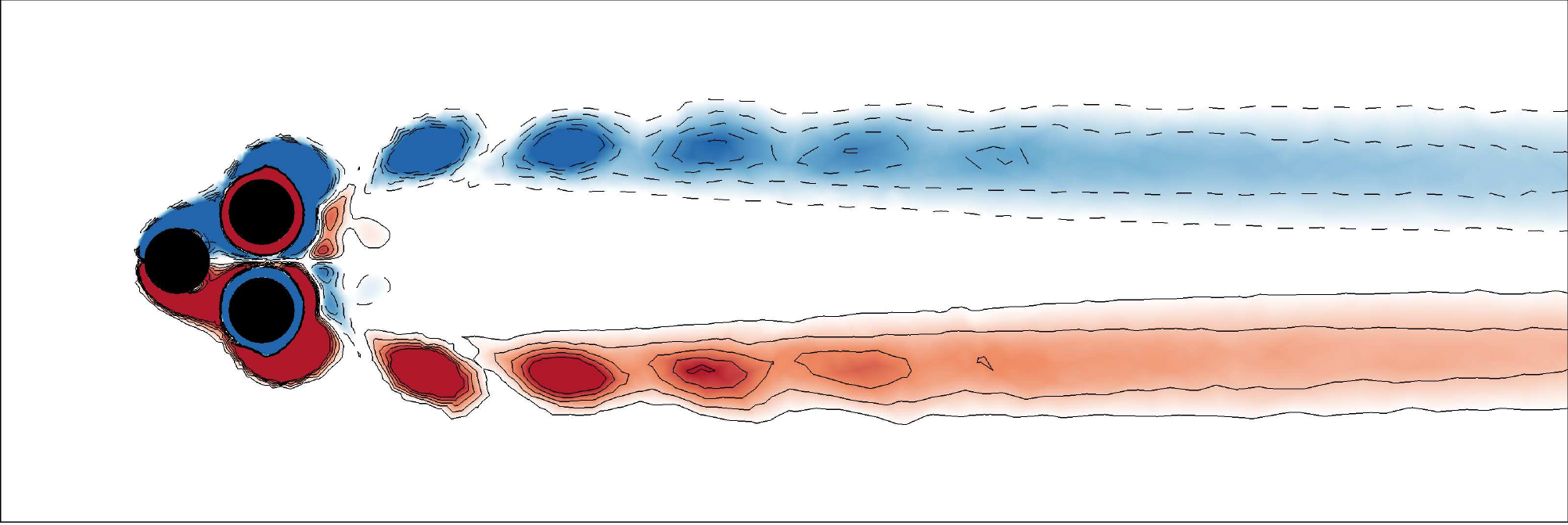}
\caption{\label{fig:EGM_Perio_T7}$t+7T_6/8$}
\end{subfigure}%
\hspace{0.5cm}
\begin{subfigure}{.45\textwidth}
  \centering
\includegraphics[width=1.\textwidth]{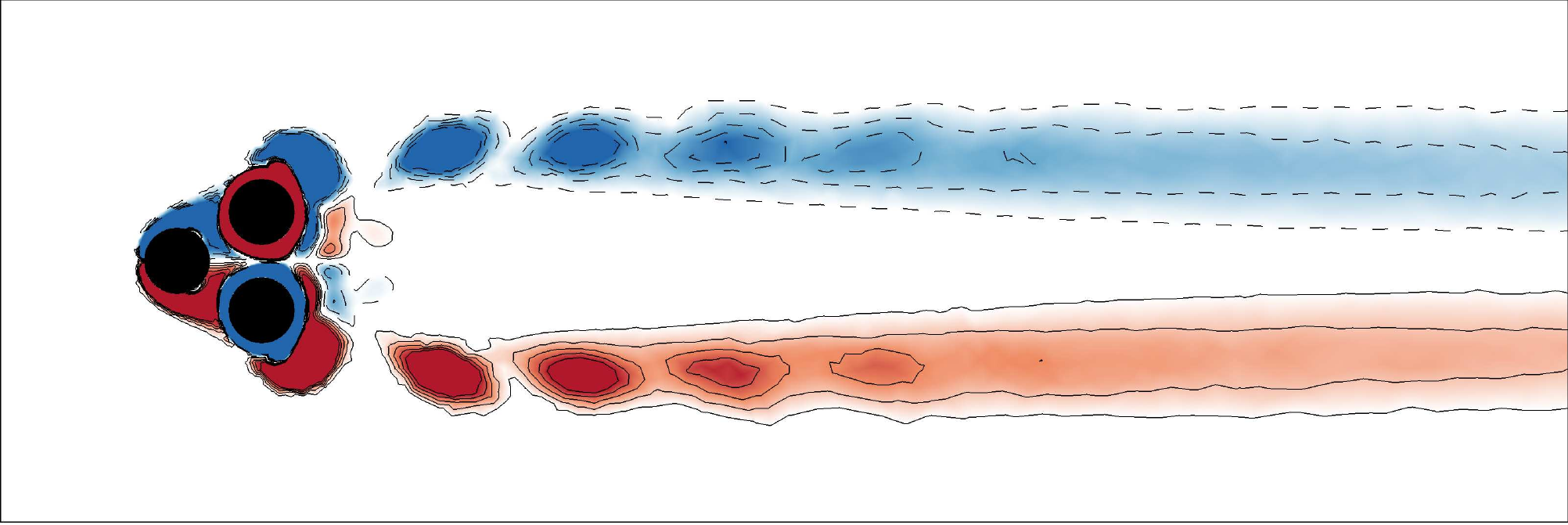}
\caption{\label{fig:EGM_Perio_T8}$t+T_6$}
\end{subfigure}

\begin{subfigure}{.45\textwidth}
  \centering
\includegraphics[width=1.\textwidth]{Figures/Snapshots/SteadySolution}
\caption{Symmetric steady solution}
\end{subfigure}%
\hspace{0.5cm}
\begin{subfigure}{.45\textwidth}
  \centering
\includegraphics[width=1.\textwidth]{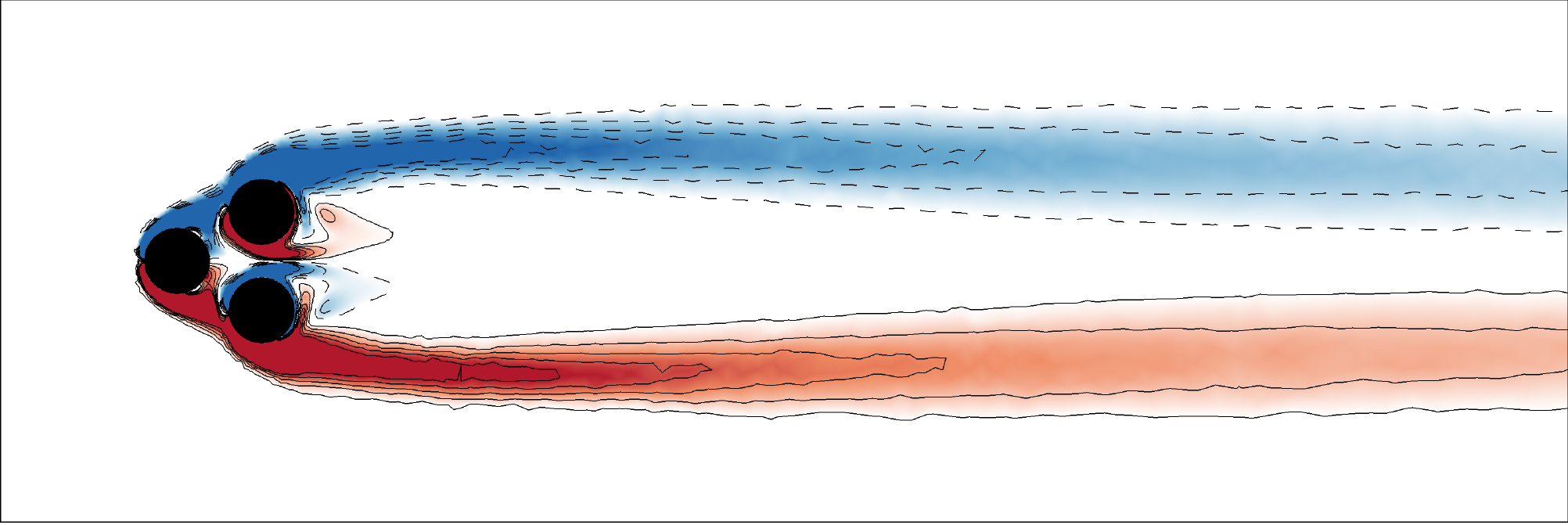}
\caption{\label{fig:Mean_EGM_Perio}Mean field}
\end{subfigure}
\caption{\label{fig:EGM_Periosnap}
Vorticity fields of the best periodic forcing found with EGM. (a)-(f) Time evolution of the vorticity field throughout the last period of the 1400 convective time units, (i) the objective symmetric steady solution and (j) the mean field of the forced flow.
The color code is the same as figure~\ref{fig:unforced_flow}.
$T_6$ is the period associated to the frequency $f_6$.
The mean field is computed by averaging 200 periods.}
\end{figure}

%% Pareto diagram ------------------------------------------------------
%\begin{figure}
%  \centerline{\includegraphics[width=0.75\linewidth]{Figures//Perio/PeriodicParetoDiagram}}
%  \caption{Pareto diagram for the periodic forcing }
%\label{fig:ParetoPeriodic}
%\end{figure}
Figure~\ref{fig:Perioprog} shows the progression of the best individual throughout the evaluations.
The EGM optimization process converges after few tests  
as $J_a/J_0$, the amplitude and the frequency reach asymptotic values, without any significant improvement afterwards.
The parameters of the best symmetric periodic forcing are denoted by the superscript `EGM' and read
\begin{equation*}
\begin{array}{cc}
B^{\rm EGM} = 3.51,\\
F^{\rm EGM}/f_0 = 3.19.
\end{array}
\end{equation*}
%% Progression
%-----------------------------------------------------------------------
The proximity between the initial values and the aimed minimum certainly accelerates the observed convergences.
Figure~\ref{fig:EGM_Perio_characteristics} shows the evolution of the lift coefficient, the phase portrait, the power spectral density and the instantaneous cost function $j_a$ for the controlled flow.
The lift coefficient presents rather symmetric low amplitude oscillations, see figure~\ref{fig:CL_Perio}.
This goes along with the flow symmetry in figure~\ref{fig:EGM_Perio_T1}-\ref{fig:EGM_Perio_T8}.
The oscillations are explained by the remaining vortex shedding on both, the upper and lower side of the fluidic pinball.
Even though the far field is close to the symmetric steady solution, this periodic solution changes radically the near field profile.
The jet is completely flattened. 
We can identify parts of the two vorticity branches close to the cylinders in figure~\ref{fig:EGM_Perio_T3}, \ref{fig:EGM_Perio_T4} and \ref{fig:EGM_Perio_T5}.
Moreover, the vorticity around the cylinders is more intense compared to the initial steady solution.
This difference is present in the final mean value $j_a$ in figure~\ref{fig:DSS_Perio} and is responsible for the high actuation power expense, $J_b=5.2799$.
The phase portrait shows a periodic regime, though deformed by the harmonics.
The mean frequency $f_6=0.398$ is slightly lower than the forcing frequency $F^{\rm EGM} = 0.37$ and much lower than the natural frequency, showing that it is not just a simple frequency locking, but a nonlinear frequency crosstalk.
The non-centered phase portrait indicates that there is still an asymmetry in the flow, that may be a residual effect of the grid's asymmetry.
The mean field in figure~\ref{fig:Mean_EGM_Perio} is similar to the symmetric steady solution, however the jet completely vanishes.
In addition, the distance between the upper and lower vorticity branches is wider compared to the symmetric steady solution.
%% Figures : EGM Periodic flow characteristics -------------------------
%-----------------------------------------------------------------------

% % gMLC flow snapshots ------------------------------------------------
 % High cost solution \label{appA}
%\input{SC} % Open-cavity experiment

\bibliographystyle{jfm}
% Note the spaces between the initials
\bibliography{Main,Main_Bernd,BIB_Cavity}

\begin{thebibliography}{64}
\expandafter\ifx\csname natexlab\endcsname\relax\def\natexlab#1{#1}\fi
\def\au#1{#1} \def\ed#1{#1} \def\yr#1{#1}\def\at#1{#1}\def\jt#1{\textit{#1}}
  \def\bt#1{#1}\def\bvol#1{\textbf{#1}} \def\vol#1{#1} \def\pg#1{#1}
  \def\publ#1{#1}\def\arxiv#1{#1}\def\org#1{#1}\def\st#1{\textit{#1}}

\bibitem[Abu-Mostafa {\em et~al.\/}(2012)Abu-Mostafa, Magndon-Ismail \&
  Lin]{AbuMostafa2012book}
{\sc \au{Abu-Mostafa, Y.~S.}, \au{Magndon-Ismail, M.} \& \au{Lin, H.-T.}}
  \yr{2012} {\em Learning from Data. A Short Course\/}.  \publ{AMLBook}.

\bibitem[Asai {\em et~al.\/}(2019)Asai, Yamato, Sunada \& Rinoie]{Asai2019aiaa}
{\sc \au{Asai, S.}, \au{Yamato, H.}, \au{Sunada, Y.} \& \au{Rinoie, K.}}
  \yr{2019} Designing machine learning control law of dynamic bubble burst
  control plate for stall suppression.  \bt{In {\em 2019 AIAA SciTech
  Forum\/}}. San Diego, CA.

\bibitem[Barros {\em et~al.\/}(2016)Barros, Bor\'ee, Noack, Spohn \&
  Ruiz]{Barros2016jfm}
{\sc \au{Barros, D.}, \au{Bor\'ee, J.}, \au{Noack, B.~R.}, \au{Spohn, A.} \&
  \au{Ruiz, T.}} \yr{2016}  \at{Bluff body drag manipulation using pulsed jets
  and {C}oanda effect}.  \jt{J.\ Fluid Mech.}  \bvol{805},  \pg{442--459}.

\bibitem[Bearman(1967)]{Bearman1967aq}
{\sc \au{Bearman, P.~W.}} \yr{1967}  \at{The effect of base bleed on the flow
  behind a two-dimensional model with a blunt trailing edge}.  \jt{Aeronautical
  Quarterly}  \bvol{18}~(03),  \pg{207--224}.

\bibitem[Belus {\em et~al.\/}(2019)Belus, Rabault, Viquerat, Che, Hachem \&
  Reglade]{Belus2019}
{\sc \au{Belus, V.}, \au{Rabault, J.}, \au{Viquerat, J.}, \au{Che, Z.},
  \au{Hachem, E.} \& \au{Reglade, U.}} \yr{2019}  \at{Exploiting locality and
  translational invariance to design effective deep reinforcement learning
  control of the 1-dimensional unstable falling liquid film}.  \jt{AIP
  Advances}  \bvol{9}~(12),  \pg{125014},  \arxiv{arXiv:
  https://doi.org/10.1063/1.5132378}.

\bibitem[Benard {\em et~al.\/}(2016)Benard, Pons-Prats, Periaux, Bugeda, Braud,
  Bonnet \& Moreau]{Benard2016ef}
{\sc \au{Benard, N.}, \au{Pons-Prats, J.}, \au{Periaux, J.}, \au{Bugeda, G.},
  \au{Braud, P.}, \au{Bonnet, J.P.} \& \au{Moreau, E.}} \yr{2016}
  \at{Turbulent separated shear flow control by surface plasma actuator:
  experimental optimization by genetic algorithm approach}.  \jt{Exp.~Fluids}
  \bvol{57}~(2),  \pg{22:1--17}.

\bibitem[Brameier \& Banzhaf(2006)]{Brameier2006book}
{\sc \au{Brameier, M.} \& \au{Banzhaf, W.}} \yr{2006} {\em Linear {G}enetic
  {P}rogramming\/}.  \publ{Springer Science \& Business Media}.

\bibitem[Brunton \& Noack(2015)]{Brunton2015amr}
{\sc \au{Brunton, S.~L.} \& \au{Noack, B.~R.}} \yr{2015}  \at{Closed-loop
  turbulence control: {P}rogress and challenges}.  \jt{Appl.\ Mech.\ Rev.}
  \bvol{67}~(5),  \pg{050801:01--48}.

\bibitem[Cattafesta \& Shelpak(2011)]{Cattafesta2011arfm}
{\sc \au{Cattafesta, L.} \& \au{Shelpak, M.}} \yr{2011}  \at{Actuators for
  active flow control}.  \jt{Ann.\ Rev.\ Fluid Mech.}  \bvol{43},
  \pg{247--272}.

\bibitem[Chen {\em et~al.\/}(2020)Chen, Ji, Alam, Williams \&
  Xu]{ChenAlam2020jfm}
{\sc \au{Chen, W.}, \au{Ji, C.}, \au{Alam, Md~M.}, \au{Williams, J.} \& \au{Xu,
  D.}} \yr{2020}  \at{Numerical simulations of flow past three circular
  cylinders in equilateral-triangular arrangements}.  \jt{Journal of Fluid
  Mechanics}  \bvol{891},  \pg{1--44}.

\bibitem[Choi {\em et~al.\/}(2008)Choi, Jeon \& Kim]{Choi2008arfm}
{\sc \au{Choi, H.}, \au{Jeon, W.-P.} \& \au{Kim, J.}} \yr{2008}  \at{Control of
  flow over a bluff body}.  \jt{Ann.\ Rev.\ Fluid Mech.}  \bvol{40},
  \pg{113--139}.

\bibitem[Choi {\em et~al.\/}(1994)Choi, Moin \& Kim]{Choi1994jfm}
{\sc \au{Choi, H.}, \au{Moin, P.} \& \au{Kim, J.}} \yr{1994}  \at{Active
  turbulence control for drag reduction in wall-bounded flows}.  \jt{J.\ Fluid
  Mech.}  \bvol{262},  \pg{75--110}.

\bibitem[Cornejo~Maceda(2021)]{cornejomacedaPhD}
{\sc \au{Cornejo~Maceda, G.~Y.}} \yr{2021}  \at{Gradient-enriched machine
  learning control exemplified for shear flows in simulations and experiments}.
  PhD thesis, Universit\'e Paris-Saclay.

\bibitem[{Cornejo Maceda} {\em et~al.\/}(2019){Cornejo Maceda}, R., Lusseyran,
  Deng, Pastur \& Morzy\'nski]{Cornejo2019pamm}
{\sc \au{{Cornejo Maceda}, G.~Y.}, \au{R., Noack~B.}, \au{Lusseyran, F.},
  \au{Deng, N.}, \au{Pastur, L.} \& \au{Morzy\'nski, M.}} \yr{2019}
  \at{Artificial intelligence control applied to drag reduction of the fluidic
  pinball}.  \jt{Proc.\ Appl.\ Math.\ Mech.}  \bvol{19}~(1),
  \pg{e201900268:1--2}.

\bibitem[Cortelezzi {\em et~al.\/}(1994)Cortelezzi, Leonard \&
  Doyle]{Cortelezzi1994jfm}
{\sc \au{Cortelezzi, L.}, \au{Leonard, A.} \& \au{Doyle, J.C.}} \yr{1994}
  \at{An example of active circulation control of the unsteady separated flow
  past a semi-infinite plate}.  \jt{J.\ Fluid Mech.}  \bvol{260},
  \pg{127--154}.

\bibitem[Debien {\em et~al.\/}(2016)Debien, von Krbek, Mazellier, Duriez,
  Cordier, Noack, Abel \& Kourta]{Debien2016ef}
{\sc \au{Debien, A.}, \au{von Krbek, K.~A.~F.~F.}, \au{Mazellier, N.},
  \au{Duriez, T.}, \au{Cordier, L.}, \au{Noack, B.~R.}, \au{Abel, M.~W.} \&
  \au{Kourta, A.}} \yr{2016}  \at{Closed-loop separation control over a
  sharp-edge ramp using genetic programming}.  \jt{Exp.\ Fluids}
  \bvol{57}~(3),  \pg{40:1--19}.

\bibitem[Deng {\em et~al.\/}(2020)Deng, Noack, Morzyński \&
  Pastur]{Deng2020jfm}
{\sc \au{Deng, N.}, \au{Noack, B.~R.}, \au{Morzyński, M.} \& \au{Pastur,
  L.~R.}} \yr{2020}  \at{Low-order model for successive bifurcations of the
  fluidic pinball}.  \jt{J.\ Fluid Mech.}  \bvol{884},  \pg{A37}.

\bibitem[Dowling \& Morgans(2005)]{Dowling2005arfm}
{\sc \au{Dowling, A.~P.} \& \au{Morgans, A.~S.}} \yr{2005}  \at{Feedback
  control of combustion oscillations}.  \jt{Annual Review of Fluid Mechanics}
  \bvol{37}~(151--182).

\bibitem[Dracopoulos(1997)]{Dracopoulos1997book}
{\sc \au{Dracopoulos, D.~C.}} \yr{1997} {\em Evolutionary Learning Algorithms
  for Neural Adaptive Control\/}.  \publ{London, etc.: Springer-Verlag}.

\bibitem[Duriez {\em et~al.\/}(2016)Duriez, Brunton \& Noack]{Duriez2016book}
{\sc \au{Duriez, T.}, \au{Brunton, S.~L.} \& \au{Noack, B.~R.}} \yr{2016} {\em
  Machine Learning Control --- Taming Nonlinear Dynamics and Turbulence\/},
  \st{Fluid Mech. its Appl.},  \vol{vol. 116}.  \publ{Springer-Verlag}.

\bibitem[Fan {\em et~al.\/}(2020)Fan, Yang, Wang, Triantafyllou \&
  Karniadakis]{Fan2020pnas}
{\sc \au{Fan, S.~L.}, \au{Yang, L.}, \au{Wang, Z.~C.}, \au{Triantafyllou,
  M.~S.} \& \au{Karniadakis, G.~M.}} \yr{2020}  \at{Reinforcement learning for
  bluff body active flow control in experiments and simulations}.  \jt{Proc.\
  Natl.\ Acad. Sci. USA}  \bvol{117}~(42),  \pg{26091--26098}.

\bibitem[Fernex {\em et~al.\/}(2020)Fernex, Semann, Albers, Meysonnat,
  Schr\"oder \& Noack]{Fernex2020prf}
{\sc \au{Fernex, D.}, \au{Semann, R.}, \au{Albers, M.}, \au{Meysonnat, P.~S},
  \au{Schr\"oder, W.} \& \au{Noack, B.~R.}} \yr{2020}  \at{Self-similar drag
  reduction formula from sparse data---{O}ptimization of turbulent
  skin-friction via spanwise travelling surface waves}.  \jt{Phys. Rev. Fluids}
   \bvol{5}~(7),  \pg{073901:1--18}.

\bibitem[Fl{\"u}gel(1930)]{Fluegel1930stg}
{\sc \au{Fl{\"u}gel, G.}} \yr{1930}  \at{Ergebnisse aus dem
  {S}tr{\"o}mungsinstitut der {T}echnischen {H}ochschule {D}anzig}.  \bt{In
  {\em Jahrbuch der Schiffbautechnischen Gesellschaft\/}},  \pg{pp. 87--113}.
  \publ{Springer}.

\bibitem[Fukagata \& Nobuhide(2003)]{Fukagata2003}
{\sc \au{Fukagata, K.} \& \au{Nobuhide, K.}} \yr{2003}  \at{Drag reduction in
  turbulent pipe flow with feedback control applied partially to wall}.
  \jt{Int. J. Heat Fluid Flow}  \bvol{24},  \pg{480--490}.

\bibitem[Gautier {\em et~al.\/}(2015)Gautier, Aider, Duriez, Noack, Segond \&
  Abel]{Gautier2015jfm}
{\sc \au{Gautier, N.}, \au{Aider, J.-L.}, \au{Duriez, T.}, \au{Noack, B.~R.},
  \au{Segond, M.} \& \au{Abel, M.~W.}} \yr{2015}  \at{Closed-loop separation
  control using machine learning}.  \jt{J.\ Fluid Mech.}  \bvol{770},
  \pg{424--441}.

\bibitem[Gelbert {\em et~al.\/}(2012)Gelbert, Moeck, Paschereit \&
  King]{Gelbert2012}
{\sc \au{Gelbert, G.}, \au{Moeck, J.~P.}, \au{Paschereit, C.~O.} \& \au{King,
  R.}} \yr{2012}  \at{Advanced algorithms for gradient estimation in one-and
  two-parameter extremum seeking controllers}.  \jt{J. Process Control}
  \bvol{22}~(4),  \pg{700--709}.

\bibitem[Gerhard {\em et~al.\/}(2003)Gerhard, Pastoor, King, Noack, Dillmann,
  Morzy\'nski \& Tadmor]{Gerhard2003aiaa}
{\sc \au{Gerhard, J.}, \au{Pastoor, M.}, \au{King, R.}, \au{Noack, B.~R.},
  \au{Dillmann, A.}, \au{Morzy\'nski, M.} \& \au{Tadmor, G.}} \yr{2003}
  Model-based control of vortex shedding using low-dimensional {G}alerkin
  models.  \bt{In {\em 33rd AIAA Fluids Conference and Exhibit\/}}. Orlando,
  Florida, USA, June 23--26, 2003, paper 2003-4262.

\bibitem[Geropp(1995)]{Geropp1995patent}
{\sc \au{Geropp, D.}} \yr{1995} Process and device for reducing the drag in the
  rear region of a vehicle, for example, a road or rail vehicle or the like.
  United States Patent {\bfseries US$\>$5407245$\>$A}.

\bibitem[Geropp \& Odenthal(2000)]{Geropp2000ef}
{\sc \au{Geropp, D.} \& \au{Odenthal, H.-J.}} \yr{2000}  \at{Drag reduction of
  motor vehicles by active flow control using the {C}oanda effect}.
  \jt{Exp.~Fluids}  \bvol{28}~(1),  \pg{74--85}.

\bibitem[Glezer {\em et~al.\/}(2005)Glezer, Amitay \& Honohan]{Glezer2005aiaaj}
{\sc \au{Glezer, A.}, \au{Amitay, M.} \& \au{Honohan, A.M.}} \yr{2005}
  \at{Aspects of low- and high-frequency actuation for aerodynamic flow
  control}.  \jt{AIAA Journal}  \bvol{43}~(7),  \pg{1501--1511}.

\bibitem[Herv{\'e} {\em et~al.\/}(2012)Herv{\'e}, Sipp, Schmid \&
  Samuelides]{Herve2012jfm}
{\sc \au{Herv{\'e}, A.}, \au{Sipp, D.}, \au{Schmid, P.~J.} \& \au{Samuelides,
  M.}} \yr{2012}  \at{A physics-based approach to flow control using system
  identification}.  \jt{J.\ Fluid Mech.}  \bvol{702},  \pg{26--58}.

\bibitem[Ishar {\em et~al.\/}(2019)Ishar, Kaiser, Morzynski, Albers, Meysonnat,
  Schr\"oder \& Noack]{Ishar2019jfm}
{\sc \au{Ishar, R.}, \au{Kaiser, E.}, \au{Morzynski, M.}, \au{Albers, M.},
  \au{Meysonnat, P.}, \au{Schr\"oder, W.} \& \au{Noack, B.~R.}} \yr{2019}
  \at{Metric for attractor overlap}.  \jt{J.~Fluid Mech.}  \bvol{874},
  \pg{720--752}.

\bibitem[Jordan \& Colonius(2013)]{Jordan2013arfm}
{\sc \au{Jordan, P.} \& \au{Colonius, T.}} \yr{2013}  \at{Wave packets and
  turbulent jet noise}.  \jt{Ann.\ Rev.\ Fluid Mech.}  \bvol{45},
  \pg{173--195}.

\bibitem[Koumoutsakos {\em et~al.\/}(2001)Koumoutsakos, Freund \&
  Parekh]{Koumoutsakos2001aiaaj}
{\sc \au{Koumoutsakos, P.}, \au{Freund, J.} \& \au{Parekh, D.}} \yr{2001}
  \at{Evolution strategies for automatic optimization of jet mixing}.  \jt{AIAA
  J.}  \bvol{39}~(5),  \pg{967--969}.

\bibitem[Li {\em et~al.\/}(2020{\natexlab{{\em a\/}}})Li, Fernex, Tan,
  Morzy\'nski \& Noack]{LiH2020jfm}
{\sc \au{Li, H.}, \au{Fernex, D.}, \au{Tan, J.}, \au{Morzy\'nski, M.} \&
  \au{Noack, B.~R.}} \yr{2020{\natexlab{{\em a\/}}}}  \at{Cluster-based network
  model of an incompressible mixing layer}.  \jt{J. Fluid Mech.}
  \bvol{submitted, http://arxiv.org/abs/2001.02911}.

\bibitem[Li {\em et~al.\/}(2019)Li, Bor{\'e}e, Noack, Cordier \&
  Harambat]{Li2019prf}
{\sc \au{Li, R.}, \au{Bor{\'e}e, J.}, \au{Noack, B.~R.}, \au{Cordier, L.} \&
  \au{Harambat, F.}} \yr{2019}  \at{Drag reduction mechanisms of a car model at
  moderate yaw by bi-frequency forcing}.  \jt{Phys.\ Rev.\ Fluids}
  \bvol{4}~(3),  \pg{034604}.

\bibitem[Li {\em et~al.\/}(2018)Li, Noack, Cordier, Bor\'ee, Kaiser \&
  Harambat]{Li2018am}
{\sc \au{Li, R.}, \au{Noack, B.~R.}, \au{Cordier, L.}, \au{Bor\'ee, J.},
  \au{Kaiser, E.} \& \au{Harambat, F.}} \yr{2018}  \at{Linear genetic
  programming control for strongly nonlinear dynamics with frequency
  crosstalk}.  \jt{Archives of Mechanics}  \bvol{70}~(6),  \pg{505--534}.

\bibitem[Li {\em et~al.\/}(2020{\natexlab{{\em b\/}}})Li, Cui, Jia, Li, Yang,
  Morzy\'nski \& Noack]{LiA2020jfm}
{\sc \au{Li, Y.}, \au{Cui, W.}, \au{Jia, Q.}, \au{Li, Q.}, \au{Yang, Z.},
  \au{Morzy\'nski, M.} \& \au{Noack, B.~R.}} \yr{2020{\natexlab{{\em b\/}}}}
  \at{Explorative gradient method for active drag reduction of the fluidic
  pinball and slanted ahmed body}.  \jt{J. Fluid Mech.}  \bvol{(in revision,
  see arXiv)}.

\bibitem[Luchtenburg {\em et~al.\/}(2009)Luchtenburg, G\"unter, Noack, King \&
  Tadmor]{Luchtenburg2009jfm}
{\sc \au{Luchtenburg, D.~M.}, \au{G\"unter, B.}, \au{Noack, B.~R.}, \au{King,
  R.} \& \au{Tadmor, G.}} \yr{2009}  \at{A generalized mean-field model of the
  natural and actuated flows around a high-lift configuration}.  \jt{J.\ Fluid
  Mech.}  \bvol{623},  \pg{283--316}.

\bibitem[Maehara \& Shimoda(2013)]{Maehara2013}
{\sc \au{Maehara, N.} \& \au{Shimoda, Y.}} \yr{2013}  \at{Application of the
  genetic algorithm and downhill simplex methods (nelder–mead methods) in the
  search for the optimum chiller configuration}.  \jt{Applied Thermal
  Engineering}  \bvol{61}~(2),  \pg{433 -- 442}.

\bibitem[McKay {\em et~al.\/}(1979)McKay, Beckman \& Conover]{McKay1979}
{\sc \au{McKay, M.~D.}, \au{Beckman, R.~J.} \& \au{Conover, W.~J.}} \yr{1979}
  \at{A comparison of three methods for selecting values of input variables in
  the analysis of output from a computer code}.  \jt{Technometrics}
  \bvol{21}~(2),  \pg{239--245}.

\bibitem[Nair {\em et~al.\/}(2019)Nair, Yeh, Kaiser, Noack, Brunton \&
  Tiara]{Nair2019jfm}
{\sc \au{Nair, A.}, \au{Yeh, C.-A.}, \au{Kaiser, E.}, \au{Noack, B.~R.},
  \au{Brunton, S.~L.} \& \au{Tiara, K.}} \yr{2019}  \at{Cluster-based feedback
  control of turbulent post-stall separated flows}.  \jt{J.~Fluid Mech.}
  \bvol{875},  \pg{345--375}.

\bibitem[Nelder \& Mead(1965)]{Nelder1965}
{\sc \au{Nelder, J.~A.} \& \au{Mead, R}} \yr{1965}  \at{A simplex method for
  function minimization}.  \jt{Computer Journal}  \bvol{7}~(4),  \pg{308--313}.

\bibitem[Noack(2019)]{Noack2019springer}
{\sc \au{Noack, B.~R.}} \yr{2019} Closed-loop turbulence control---{F}rom human
  to machine learning (and retour).  \bt{In {\em Proceedings of the 4th
  Symposium on Fluid Structure-Sound Interactions and Control (FSSIC), Tokyo,
  Japan\/} (ed. \ed{Y.~Zhou, M.~Kimura, G.~Peng, A.~D. Lucey \& L.~Hung})},
  \pg{pp. 23--32}.  \publ{Springer}.

\bibitem[Noack {\em et~al.\/}(2003)Noack, Afanasiev, Morzy\'nski, Tadmor \&
  Thiele]{Noack2003jfm}
{\sc \au{Noack, B.~R.}, \au{Afanasiev, K.}, \au{Morzy\'nski, M.}, \au{Tadmor,
  G.} \& \au{Thiele, F.}} \yr{2003}  \at{A hierarchy of low-dimensional models
  for the transient and post-transient cylinder wake}.  \jt{J.\ Fluid Mech.}
  \bvol{497},  \pg{335--363}.

\bibitem[Noack {\em et~al.\/}(2016)Noack, Stankiewicz, Morzy\'nski \&
  Schmid]{Noack2016jfm}
{\sc \au{Noack, B.~R.}, \au{Stankiewicz, W.}, \au{Morzy\'nski, M.} \&
  \au{Schmid, P.~J.}} \yr{2016}  \at{Recursive dynamic mode decomposition of
  transient and post-transient wake flows}.  \jt{J.~Fluid Mech.}  \bvol{809},
  \pg{843--872}.

\bibitem[Parezanovi\'c {\em et~al.\/}(2016)Parezanovi\'c, Cordier, Spohn,
  Duriez, Noack, Bonnet, Segond, Abel \& Brunton]{Parezanovic2016jfm}
{\sc \au{Parezanovi\'c, V.}, \au{Cordier, L.}, \au{Spohn, A.}, \au{Duriez, T.},
  \au{Noack, B.~R.}, \au{Bonnet, J.-P.}, \au{Segond, M.}, \au{Abel, M.} \&
  \au{Brunton, S.~L.}} \yr{2016}  \at{Frequency selection by feedback control
  in a turbulent shear flow}.  \jt{J.\ Fluid Mech.}  \bvol{797},
  \pg{247--283}.

\bibitem[Pastoor {\em et~al.\/}(2008)Pastoor, Henning, Noack, King \&
  Tadmor]{Pastoor2008jfm}
{\sc \au{Pastoor, M.}, \au{Henning, L.}, \au{Noack, B.~R.}, \au{King, R.} \&
  \au{Tadmor, G.}} \yr{2008}  \at{Feedback shear layer control for bluff body
  drag reduction}.  \jt{J.\ Fluid Mech.}  \bvol{608},  \pg{161--196}.

\bibitem[Pfeiffer \& King(2012)]{Pfeiffer2012aiaa}
{\sc \au{Pfeiffer, J.} \& \au{King, R.}} \yr{2012} Multivariable closed-loop
  flow control of drag and yaw moment for a 3d bluff body.  \bt{In {\em 6th
  AIAA Flow Control Conference\/}},  \pg{pp. 1--14}. Atlanta, Georgia, USA.

\bibitem[Protas(2004)]{Protas2004pf}
{\sc \au{Protas, B.}} \yr{2004}  \at{Linear feedback stabilization of laminar
  vortex shedding based on a point vortex model}.  \jt{Phys.\ Fluids}
  \bvol{16}~(12),  \pg{4473--4488}.

\bibitem[Rabault {\em et~al.\/}(2019)Rabault, Kuchta, Jensen, R{\'e}glade \&
  Cerardi]{Rabault2019jfm}
{\sc \au{Rabault, J.}, \au{Kuchta, M.}, \au{Jensen, A.}, \au{R{\'e}glade, U.}
  \& \au{Cerardi, N.}} \yr{2019}  \at{Artificial neural networks trained
  through deep reinforcement learning discover control strategies for active
  flow control}.  \jt{J. Fluid Mech.}  \bvol{865},  \pg{281--302}.

\bibitem[Raibaudo {\em et~al.\/}(2019)Raibaudo, Zhong, Noack \&
  Martinuzzi]{Raibaudo2020pf}
{\sc \au{Raibaudo, C.}, \au{Zhong, P.}, \au{Noack, B.~R.} \& \au{Martinuzzi,
  R.~J.}} \yr{2019}  \at{Machine learning strategies applied to the control of
  a fluidic pinball}.  \jt{Phys. Fluids}  \bvol{32},  \pg{015108}.

\bibitem[Ren {\em et~al.\/}(2020)Ren, Hu \& Tang]{Ren2020jh}
{\sc \au{Ren, F.}, \au{Hu, H.-B.} \& \au{Tang, H.}} \yr{2020}  \at{Active flow
  control using machine learning: A brief review}.  \jt{J. Hydrodyn.}
  \bvol{32}~(2),  \pg{247--253}.

\bibitem[Ren {\em et~al.\/}(2019)Ren, Wang \& Tang]{Ren2019pof}
{\sc \au{Ren, F.}, \au{Wang, C.} \& \au{Tang, H.}} \yr{2019}  \at{Active
  control of vortex-induced vibration of a circular cylinder using machine
  learning}.  \jt{Physics of Fluids}  \bvol{31}~(9),  \pg{093601},
  \arxiv{arXiv: https://doi.org/10.1063/1.5115258}.

\bibitem[Roussopoulos(1993)]{Roussopoulos1993jfm}
{\sc \au{Roussopoulos, K.}} \yr{1993}  \at{Feedback control of vortex shedding
  at low {R}eynolds numbers}.  \jt{J.\ Fluid Mech.}  \bvol{248},
  \pg{267--296}.

\bibitem[Rowan(1990)]{Rowan1990phd}
{\sc \au{Rowan, T.}} \yr{1990}  \at{The subplex method for unconstrained
  optimization}. PhD thesis, PhD thesis, Department of Computer Sciences,
  University of Texas.

\bibitem[Rowley \& Williams(2006)]{Rowley2006arfm}
{\sc \au{Rowley, C.~W.} \& \au{Williams, D.~R.}} \yr{2006}  \at{Dynamics and
  control of high-{R}eynolds number flows over open cavities}.  \jt{Ann.\ Rev.\
  Fluid Mech.}  \bvol{38},  \pg{251--276}.

\bibitem[Seifert(2012)]{Seifert2012review}
{\sc \au{Seifert, J.}} \yr{2012}  \at{A review of the {M}agnus effect in
  aeronautics}.  \jt{Prog. Aerosp. Sci.}  \bvol{55},  \pg{17--45}.

\bibitem[Semaan {\em et~al.\/}(2016)Semaan, Kumar, Burnazzi, Tissot, Cordier \&
  Noack]{Semaan2016jfm}
{\sc \au{Semaan, R.}, \au{Kumar, P.}, \au{Burnazzi, M.}, \au{Tissot, G.},
  \au{Cordier, L.} \& \au{Noack, B.~R.}} \yr{2016}  \at{Reduced-order modeling
  of the flow around a high-lift configuration with unsteady {C}oanda blowing}.
   \jt{J.~Fluid Mech.}  \bvol{800},  \pg{71--110}.

\bibitem[Tang {\em et~al.\/}(2020)Tang, Rabault, Kuhnle, Wang \&
  Wang]{Tang2020}
{\sc \au{Tang, H.}, \au{Rabault, J.}, \au{Kuhnle, A.}, \au{Wang, Y.} \&
  \au{Wang, T.}} \yr{2020}  \at{Robust active flow control over a range of
  reynolds numbers using an artificial neural network trained through deep
  reinforcement learning}.  \jt{Physics of Fluids}  \bvol{32}~(5),
  \pg{053605},  \arxiv{arXiv: https://doi.org/10.1063/5.0006492}.

\bibitem[Thiria {\em et~al.\/}(2006)Thiria, Goujon-Durand \&
  Wesfreid]{Thiria2006jfm}
{\sc \au{Thiria, B.}, \au{Goujon-Durand, S.} \& \au{Wesfreid, J.~E.}} \yr{2006}
   \at{The wake of a cylinder performing rotary oscillations}.  \jt{J.\ Fluid
  Mech.}  \bvol{560},  \pg{123--147}.

\bibitem[Wood(1964)]{Wood1964jras}
{\sc \au{Wood, C.~J.}} \yr{1964}  \at{The effect of base bleed on a periodic
  wake}.  \jt{J. R. Soc. Interface}  \bvol{68}~(643),  \pg{477--482}.

\bibitem[Wu {\em et~al.\/}(2018)Wu, Fan, Zhou, Li \& Noack]{Wu2018ef}
{\sc \au{Wu, Z.}, \au{Fan, D.}, \au{Zhou, Y.}, \au{Li, R.} \& \au{Noack,
  B.~R.}} \yr{2018}  \at{Jet mixing enhancement using machine learning
  control}.  \jt{Exp.~Fluids}  \bvol{59},  \pg{131:1--17}.

\bibitem[Zhou {\em et~al.\/}(2020)Zhou, D., Zhang, , Li \& Noack]{Zhou2020jfm}
{\sc \au{Zhou, Y.}, \au{D., Fan}, \au{Zhang, B.}, , \au{Li, R.} \& \au{Noack,
  B.~R.}} \yr{2020}  \at{Artificial intelligence control of a turbulent jet}.
  \jt{J. Fluid Mech.}  \bvol{897},  \pg{1--46}.

\end{thebibliography}

\end{document}